\documentclass{article}
\usepackage[a4paper, centering, margin=2cm]{geometry}
\usepackage{amsmath,algorithmic,graphicx,charter}

\newtheorem{Definition}{Definition}
\newtheorem{validity test}{Validity Test}
\newtheorem{Lemma}{Lemma} 
\newtheorem{proof}{Proof} 
\newtheorem{Corollary}{Corollary}
\newtheorem{Property}{Property}
\newtheorem{Example}{Example}
\newtheorem{KeyResult}{Key Result}
\newtheorem{Observation}{Observation}

\newcommand{\cpu}{\operatorname{CPU}}
\newcommand{\avl}{\operatorname{avl}}

\newcommand{\equals}{\stackrel{\mathrm{def}}{=}}

\newcommand{\dm}{\mbox{\textsf{DM}}} 
\newcommand{\edf}{\operatorname{\sc EDF}}
\newcommand{\edfmin}{\tiny{\operatorname{\sc EDF}}}

\newcommand{\cumulspeed}[1]{s(#1)}
\newcommand{\totalspeed}{\cumulspeed{1}}
\newcommand{\schedulable}{\operatorname{sched}}
\newcommand{\old}{\operatorname{old}}
\newcommand{\new}{\operatorname{new}}
\newcommand{\trans}{\operatorname{trans}}
\newcommand{\mode}{M}

\newcommand{\act}{\operatorname{active}}
\newcommand{\wcremjobs}[1]{{\cal J}^{\operatorname{wc}}_{#1}}
\newcommand{\idle}[1]{{\operatorname{idle}}_{#1}}
\newcommand{\maxidle}[1]{{\overline{\operatorname{idle}}}_{#1}}
\newcommand{\minidle}[1]{{\underline{\operatorname{idle}}}_{#1}}

\newcommand{\comp}[1]{\operatorname{comp}_{#1}}

\newcommand{\work}[1]{w_{#1}}
\newcommand{\procwork}[2]{\operatorname{Work}_{#1}^{#2}}
\newcommand{\enabled}{\operatorname{enabled}}
\newcommand{\disabled}{\operatorname{disabled}}
\newcommand{\MCR}{\operatorname{MCR}}
\newcommand{\maxmakespan}{\overline{\operatorname{ms}}}
\newcommand{\makespan}{\operatorname{ms}}

\newcommand{\minmakespanIdent}{\underline{\operatorname{ms}}^{\operatorname{ident}}}
\newcommand{\maxmakespanIdent}{\overline{\operatorname{ms}}^{\operatorname{ident}}}
\newcommand{\maxmakespanUnifZero}{\overline{\operatorname{ms}}^{\operatorname{unif}}_0}
\newcommand{\maxmakespanUnifOne}{\overline{\operatorname{ms}}^{\operatorname{unif}}_1}
\newcommand{\maxmakespanUnifTwo}{\overline{\operatorname{ms}}^{\operatorname{unif}}_2}
\newcommand{\maxmakespanUnifThree}{\overline{\operatorname{ms}}^{\operatorname{unif}}_3}
\newcommand{\maxmakespanUnifMin}{\overline{\operatorname{ms}}^{\operatorname{unif}}_{\operatorname{min}}}

\newcommand{\mkerrorUnifOne}{E^{\operatorname{unif}}_1}
\newcommand{\mkerrorUnifTwo}{E^{\operatorname{unif}}_2}
\newcommand{\mkerrorUnifThree}{E^{\operatorname{unif}}_3}
\newcommand{\mkerrorUnifMin}{E^{\operatorname{unif}}_{\operatorname{min}}}

\newcommand{\any}{\operatorname{any}}
\newcommand{\low}{\operatorname{low}}

\newcommand{\MSO}{\operatorname{MSO}}
\newcommand{\SMMSO}{\operatorname{SM-MSO}}
\newcommand{\AMMSO}{\operatorname{AM-MSO}}

\begin{document}
%
% paper title
% can use linebreaks \\ within to get better formatting as desired
\title{Global Scheduling of Multi-Mode Real-Time Applications upon Multiprocessor Platforms}
%
%
% author names and IEEE memberships
% note positions of commas and nonbreaking spaces ( ~ ) LaTeX will not break
% a structure at a ~ so this keeps an author's name from being broken across
% two lines.
% use \thanks{} to gain access to the first footnote area
% a separate \thanks must be used for each paragraph as LaTeX2e's \thanks
% was not built to handle multiple paragraphs
%

\author{Vincent~N\'elis \and  Patrick~Meumeu~Yomsi \and
          Bj{\"{o}}rn Andersson
         \and
         Jo\"el~Goossens
        }% <-this % stops a space
%\thanks{Vincent~Nelis is with the Department
%of Electrical and Computer Engineering, Georgia Institute of Technology, Atlanta,
%GA, 30332 USA e-mail: (see http://www.michaelshell.org/contact.html).}% <-this % stops a space
%\thanks{J. Doe and J. Doe are with Anonymous University.}% <-this % stops a space
%\thanks{Manuscript received April 19, 2005; revised January 11, 2007.}

% note the % following the last \IEEEmembership and also \thanks - 
% these prevent an unwanted space from occurring between the last author name
% and the end of the author line. i.e., if you had this:
% 
% \author{....lastname \thanks{...} \thanks{...} }
%                     ^------------^------------^----Do not want these spaces!
%
% a space would be appended to the last name and could cause every name on that
% line to be shifted left slightly. This is one of those "LaTeX things". For
% instance, "\textbf{A} \textbf{B}" will typeset as "A B" not "AB". To get
% "AB" then you have to do: "\textbf{A}\textbf{B}"
% \thanks is no different in this regard, so shield the last } of each \thanks
% that ends a line with a % and do not let a space in before the next \thanks.
% Spaces after \IEEEmembership other than the last one are OK (and needed) as
% you are supposed to have spaces between the names. For what it is worth,
% this is a minor point as most people would not even notice if the said evil
% space somehow managed to creep in.

% The paper headers
\markboth{IEEE Transactions on Industrial Informatics}
{Shell \MakeLowercase{\textit{et al.}}: Bare Demo of IEEEtran.cls for Journals}
% The only time the second header will appear is for the odd numbered pages
% after the title page when using the twoside option.
% 
% *** Note that you probably will NOT want to include the author's ***
% *** name in the headers of peer review papers.                   ***
% You can use \ifCLASSOPTIONpeerreview for conditional compilation here if
% you desire.

% If you want to put a publisher's ID mark on the page you can do it like
% this:
%\IEEEpubid{0000--0000/00\$00.00~\copyright~2007 IEEE}
% Remember, if you use this you must call \IEEEpubidadjcol in the second
% column for its text to clear the IEEEpubid mark.

% use for special paper notices
%\IEEEspecialpapernotice{(Invited Paper)}

\date{}

% make the title area
\maketitle

\begin{abstract}
\boldmath
Multi-mode real-time systems are those which support applications with different modes of operation, where each mode is characterized by a specific set of tasks. At run-time, such systems can, at any time, be requested to switch from its current operating mode to another mode (called ``new mode'') by replacing the current set of tasks with that of the new-mode. Thereby, ensuring that all the timing requirements are met not only requires that a schedulability test is performed on the tasks of each mode but also that \mbox{(i) a}~protocol for transitioning from one mode to another is specified and \mbox{(ii) a}~schedulability test for each transition is performed. We propose two distinct protocols that manage the mode transitions upon uniform and identical multiprocessor platforms at run-time, each specific to distinct task requirements. For each protocol, we formally establish schedulability analyses that indicate beforehand whether all the timing requirements will be met during any mode transition of the system. This is performed assuming both Fixed-Task-Priority and Fixed-Job-Priority schedulers. 
\end{abstract}
% IEEEtran.cls defaults to using nonbold math in the Abstract.
% This preserves the distinction between vectors and scalars. However,
% if the journal you are submitting to favors bold math in the abstract,
% then you can use LaTeX's standard command \boldmath at the very start
% of the abstract to achieve this. Many IEEE journals frown on math
% in the abstract anyway.

% Note that keywords are not normally used for peerreview papers.

\section{Introduction}
\label{sec:Introduction} 

Hard real-time systems require both functionally correct executions and \emph{results that are produced on time}. Control of the traffic (ground or air), control of engines, control of chemical and nuclear power plants are just some examples of such systems. Currently, numerous techniques exist that enable engineers to design real-time systems while guaranteeing that all the temporal requirements are met. These techniques generally model each functionality of the application by a \emph{recurrent} task, characterized by a computing requirement, a temporal deadline and an activation rate. Commonly, real-time applications are simply modeled by a single and finite set of such tasks. However, practical applications often exhibit multiple behaviors issued from several operating modes (e.g.,~an initialization mode, an emergency mode, a fault recovery mode, etc.), where each mode is characterized by its own set of functionalities, i.e., its set of tasks. During the execution of such \emph{multi-mode} real-time applications, switching from the current mode (called the \emph{old-mode}) to any other mode (called the \emph{new-mode}) requires to substitute the currently executing task set with the set of tasks of the new-mode. This substitution introduces a {\em transient phase}, where tasks of both the old- and new-mode may be scheduled \emph{simultaneously}, thereby leading to a possible overload that can compromise the system schedulability---indeed it can be the case that both the old- and new-mode have been asserted schedulable by the schedulability analysis but the transition between them fails at run-time.

The scheduling problem during a transition between two modes has multiple aspects, depending on the behavior and requirements of the old- and new-mode tasks when a mode change is initiated. Upon a mode change request: 
\begin{itemize}
\item an \emph{old-mode} task may be allowed to be immediately aborted or, on the contrary, can be required to complete the execution of its current active job (so that it preserves data consistency for instance). Using scheduling algorithms such as the one considered in this study, we will prove in Section~\ref{sec:Multimode:prelim_validity_tests} that aborting tasks upon a mode change request does not jeopardize the schedulability of the mode transitions. Hence, we assume in this paper the most problematic scenario in which \emph{every old-mode task must complete its current active job} (if any) when a mode change is requested.
\item a \emph{new-mode} task either requires to be activated as soon as possible when a mode change is requested or requires to be activated only when all the active jobs issued from the old-mode have totally completed their execution. 
\end{itemize}

Finally, there may be some tasks (called mode-independent tasks in the literature) that belong to more than one mode and such that their activation pattern must not be jeopardized during the transition between those modes\footnote{In practice, mode-independent tasks typically allow to model daemon functionalities.}. However this paper will  consider only systems that do \emph{not} include such tasks. 

Transition scheduling protocols for tasks without mode-independent tasks are often classified with respect to the way they schedule the old- and new-mode tasks during the transitions. In the literature (see for instance~\cite{JoAlfons:04} which considers uniprocessor systems), the following definitions are used. 

\begin{Definition}[Synchronous/asynchronous protocol~\cite{JoAlfons:04}] 
\label{def:Multimode:synchronous_asynchronous}
A transition protocol is said to be synchronous if it schedules new-mode tasks only when all the old-mode tasks have completed. Otherwise, it is said to be asynchronous. 
\end{Definition}

\begin{Definition}[Protocol with/without periodicity~\cite{JoAlfons:04}]
\label{def:Multimode:with_without_periodicity}
A transition protocol is said to be ``with periodicity'' if and only if it is able to deal with mode-independent tasks. Otherwise, it is said to be ``without periodicity''.
\end{Definition}

\subsection{Related work}

Numerous transition protocols have been proposed for \emph{uni}processor platforms (a survey about this concern is presented in~\cite{JoAlfons:04}). In such environments, existing researches \cite{JoAlfons:04, Henia:07, Pedro:98} have shown that even if two modes of the application have been proven feasible, the transition between the two modes can cause violation of timing constraints, hence needing explicit analyses. Such analyses have been proposed in~\cite{Sha:89}, considering the popular Rate Monotonic Algorithm. Unfortunately three years later, this analysis was shown optimistic~\cite{Tindell:92} in the sense that some unfeasible task sets could be asserted schedulable. In the same paper~\cite{Tindell:92}, the authors improved the previous analysis and proposed a new one which considers the popular Deadline Monotonic Algorithm. An analysis of sporadic tasks scheduled on EDF is known as well~\cite{Andersson:08}. In \cite{Stoimenov:09}, the authors proposed an analysis which considers Fixed-Task-Priority scheduling (FTP), Earliest-Deadline-First~\cite{Liu:73} scheduling and arbitrary task activation pattern. Furthermore, for applications that were initially proven not schedulable during the transition phases, they derived the required offsets for delaying the initialization of transition between two modes in order to make the application schedulable. \\
\noindent Among the uniprocessor synchronous protocols, the authors of~\cite{Bailey:93, Tindell:96, JoAlfons:04} proposed the following protocols.
\begin{itemize}
\renewcommand{\labelitemi}{$\triangleright$}
\item The \emph{Minimum Single Offset Protocol} (MSO)~\cite{JoAlfons:04} where the last activation of each old-mode task completes and then, the new-mode tasks are released. 

\item The \emph{Idle Time Protocol} (IT)~\cite{Tindell:96} where the periodic activations of the old-mode tasks are suspended at the first idle time-instant occurring during the transition and then, the new-mode tasks are released.

\item The~\emph{Maximum-Period Offset Protocol} (MPO)~\cite{Bailey:93} where the delays of first activation of each new-mode task is equal to the period of the less frequent task in both modes, %This is a protocol \emph{with periodicity}.
\end{itemize}

\noindent Among the uniprocessor \emph{asynchronous} protocols, the authors of~\cite{Tindell:92,Pedro:99,Andersson:08} proposed the following protocols.
\begin{itemize}
\renewcommand{\labelitemi}{$\triangleright$}
\item A protocol \emph{without periodicity}~\cite{Pedro:99} where tasks are assigned priorities according to the Deadline Monotonic Scheduling algorithm and are scheduled with time offsets during the mode change only. 

\item A protocol \emph{with periodicity} has been introduced by Sha et al. in~\cite{Sha:88}, assuming Fixed-Task-Priority scheduling. Then, the authors of~\cite{Andersson:08} extended this protocol to the Earliest Deadline First~\cite{Liu:73} scheduling algorithm.

\item The authors of~\cite{Tindell:92} introduced a particular protocol which allows tasks to modify their parameters (period, execution time, etc.) during the mode changes.  As in~\cite{Pedro:99}, this study assumes that the tasks are scheduled according to the Deadline Monotonic scheduling algorithm.
\end{itemize}

\subsection{Contribution and paper organization}

In this paper we propose two protocols \emph{without periodicity} ($\SMMSO$ which is \emph{synchronous} and $\AMMSO$ which is \emph{asynchronous}) for managing mode transitions during the execution of multi-mode real-time applications on \emph{multi}processor platforms. Both protocols can be considered as a generalization to multiprocessors of the MSO protocol proposed in~\cite{JoAlfons:04}. We assume that every operating mode of the application is scheduled by a \emph{global}, \emph{work-conserving}, \emph{preemptive} and \emph{Fixed-Job-Priority} (FJP) scheduling algorithm (formal definitions are given in Section~\ref{sec:Multimode:scheduler_specifications}). Some of the results presented here have already been published (see~\cite{MeumeuNelisGoossens:10, NelisAnderssonGoossens:09, NelisGoossensAndersson:09, NelisGoossens:08}). It is worth noticing that the problem of scheduling multi-mode applications upon multiprocessor platforms is much more complex than upon uniprocessor platforms, especially due to the presence of scheduling anomalies (see Chapter 5 of~\cite{Andersson:03} for a definition) and it is now well known that real-time multiprocessor scheduling problems are typically not solved by applying straightforward extensions of techniques used for solving similar uniprocessor problems. 

The paper is organized as follows. Section~\ref{sec:Multimode:Models_of_computation} defines the computational model used throughout the paper. Sections~\ref{sec:Multimode:SMMSO} and~\ref{sec:Multimode:AMMSO} describe the synchronous and asynchronous protocols $\SMMSO$ and $\AMMSO$, respectively. Section~\ref{sec:Multimode:prelim_validity_tests} introduces some definitions and basic results necessary for the establishment of our schedulability analyses. These four first Sections~\ref{sec:Multimode:Models_of_computation}--\ref{sec:Multimode:prelim_validity_tests} are a common base of the paper, in the sense that these~\pageref{sec:Multimode:ident_FJP} pages describe both the models of computation and protocols independently of the platform and scheduler characteristics. Then, the four next Sections~\ref{sec:Multimode:ident_FJP}--\ref{sec:Multimode:unif_FTP} are each specific to a platform and scheduler model. More precisely, they provide a schedulability analysis for both $\SMMSO$ and $\AMMSO$, assuming in turn identical platforms and Fixed-Job-Priority schedulers (in Section~\ref{sec:Multimode:ident_FJP}), identical platforms and Fixed-Task-Priority schedulers (in Section~\ref{sec:Multimode:ident_FTP}), uniform platforms and Fixed-Job-Priority schedulers (in Section~\ref{sec:Multimode:unif_FJP}) and uniform platforms and Fixed-Task-Priority schedulers (in Section~\ref{sec:Multimode:unif_FTP})\footnote{Even though Fixed-Job-Priority schedulers encompass the family of Fixed-Task-Priority schedulers, the particular case of Fixed-Task-Priority schedulers is treated separately so that the schedulability analyses are more specific and therefore more accurate.}. Finally, Section~\ref{sec:Multimode:Conclusion} gives our conclusions and future work, together with some remaining open problems.

% An example of a floating table. Note that, for IEEE style tables, the 
% \caption command should come BEFORE the table. Table text will default to
% \footnotesize as IEEE normally uses this smaller font for tables.
% The \label must come after \caption as always.
%
%\begin{table}[!t]
%% increase table row spacing, adjust to taste
%\renewcommand{\arraystretch}{1.3}
% if using array.sty, it might be a good idea to tweak the value of
% \extrarowheight as needed to properly center the text within the cells
%\caption{An Example of a Table}
%\label{table_example}
%\centering
%% Some packages, such as MDW tools, offer better commands for making tables
%% than the plain LaTeX2e tabular which is used here.
%\begin{tabular}{|c||c|}
%\hline
%One & Two\\
%\hline
%Three & Four\\
%\hline
%\end{tabular}
%\end{table}

% Note that IEEE does not put floats in the very first column - or typically
% anywhere on the first page for that matter. Also, in-text middle ("here")
% positioning is not used. Most IEEE journals use top floats exclusively.
% Note that, LaTeX2e, unlike IEEE journals, places footnotes above bottom
% floats. This can be corrected via the \fnbelowfloat command of the
% stfloats package.

\section{Models of computation and specifications}
\label{sec:Multimode:Models_of_computation}

\subsection{Application specifications}

We define a multi-mode real-time application $\tau$ as a set of $x$ operating modes denoted by $\mode^1, \mode^2, \ldots, \mode^x$ where each mode $\mode^k$ has to execute its associated task set $\tau^k \equals \{\tau^k_1, \tau^k_2, \ldots, \tau^k_{n_k}\}$ composed of $n_k$ tasks by following the scheduler ${\cal S}^k$. At run-time, the application is either running in one and only one mode, i.e., it is executing only the set of tasks associated to that mode, or it is switching from one mode to another one. Since we do \emph{not} consider mode-independent tasks in this study, it holds that $\tau^k \cap \tau^j = \emptyset, \:\: \forall k \neq j$. 

Each task $\tau^k_i$ is modeled by a \emph{sporadic} and \emph{constrained-deadline} task characterized by three parameters $\left<C^k_i, D^k_i, T^k_i\right>$---a worst-case execution time $C_{i}^k$, a minimum inter-arrival time $T_{i}^k$ and a relative deadline $D_{i}^k \leq T_i^k$---with the interpretation that, during the execution in mode $\mode^k$, task $\tau_i^k$ generates successive \emph{jobs} $\tau_{i,j}^k$ (with $j = 1, \ldots, \infty$) released at times $a_{i,j}^k$ such that $a_{i,j}^k \geq a_{i,j-1}^k + T_i^k$ (with $a_{i,1}^k \geq 0$), each such job has an execution requirement of at most $C_{i}^k$, and must be completed at (or before) its absolute deadline noted $d_{i,j}^k \equals a_{i,j}^k + D_i^k$. In the particular case where $a_{i,j}^k = a_{i,j-1}^k + T_i^k, \:\: \forall j > 1$, the task $\tau^k_i$ is said to be {\em periodic} instead of sporadic. In the same vein, if $D_i^k = T_i^k$ then the task $\tau^k_i$ is said to be {\em implicit-deadline} instead of constrained-deadline. 

\begin{Definition}[Active job]
We say that a job $\tau_{i,j}^k$ is \emph{active} at time $t$ if it has been already released (i.e., $t \leq a_{i,j}^k$) and it is not completed yet. 
\end{Definition}

%More precisely, an active task is said to be {\em running} at time $t$ if it is allocated to a processor ($\cpu$) and is being executed. Otherwise the active task is in the ready queue of the operating system and it is said to be {\em ready}. We denote by {\em active}$(\tau^k, t)$, {\em run}$(\tau^k, t)$ and {\em ready}$(\tau^k, t)$ the subsets of active, running and ready tasks of $\tau^k$ at time $t$, respectively. It holds that {\em active}$(\tau^k, t)$ $\equals$ {\em run}$(\tau^k, t)$ $\cup$ {\em ready}$(\tau^k, t)$. 
Since we assume $D_i^k \leq T_i^k$, there cannot be two jobs of a same task $\tau_i^k$ active at a same time in any feasible schedule. All the tasks are assumed to be independent, i.e., there is no communication, no precedence constraint and no shared resource (except the processors) between them. In~\cite{NelisGoossensAndersson:09}, we introduced the following concept of enabled/disabled tasks.

\begin{Definition}[Enabled/disabled tasks~\cite{NelisGoossensAndersson:09}]
\label{def:Multimode:enabled_disabled}
At run-time, any task $\tau^i_k$ of the application can generate jobs if and only if $\tau^i_k$ is enabled. Symmetrically, a disabled task cannot generate jobs. 
\end{Definition}

As such, disabling a task $\tau^i_k$ prevents future job releases from $\tau^i_k$. When all the tasks of any mode $\tau^i$ are \emph{enabled} and all the tasks of all the other modes are {\em disabled}, the application is said to be running in mode $\mode^i$ (since only the tasks of mode $\tau^i$ can release jobs). We denote by $\enabled(\tau^i, t)$ and $\disabled(\tau^i, t)$ the subsets of {\em enabled} and {\em disabled} tasks of $\tau^i$ at time $t$, respectively.  

\subsection{Platform specifications}

Many recent embedded systems are built upon multiprocessor platforms in order to fulfill the high computational requirements of applications. As pointed out in~\cite{Baruah:03, Baruah:04}, another advantage of such a choice is the fact that multiprocessor systems are more energy efficient than equally powerful uniprocessor platforms. Indeed, raising the frequency of a single $\cpu$ results in a \emph{multiplicative} increase of the consumption while adding $\cpu$s results in an \emph{additive} increase. Two distinct multiprocessor architectures are commonly used in the industrial world and thus, are considered in this paper: identical and uniform platforms.

\textit{Identical platform.} In such multiprocessor platforms, all the $\cpu$s have the same computational capabilities, with the interpretation that in any interval of time two $\cpu$s execute the same amount of work (assuming that none of them is idling). In the remainder of this paper, any platform composed of $m$ identical $\cpu$s will be modeled by $\pi \equals \left\{ \pi_1, \pi_2, \ldots, \pi_m \right\}$ where $\pi_i$ denotes the $i^{\operatorname{th}}$ $\cpu$ of the platform.

\textit{Uniform platform.} In such multiprocessor platforms, the $\cpu$s are allowed to have different computational capabilities. That is, a \emph{parameter} $s_i$ is associated to every $\cpu$ $\pi_i$ with the interpretation that in any time interval of length $t$, $\cpu$ $\pi_i$ executes $s_i \cdot t$ units of execution (if it is not idling). This parameter can be seen as the \emph{execution speed} of the $\cpu$. In the remainder of this paper, any platform composed of $m$ uniform $\cpu$s is modeled by $\pi \equals \left\{ s_1, s_2, \ldots, s_m \right\}$, where $s_i$ is the execution speed of $\cpu$ $\pi_i$. Without loss of generality, we assume that $s_i \geq s_{i-1}$ $\forall i=2, 3, \ldots, m$, meaning that $\cpu$ $\pi_m$ is the fastest $\cpu$ while $\pi_1$ is the slowest one. For all $k \in \left[ 1, m \right]$, we denote by $\cumulspeed{k}$ the cumulated speed of the $(m-k+1)$ fastest $\cpu$s, i.e., 
\begin{equation}
\label{equ:Multimode:cumul_speed}
\cumulspeed{k} \equals \sum_{i=k}^m s_i
\end{equation}

Notice that identical platforms are a particular case of uniform platforms where $s_i = s_j$ $\forall i,j \in \left[ 1, m\right]$. In this particular case we assume without any loss of generality that $\forall i$: $s_i = 1$. 

\subsection{Mode transition specifications}
\label{sec:Multimode:mode_transition_specifications}

While the application is running in any mode $\mode^i$, a mode change can be initiated by any task of $\tau^i$ or by the system itself, whenever it detects a change in the environment or in its internal state for instance. This is performed by invoking a $\MCR(j)$ (i.e., a Mode Change Request), where $\mode^{j}$ is the destination mode. We denote by $t_{\MCR(j)}$ the invoking time of the \emph{last} $\MCR(j)$. From the time at which a mode change is requested to the time at which the transition phase ends, $\mode^i$ and $\mode^j$ are referred to as the old- and new-mode, respectively. 

At run-time, mode transitions are managed as follows. Suppose that the application is running in mode $\mode^i$ and the system (or any task of $\tau^i$) comes to request a mode change to mode $\mode^j$, with $j \neq i$. At time $t_{\MCR(j)}$, the system entrusts the scheduling decisions to a transition protocol which \emph{immediately} disables all the old-mode tasks, thus preventing them from releasing new jobs. At this time, the active jobs issued from these disabled tasks, henceforth called the \emph{rem-jobs} (for ``\emph{rem}aining \emph{jobs}''), may have two distinct behaviors: either \emph{they can be aborted} upon the $\MCR(j)$ or \emph{they can complete their execution}. From the schedulability point of view, we will show that aborting some (or all) rem-jobs upon a mode change request does not jeopardize the system schedulability during the transition phase. Consequently, we assume the worst-case scenario for every mode transition, i.e., the scenario in which \emph{every old-mode task has to complete its last released job (if any) during every mode transition}\footnote{Aborting a job consists in suddenly stopping its execution and removing it from the system memory. But in the real world, suddenly killing a process may cause system failures and the rem-jobs often have to complete their execution.}. The fact that the rem-jobs have to complete their execution upon the $\MCR(j)$ brings the following problem. Even if both task sets $\tau^i$ and $\tau^j$ (from the old- and new-mode, respectively) have been asserted to be schedulable upon the $m$ $\cpu$s at system design-time, the presence of the rem-jobs may cause an \emph{overload} during the transition phase (at run-time) if all the new-mode tasks of $\tau^j$ are enabled \emph{immediately} upon the mode change request. Indeed, the schedulability analysis performed beforehand on $\tau^j$ did not take into account the additional work generated by the rem-jobs. To solve this problem, transition protocols usually delay the enablement of each new-mode task until it is safe to do so. However, these delays are also subject to hard constraints. More precisely, we denote by ${\cal D}_k^j(\mode^i)$ \emph{the relative transition deadline of task $\tau_k^j$ during every transition from mode $\mode^i$ to mode $\mode^j$}, with the following interpretation: the transition protocol must ensure that $\tau_k^j$ is enabled not later than time $t_{\MCR(j)} + {\cal D}_k^j(\mode^i)$. Finally, when all the rem-jobs are completed and all the new-mode tasks of $\tau^j$ are enabled, the system entrusts the scheduling decisions to the scheduler ${\cal S}^j$ of the new-mode $\mode^j$ and the transition phase ends. 

In short, the goal of any transition protocol is to fulfill the following requirements during every mode change:
\begin{enumerate}
\item Complete each rem-job $\tau^i_{a,b}$ by its absolute deadline $d^i_{a,b}$.
\item Enable each new-mode task $\tau^j_k$ by its absolute transition deadline $t_{\MCR(j)} + {\cal D}_k^j(\mode^i)$.
\item Complete each new-mode job\footnote{This requirement is automatically fulfilled for synchronous protocols since no new-mode jobs are scheduled during the mode transitions.} $\tau^j_{a,b}$ by its absolute deadline $d^j_{a,b}$.

\end{enumerate}

\begin{Definition}[Valid protocol~\cite{NelisGoossensAndersson:09}]
A transition protocol ${\cal A}$ is said to be \emph{valid} for a given application $\tau$ and platform $\pi$ if and only if ${\cal A}$ meets all the job and transition deadlines during every transition between every pair of operating modes of $\tau$. 
\end{Definition}

This notion of ``valid protocol'' is directly related to that of a ``validity test'' defined as follows. 

\begin{Definition}[Validity test~\cite{NelisGoossensAndersson:09}]
For a given transition protocol ${\cal A}$, a validity test is a condition based on the tasks and platform characteristics that indicates \emph{a priori} whether ${\cal A}$ is valid for a given application $\tau$ and platform $\pi$. 
\end{Definition}

\subsection{Scheduler specifications}
\label{sec:Multimode:scheduler_specifications}

We consider the \emph{global} preemptive scheduling problem of sporadic constrained-deadline tasks upon multiprocessor platforms. ``Global'' schedulers, in contrast to partitioned ones, allow different tasks and \emph{different} jobs of the same task to be executed upon \emph{different} $\cpu$s. When preemptive, global schedulers allow any job to be interrupted at any time prior to completion on any $\cpu$ and resumed (possibly later) on any other $\cpu$. We consider that every mode $\mode^k$ uses its own scheduler denoted by ${\cal S}^k$ which can be either {\em Fixed-Task-Priority (FTP)} or {\em Fixed-Job-Priority (FJP)} according to the following interpretations.
\begin{itemize}
\item FTP schedulers assign a priority to each task at system design-time (i.e., before the execution of the application) and then at run-time, every released job uses the priority of its task  and the priority of a job is kept constant until it completes.  
\item FJP schedulers assign a priority to each job at run-time (i.e., as soon as it arrives in the system) and every job keeps its priority constant until it completes. As such, different jobs issued from the same task may have different priorities\footnote{According to these interpretations, FTP schedulers are a particular case of FJP schedulers in which all the jobs issued from a same task receive the same priority determined beforehand.}. 
\end{itemize}

\noindent Without loss of generality we assume that, at any time, two active jobs cannot have the same priority. Furthermore, we consider \emph{work-conserving} schedulers according to the following definition.

\begin{Definition}[Work-conserving global scheduler]
\label{def:work-conserving}
A $\cpu$ cannot be idle if there is a job awaiting execution. Usually, priority-based schedulers assign at each instant in time the $m$ highest priority active jobs (if any) to the $m$ $\cpu$s.
\end{Definition}

The above definition of work-conserving schedulers encompasses a large family of schedulers, but suffers from an important lack of \emph{determinism}. Indeed for a given set of jobs, multiple (and different) schedules can sometimes be derived from the same work-conserving scheduler (and thus from the same job priority assignment). The following example illustrates this drawback.

\begin{Example}\label{scheduler_example}
Let us consider the set $J$ of $5$ jobs with respective processing time 4, 8, 4, 4 and 6.  Suppose that $J$ is scheduled on a $2$-processors identical platform $\pi$ by an FTP, global, preemptive and \emph{work-conserving} scheduler such that $J_1 > J_2 > J_3 > J_4 > J_5$. According to Definition~\ref{def:work-conserving}, Figures~\ref{fig:Multimode:ident_work_conserving_1} and~\ref{fig:Multimode:ident_work_conserving_2} depict two possible different schedules corresponding to this priority assignment. 
\begin{figure}[h]
\begin{center}
\includegraphics*[width=0.5\linewidth, viewport=0 0 600 200]{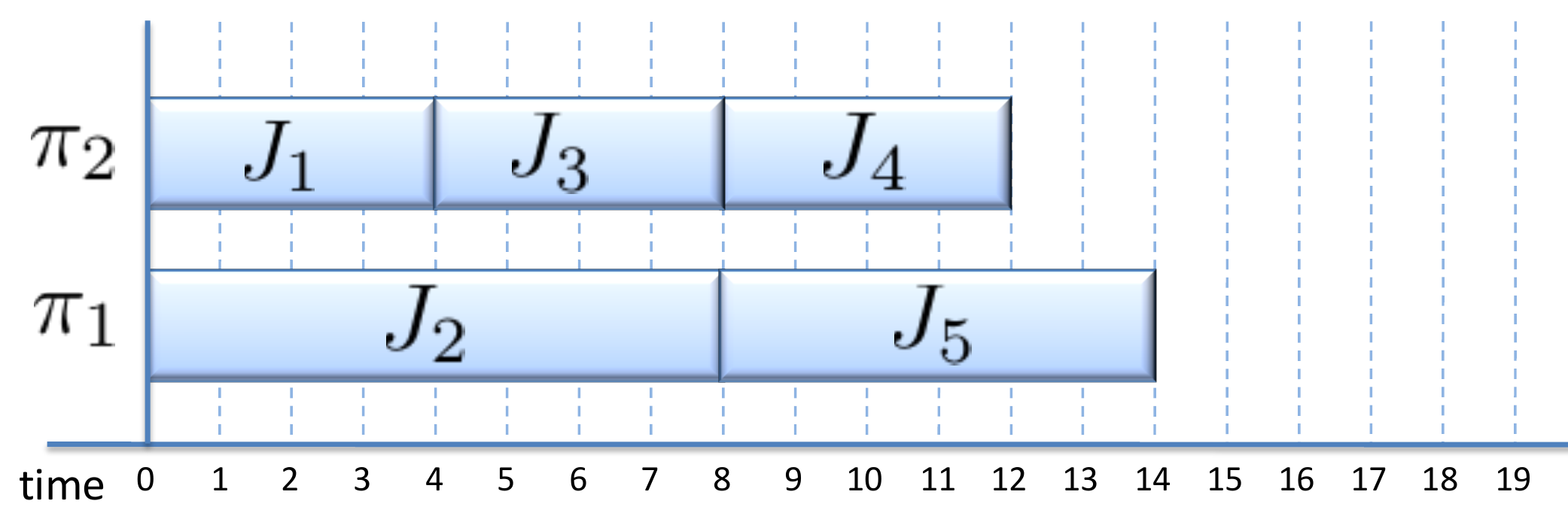}
\caption{A possible schedule of $J_1, J_2, J_3, J_4$ and $J_5$.}
\label{fig:Multimode:ident_work_conserving_1}
\end{center}
\end{figure}

\begin{figure}[h]
\begin{center}
\includegraphics*[width=0.5\linewidth, viewport=0 0 600 200]{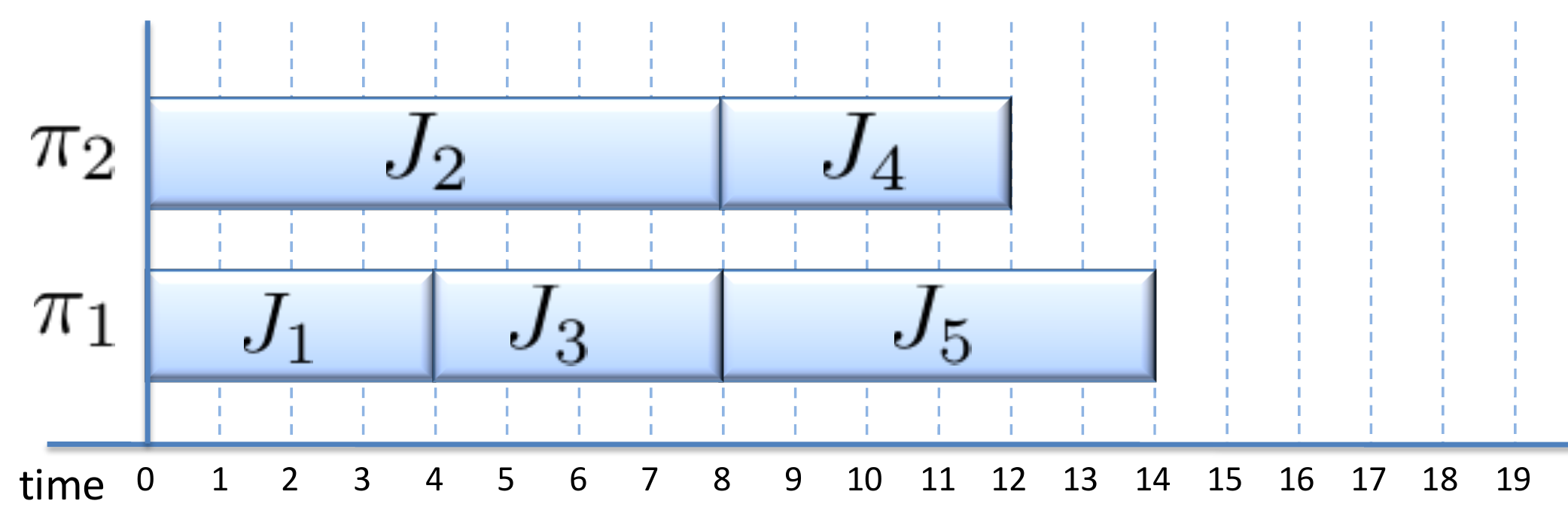}
\caption{Another possible schedule of $J_1, J_2, J_3, J_4$ and $J_5$.}
\label{fig:Multimode:ident_work_conserving_2}
\end{center}
\end{figure}
\end{Example}

In order to get around this lack of determinism, we introduce two refinements of Definition~\ref{def:work-conserving} that we name \emph{weakly} and \emph{strongly} work-conserving schedulers, respectively. Weakly work-conserving schedulers concern only identical platforms whereas strongly work-conserving schedulers concern only in uniform (and non-identical) platforms. The rationale for introducing these two refinements is to have \emph{one and only one possible schedule} for any given set of synchronous\footnote{The term ``synchronous'' jobs is commonly used in the literature to refer to jobs that are all ready for execution at the same time.} jobs, multiprocessor platform and job priority assignment.  

\begin{Definition}[Weakly work-conserving scheduler]
\label{def:Multimode:weakly_work_conserving}
A scheduler ${\cal S}$ is weakly work-conserving if and only if:
\begin{itemize}
\item no $\cpu$ idles while there are active jobs awaiting execution, and
\item if there are more than one job awaiting execution and more than one CPU available for the execution of those jobs then {\cal S} assigns the highest priority waiting job to the available $\cpu$ with the highest index. 
\end{itemize}
\end{Definition}

\begin{Property}[Unique schedule]
For any given finite set $J$ of jobs, any \emph{weakly} work-conserving scheduler ${\cal S}$ and any \emph{identical} multiprocessor platform $\pi$, there exists \emph{one and only one} possible schedule of $J$ upon $\pi$ following ${\cal S}$. 
\end{Property}

In order to illustrate this property, let us consider the set of $5$ jobs used in Example~\ref{scheduler_example}, a 2-processors identical platform $\pi$ and any \emph{weakly} work-conserving scheduler assigning priorities such that $J_1 > J_2 > J_3 > J_4 > J_5$. The unique possible schedule of $J$ upon $\pi$ is the one depicted in Figure~\ref{fig:Multimode:ident_work_conserving_1}. Indeed at time $0$, $\cpu$s $\pi_1$ and $\pi_2$ are idle and the second condition of Definition~\ref{def:Multimode:weakly_work_conserving} imposes $J_1$ to execute on $\pi_2$. From the same rule, $J_4$ must execute on $\pi_2$ at time $8$. Notice that the refinement of ``weakly'' work-conserving scheduler clarifies only the job-to-$\cpu$ assignment rule when the highest-priority waiting job has to be dispatched to a $\cpu$. 

\begin{Definition}[Strongly work-conserving scheduler]
\label{def:Multimode:strongly_work_conserving}
A scheduler ${\cal S}$ is strongly work-conserving if and only if:
\begin{itemize}
\item no $\cpu$ idles while there are active jobs awaiting execution, and 
\item at \emph{every time} during the system execution, the job-to-$\cpu$ assignment uses the rule: highest priority active job upon highest indexed $\cpu$. 
\end{itemize}
\end{Definition}

In contrast to the refinement of ``weakly'' work-conserving schedulers, the ``strongly''-refinement clarifies the job-to-$\cpu$ assignment rule \emph{at each time-instant} during the system execution. It is \emph{essential} to keep in mind that in our study weakly work-conserving schedulers will be used \emph{only} on \emph{identical} platforms whereas strongly work-conserving schedulers will be used \emph{only} on \emph{uniform} and \emph{non-identical} platforms. For strongly work-conserving schedulers, the concept of migrating jobs to faster $\cpu$s as soon as possible (as specified by the second condition of Definition~\ref{def:Multimode:strongly_work_conserving}) has been widely used over the years on uniform platforms (see~\cite{BaruahGoossens:08:2,BaruahGoossens:03,GoossensFunkBaruah:02,BaruahGoossens:08,CucuGoossens:10,FunkGoossensBaruah:01}). This refinement is extremely important, especially because it yields the following property. 

\begin{Property}[Staircase property]
Let $J$ denote any finite set of synchronous jobs, $\pi$ any \emph{uniform} multiprocessor platform and ${\cal S}$ any \emph{strongly} work-conserving scheduler. In the schedule of $J$ upon $\pi$ by ${\cal S}$, $\cpu~\pi_\ell$ idles before or at the same time-instant as $\cpu~\pi_{\ell + 1}$ for all $\ell < m$. 
\end{Property}

\begin{figure}
\begin{center}
\includegraphics*[width=0.5\linewidth, viewport=0 0 600 400]{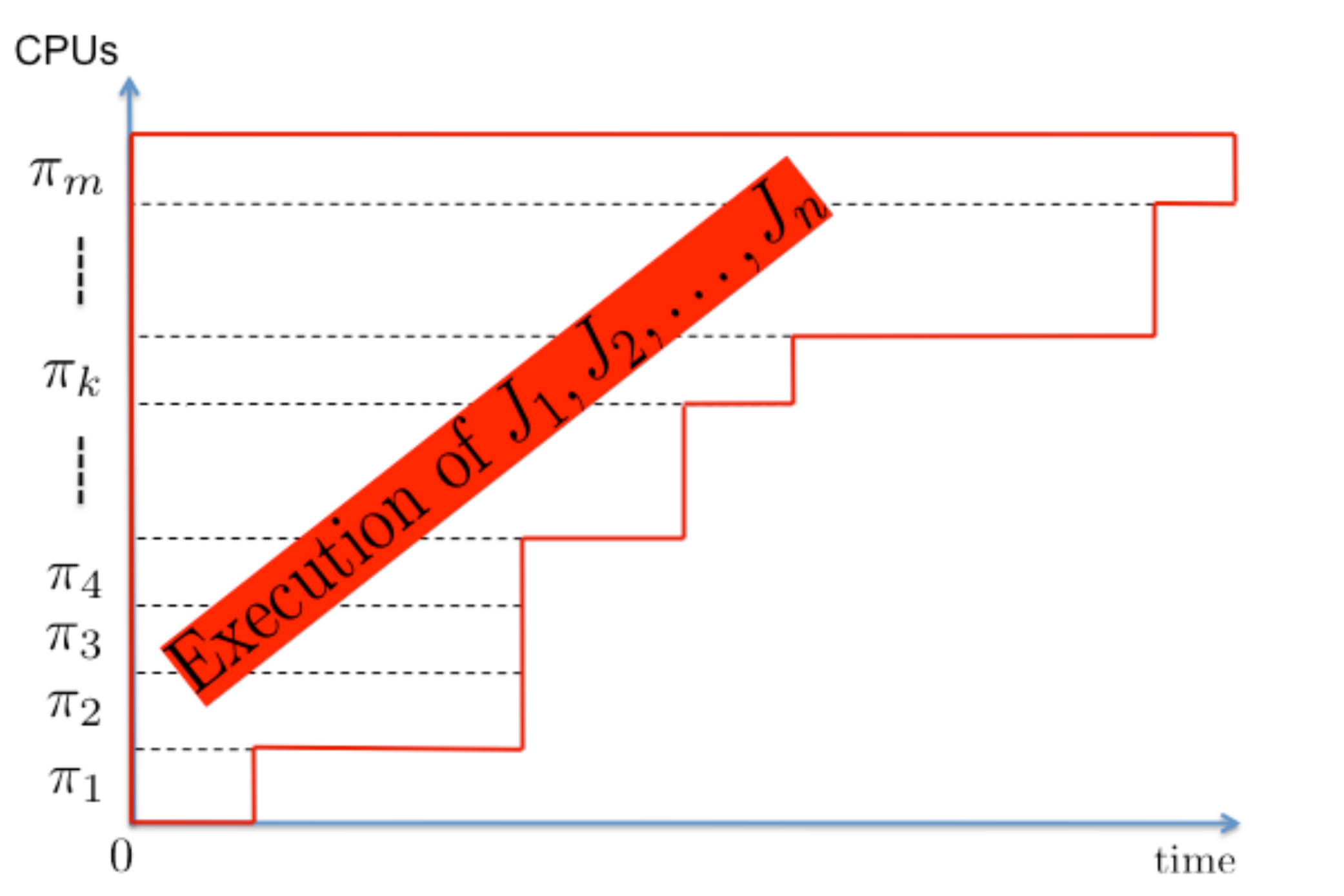}
\caption{For any fixed set of jobs and uniform platform, the schedule generated by any strongly work-conserving scheduler forms a staircase.}
\label{fig:Multimode:unif_staircase}
\end{center}
\end{figure}

Informally speaking, the schedule of $J$ upon $\pi$ by ${\cal S}$ forms a \emph{staircase} (see Figure~\ref{fig:Multimode:unif_staircase}).

This property stems from the fact that the $\cpu$s are indexed in such a manner that $s_i \geq s_j$ $\forall i > j$. Thus, it holds from the second condition of Definition~\ref{def:Multimode:strongly_work_conserving} that at any instant $t$, if ${\cal S}$ idles the $i^{\operatorname{th}}$-slowest $\cpu$ then ${\cal S}$ also idles the $j^{\operatorname{th}}$ slowest $\cpu$s for all $j < i$. Also, it results from the same condition that the $i^{\operatorname{th}}$ $\cpu$ that starts idling is always $\pi_i$. 

The following definition introduces the fundamental notion of \emph{predictability}, and Lemmas \ref{lem:Multimode:weakly_schedulers_predictable} and \ref{lem:Multimode:strongly_schedulers_predictable} are essential for the rest of the paper. 

\begin{Definition}[Predictability~\cite{HaLiu:94}]
\label{def:Multimode:predictability}
Let $A$ denote a scheduler, and let $J = \left\{ J_1, J_2, J_3, \ldots \right\}$ be a potentially infinite set of jobs, where each job $J_i = (a_i, c_i, d_i)$ is characterized by an arrival time $a_i$, a computing requirement $c_i$ and an absolute deadline $d_i$. Let $r_i$ and $f_i$ denote the time at which job $J_i$ starts and completes its execution (respectively) when $J$ is scheduled by $A$. Now, consider any set $J' = \left\{ J_1', J_2', J_3', \ldots  \right\}$ of jobs obtained from $J$ as follows. Job $J_i'$ has an arrival time $a_i$, an execution requirement $c_i' \leq c_i$, and a deadline $d_i$. Let $r_i'$ and $f_i'$ denote the time at which job $J_i'$ starts and completes its execution (respectively) when $J'$ is scheduled by $A$. Algorithm $A$ is said to be predictable if and only if for any set of jobs $J$ and for any such $J'$ obtained from $J$, it is the case that $r_i' \leq r_i$ and $f_i' \leq f_i$ $\forall i$. 
\end{Definition}

Informally speaking, Definition~\ref{def:Multimode:predictability} claims that an upper-bound on the starting time and on the completion time of each job can be determined by analyzing the situation under the assumption that each job executes for its WCET. The result from~\cite{Ha:95, HaLiu:94, HaLiu:93} that we will be using can be stated as follows. 

\begin{Lemma}[See~\cite{Ha:95, HaLiu:94, HaLiu:93}]
\label{lem:Multimode:weakly_schedulers_predictable}
On identical multiprocessor platforms, any FJP, global, preemptive and weakly work-conserving scheduler is predictable. 
\end{Lemma}

In the same vein, the result from~\cite{CucuGoossens:10} that we will be using can be stated as follows. 

\begin{Lemma}[See~\cite{CucuGoossens:10}]
\label{lem:Multimode:strongly_schedulers_predictable}
On uniform multiprocessor platforms, any FJP, global, preemptive and strongly work-conserving scheduler is predictable.  
\end{Lemma}

We use the notation ${\cal P}$ to refer to a specific job priority assignment. A job priority assignment can be seen as a key component of any scheduler, but the definition of a scheduler is more general since, in addition to a job priority assignment, a scheduler must also provide specifications like ``global or partitioned'', ``preemptive or non-preemptive'', etc. For any job priority assignment ${\cal P}$, we denote by $J_i >_{\cal P} J_j$ the fact that job $J_i$ has a higher priority than $J_j$ according to ${\cal P}$, and we assume that every assigned priority is distinct from the others. That is, $\forall {\cal P},i,j$ such that $i \neq j$ we have either $J_i >_{\cal P} J_j$ xor $J_i <_{\cal P} J_j$. Similarly, and without any distinction with the interpretation given above, we will sometimes use the notations $J_i >_{{\cal S}^k} J_j$ and $J_i <_{{\cal S}^k} J_j$ where ${\cal S}^k$ is the scheduler of mode $\mode^k$, and we will sometimes use the notations $J_i > J_j$ and $J_i < J_j$ when the job priority assignment has no label (for instance, when we will depict some examples of schedules, we will just say ``$J_i > J_j$'' without giving a name to the job priority assignment). Finally, the problems and solutions presented in this paper are addressed under the following assumptions:

\begin{itemize}
\renewcommand{\labelitemi}{$\triangleright$}
\label{null:assumption1}
\item \textbf{Assumption~1.} The set $\tau^k$ of tasks of every mode $\mode^k$ can be scheduled by ${\cal S}^k$ on $m$ $\cpu$s without missing any deadline. 

\item \textbf{Assumption~2.} Job migrations and preemptions are permitted and are carried out at no loss or penalty.

\item \textbf{Assumption~3.} Job parallelism is forbidden, i.e., jobs execute on at most one $\cpu$ at any instant in time. 

\item \textbf{Assumption~4.} For every mode $\mode^i$ it holds that $m \leq n_i$, where $n_i$ is the number of tasks in mode $\mode^i$. 
\end{itemize}

Regarding Assumption~1, it allows us to focus only on the schedulability of the application during the {\em transient phases} corresponding to mode transitions, rather than on the schedulability of the application during the execution in a given mode. 

Regarding Assumption~4, it is worth noticing that since job parallelism is forbidden and tasks are assumed to be constrained-deadline, there are at most $n_i$ jobs active at a same time during the execution of any mode $\mode^i$. As a result, it holds for each mode $\mode^i$ that in every schedulable application where $m > n_i$, there are always $m-n_i$ $\cpu$s that constantly idle. We will see later that these $m - n_i$ idling $\cpu$s are the slowest ones and the problem in that case thereby reduces to the same problem upon the subset of the $n_i$ fastest $\cpu$s among these $m$ $\cpu$s.

%%%%%%%%%%%%%%%%%%%%%%%%%%%%%%%%%%%%%%%%%%%%%%%%%%%%%%%%%%%%%%%%%%%%%%%%%%%%%%%%%%%%%%%%%%%%%%%%%%%%%%%%%%%%%%%%%%%%%%%
%%%%%%%%%%%%%%%%%%%%%%%%%%%%%%%%%%%%%%%%%%%%%%%%%%%%%%%%%%%%%%%%%%%%%%%%%%%%%%%%%%%%%%%%%%%%%%%%%%%%%%%%%%%%%%%%%%%%%%%
%%%%%%%%%%%%%%%%%%%%%%%%%%%%%%%%%%%%%%%%%%%%%%%%%%%%%%%%%%%%%%%%%%%%%%%%%%%%%%%%%%%%%%%%%%%%%%%%%%%%%%%%%%%%%%%%%%%%%%%
%%%%%%%%%%%%%%%%%%%%%%%%%%%%%%%%%%%%%%%%%%%%%%%%%%%%%%%%%%%%%%%%%%%%%%%%%%%%%%%%%%%%%%%%%%%%%%%%%%%%%%%%%%%%%%%%%%%%%%%
%%%%%%%%%%%%%%%%%%%%%%%%%%%%%%%%%%%%%%%%%%%%%%%%%%%%%%%%%%%%%%%%%%%%%%%%%%%%%%%%%%%%%%%%%%%%%%%%%%%%%%%%%%%%%%%%%%%%%%%

\section{The synchronous protocol SM-MSO}
\label{sec:Multimode:SMMSO}

\subsection{Description of the protocol}

The protocol $\SMMSO$ (which stands for ``Synchronous Multiprocessor Minimum Single Offset'' protocol) is an extension to multiprocessor platforms of the protocol $\MSO$ defined in~\cite{JoAlfons:04} for uniprocessor platforms. This protocol supports both uniform and identical platforms. The main idea of $\SMMSO$ is the following: upon a MCR($j$), $\forall j$, all the tasks of the old-mode (say $\mode^i$) are disabled and the rem-jobs continue to be scheduled by the old-mode scheduler ${\cal S}^i$ upon the $m$ $\cpu$s. Once \emph{all} the rem-jobs are completed, \emph{all} the new-mode tasks (i.e., the tasks of $\tau^j$) are simultaneously enabled. Algorithm~\ref{algo:SMMSO} gives the pseudo-code of this protocol and Example~\ref{example_protocol_smmso} illustrates how $\SMMSO$ handles the mode transitions. 

\begin{figure}
\begin{center}
\begin{minipage}{10cm}
\begin{algorithmic}[1]
\REQUIRE $M^i$: the old mode
\REQUIRE $M^j$: the new-mode
\REQUIRE the rem-jobs
\WHILE{\TRUE}
	\STATE Schedule the rem-jobs according to ${\cal S}^i$
	\IF{(any rem-job $J_k$ completes at time $t$)} 
		\IF{($\act(\tau^{i},t) = \phi$)}
			\STATE enable all the new-mode tasks of $\tau^{j}$
			\STATE enter the new-mode $\mode^{j}$
		\ENDIF
	\ENDIF
\ENDWHILE
\end{algorithmic}
\end{minipage}
\end{center}
\caption{$\SMMSO$ protocol}
\label{algo:SMMSO}
\end{figure}

\begin{Example}\label{example_protocol_smmso}
Let us consider a platform $\pi$ composed of only 2 \emph{identical} $\cpu$s and an application composed of 2 modes $\mode^i$ and $\mode^j$ depicted in blue and red, respectively. We assume that these two modes contain only synchronous implicit-deadline periodic tasks. The old-mode $\mode^i$ contains 4 tasks with characteristics given in Table~\ref{tab:scheduler_example_smmso1} and uses an FTP scheduler ${\cal S}^i$ such that $\tau^i_1 >_{{\cal S}^i} \tau^i_2 >_{{\cal S}^i} \tau^i_3 >_{{\cal S}^i} \tau^i_4$.

\begin{table}[h!]
\centering
\begin{tabular}{| c | c | c |}
\hline
Tasks & $C^i_k$ & $D^i_k = T^i_k$  \\
\hline
$\tau^i_1$ & 40 & 120 \\
\hline
$\tau^i_2$ & 20 & 120 \\
\hline
$\tau^i_3$ & 40 & 120 \\
\hline
$\tau^i_4$ & 60 & 120 \\
\hline
\end{tabular}
\caption{Characteristics of the tasks in $\mode^i$.}
\label{tab:scheduler_example_smmso1}
\end{table}

\noindent The new-mode $\mode^j$ contains 3 tasks $\tau^j_1, \tau^j_2, \tau^j_3$ and uses an FTP scheduler ${\cal S}^j$ such that $\tau^j_1 >_{{\cal S}^j} \tau^j_2 >_{{\cal S}^j} \tau^j_3$. The characteristics of these tasks are: $C^j_1 = 100$ and $C^j_2 = C^j_3 = 40$. The deadline and period of these new-mode tasks do not have any importance in this example and we intentionally omitted to specify them. Figure~\ref{fig:Multimode:SMMSO_example} illustrates the $\SMMSO$ transition protocol between these two modes. 

\begin{figure}
\begin{center}
\includegraphics*[width=0.5\linewidth, viewport=0 0 750 430]{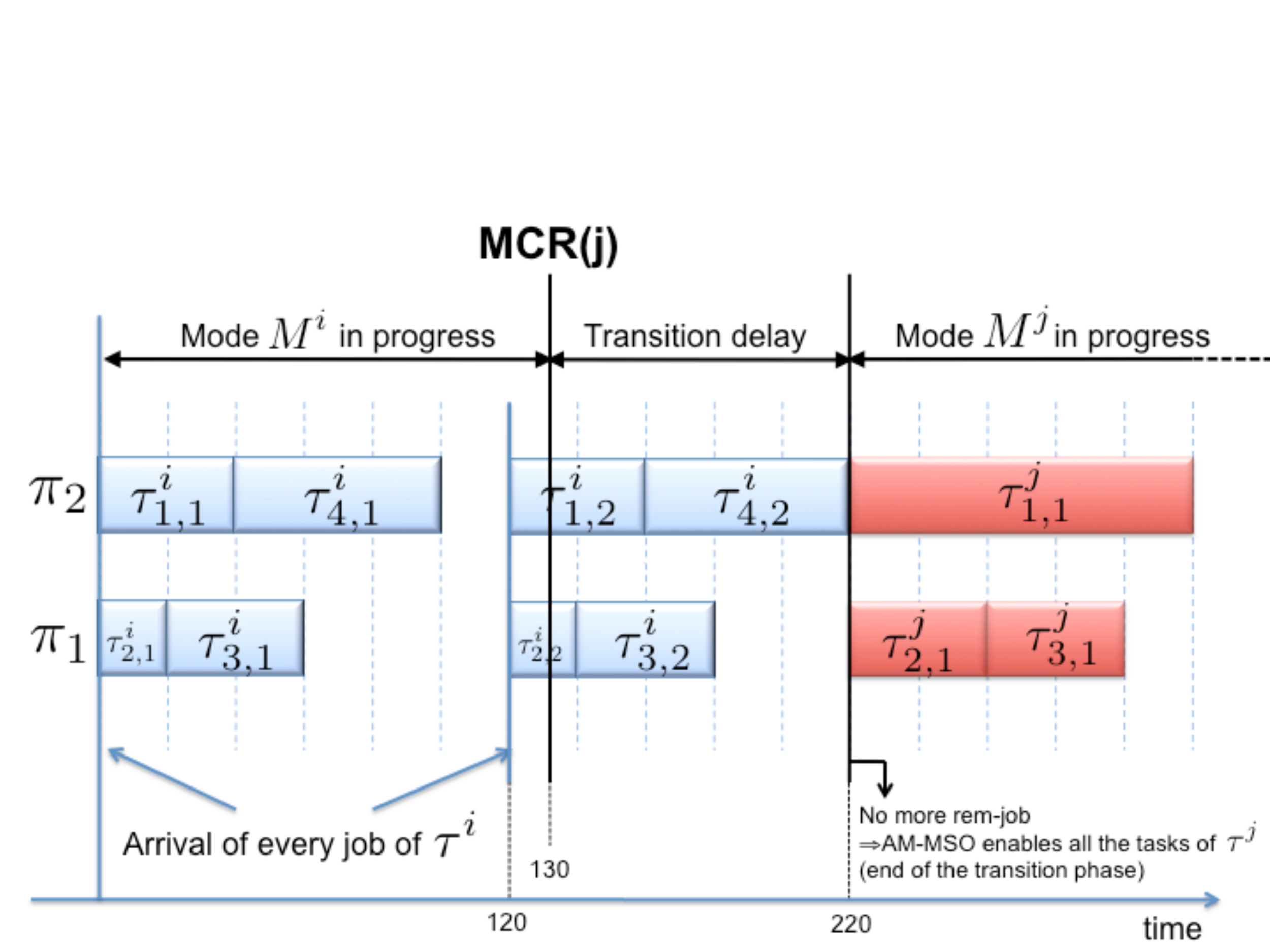}
\caption{Illustration of a mode transition handled by $\SMMSO$.}
\label{fig:Multimode:SMMSO_example}
\end{center}
\end{figure}

At time 120, every task of $\mode^i$ releases its second job and the scheduler ${\cal S}^i$ starts the execution of $\tau_{1,2}^i$ and $\tau_{2,2}^i$ on $\cpu$ $\pi_2$ and $\pi_1$, respectively. Then, suppose that the system requests a mode change at time 130. Here starts the transition phase from mode $\mode^i$ to mode $\mode^j$. As specified by the protocol $\SMMSO$, all the old-mode tasks are immediately disabled and the remaining active jobs $\tau_{1,2}^i, \tau_{2,2}^i, \tau_{3,2}^i$ and $\tau_{4,2}^i$ (named the rem-jobs from this point forward) continue to be scheduled according to the old-mode scheduler ${\cal S}^i$. These rem-jobs execute until time 220, time at which they are all completed. At this instant 220, the condition at line 4 of Algorithm~\ref{algo:SMMSO} is verified. Thus, $\SMMSO$ enables all the new-mode tasks and starts scheduling the incoming new-mode jobs according to the new-mode scheduler ${\cal S}^j$. Notice that at any time \emph{during every transition phase}, our protocol $\SMMSO$ allows the system (or any task) to request any other mode change. At the very end of the current transition phase (at time 220 in this example), $\SMMSO$ enables all the tasks of the mode $\mode^z$ assuming that $\MCR(z)$ is the last mode change that has been requested.
\end{Example}

\subsection{Design of a validity test}
\label{sec:Multimode:SMMSO_main_idea}

In order to establish a {\em validity test} for the protocol $\SMMSO$, two key results are required:

\begin{enumerate}
\item It must be proved for every mode transition that disabling the old-mode tasks upon a MCR does not jeopardize the schedulability of the rem-jobs when they continue to be scheduled by the old-mode scheduler. That is, it must be guaranteed that the absolute deadline $d^i_{a,b}$ of every rem-job $\tau^i_{a,b}$ is met during every mode transition from every mode $\mode^i$.

\item It must be proved for every mode transition that the length of the transition phase \emph{can never be larger} than the minimum transition deadline of all new-mode tasks. Indeed, it follows from this statement and the definition of $\SMMSO$ that all the transition deadlines would be met during every mode transition. 
\end{enumerate} 

We provided a proof for the first key result in~\cite{NelisGoossensAndersson:09} (the proof is replicated in Section~\ref{sec:Multimode:prelim_validity_tests}, page~\pageref{lem:Multimode:remjobs_meet_deadline}), and this result holds for any uniform platform (including identical platforms). About the second key result, it is worth noticing that there is \emph{no job release} (and therefore no preemption) during every transition phase since we consider only FJP schedulers and all the old-mode tasks are disabled upon any mode change request. As a consequence, the length of any transition phase corresponds to \emph{the time needed to complete all the rem-jobs} (this clearly appears in Figure~\ref{fig:Multimode:SMMSO_example}). In the literature (and hereafter as well), the time needed to complete a given set of synchronous jobs upon a given platform is called the \emph{makespan} defined as follows.

\begin{Definition}[Makespan]
\label{def:Multimode:makespan}
Let $J = \left\{J_1, J_2, \ldots, J_n\right\}$ denote any set of $n$ jobs of processing times $c_1, c_2, \ldots, c_n$. Let $\pi$ denotes any uniform multiprocessor platform composed of $m$ $\cpu$s. Let ${\cal P}$ denote any job priority assignment and ${\cal S}$ denotes the schedule of $J$ upon $\pi$ by any work-conserving scheduler (including weakly and strongly work-conserving schedulers) using the priority assignment ${\cal P}$. The makespan denoted by $\makespan(J, \pi, {\cal P})$ is the earliest instant in ${\cal S}$ such that the $n$ jobs of $J$ are completed. 
\end{Definition}

According to Definition~\ref{def:Multimode:makespan}, the length of any transition phase corresponds to the makespan generated by the set of jobs that are active in the system when the mode change is requested, i.e., the set of rem-jobs. Since the value of the makespan obviously depends on the number and processing times of the jobs (as well as on the $\cpu$ speeds), then the length of any transition phase from any mode $\mode^i$ to any other mode $\mode^j$ depends on both the number of rem-jobs and their remaining processing time at time $t_{\MCR(j)}$. From this observation, determining an \emph{upper-bound} on the makespan requires to consider the worst-case scenario, i.e., the scenario in which the number and the remaining processing time of the rem-jobs at time $t_{\MCR(j)}$ is such that the generated makespan is maximum. This worst-case scenario is thus entirely defined by a specific set of rem-jobs that we name the \emph{critical rem-job set} defined as follows.

\begin{Definition}[Critical rem-job set $\wcremjobs{i}$]
\label{def:Multimode:worst_case_configuration}
Assuming any transition from a specific mode $\mode^i$ to any other mode $\mode^j$, the critical rem-job set $\wcremjobs{i}$ is the set of jobs issued from the tasks of $\tau^i$ that leads to the largest makespan. 
\end{Definition}

For any work-conserving FJP scheduler (including FTP schedulers) and uniform platform (including identical platform), we will show that the critical rem-job set $\wcremjobs{i}$ of every transition from mode $\mode^i$ to mode $\mode^j$ is the one where each task $\tau_k^i$ has a rem-job at time $t_{\MCR(j)}$ with a remaining processing time equals to $C_k^i$ (i.e., the WCET of $\tau_k^i$). This result is very intuitive: the makespan is as large as the number and processing times of the rem-jobs are large. 

In this paper we address the problem of establishing mathematical expressions that provide the \emph{maximum makespan} for any given set of \emph{synchronous}\footnote{During every mode transition, the considered jobs are assumed to be synchronous because every rem-job is active and ready to execute upon the mode change request.} jobs and especially for the critical rem-job set during each mode transition. This intention stems from the fact that the knowledge of the maximum makespan allows us to assert (or refute) that every new-mode task will meet its transition deadline during any mode transition using $\SMMSO$, thus ensuring the validity of $\SMMSO$ for a given application $\tau$ and platform $\pi$ as follows. 

\begin{validity test}[For protocol $\SMMSO$]
\label{test:Multimode:SMMSO_first}
For any multi-mode real-time application $\tau$ and any uniform multiprocessor platform $\pi$, protocol $\SMMSO$ is valid provided that, for every mode $\mode^i$,
\begin{equation}
\label{equ:basic_SMMSO_validity_test}
\maxmakespan(\wcremjobs{i}, \pi, {\cal P}^i) \leq \min_{j \neq i} \left\{ \min_{1 \le k \le n_j} \left\{{\cal D}_k^j(M^i)\right\}\right\}
\end{equation}
\noindent where ${\cal P}^i$ is the job priority assignment derived from the old-mode scheduler ${\cal S}^i$ and $\maxmakespan(\wcremjobs{i}, \pi, {\cal P}^i)$ is an upper-bound on the makespan, considering the set $\wcremjobs{i}$ of jobs, the platform $\pi$ and the job priority assignment ${\cal P}^i$. 
\end{validity test}

The above expression can be interpreted as follows: all the transition deadlines will be met during the execution of the system if, for every mode $\mode^i$, the maximum makespan (i.e., the maximum transition latency) generated by the rem-jobs issued from the tasks of $\tau^i$ cannot be larger than the minimum transition deadline of every task of every mode $\mode^j$. 

This validity test is a \emph{sufficient} condition that indicates, a priori, if all the deadlines will be met during all possible mode changes using the protocol $\SMMSO$. Unfortunately, to the best of our knowledge, the problem of determining the maximum makespan has never been studied in the literature. Rather, authors usually address the problem of determining a job priority assignment that minimizes the makespan~\cite{Goyal:05, Garey:90}. The goal in that framework being to ultimately reduce the completion times of the jobs as much as possible. This problem of finding priorities that minimize the makespan can be cast as a strongly NP-hard bin-packing problem~\cite{Goyal:05, Garey:90} for which numerous heuristics have been proposed in the literature. On the contrary, we provide in Sections~\ref{sec:Multimode:ident_FJP}--\ref{sec:Multimode:unif_FTP} different \emph{upper-bounds} on the makespan, assuming in turn {\em identical platforms and FJP schedulers}, {\em identical platforms and FTP schedulers}, {\em uniform platforms and FJP schedulers} and finally, {\em uniform platforms and FTP schedulers}. 

\subsection{FTP schedulers vs. FJP schedulers}

As mentioned in Section \ref{sec:Multimode:scheduler_specifications}, FTP schedulers are a particular case of FJP schedulers. However the remainder of this study distinguishes between these two scheduler families because FTP schedulers allow to determining a more precise upper-bound $\maxmakespan(\wcremjobs{i}, \pi, {\cal P}^i)$ than FJP schedulers. The reason of this stems from the fact that the priority of each task (and thus the priority of every job) is known \emph{at system design-time} for FTP schedulers whereas it is \emph{un}known beforehand for FJP schedulers. 

At first blush, assuming that the job priority assignment ${\cal P}^i$ is unknown for FJP schedulers can seem inconsistent since during every mode transition, we consider the critical rem job set in the computation of $\maxmakespan(\wcremjobs{i}, \pi, {\cal P}^i)$ (and this critical rem-job set is determined \emph{at system design-time}). Therefore, it could be thought that ${\cal P}^i$ can simply be derived from $\wcremjobs{i}$. But this intuition is \emph{erroneous} because for a given FJP scheduler, several job priority assignments can be derived from the same critical rem-job set as shown in the following example. Actually, given set of jobs, we are not aware of any job priority assignment leading to the maximum makespan.

\begin{Example}
\label{lem:Multimode:FJP_priority_assignment_not_unique}

Let us consider a platform $\pi$ composed of only 2 \emph{identical} $\cpu$s and an application $\tau$ composed of 2 modes $\mode^i$ and $\mode^j$. Suppose that a mode change is requested from $\mode^i$ to $\mode ^j$ and the old-mode scheduler ${\cal S}^i$ is $\edf$. The old-mode $\mode^i$ contains 3 tasks with characteristics given in Table~\ref{tab:priority_example1}.

\begin{table}[h!]
\centering
\begin{tabular}{| c | c | c |}
\hline
Tasks & $C^i_k$ & $D^i_k = T^i_k$  \\
\hline
$\tau^i_1$ & 5 & 15 \\
\hline
$\tau^i_2$ & 5 & 16 \\
\hline
$\tau^i_3$ & 7 & 18 \\
\hline
\end{tabular}
\caption{Characteristics of the tasks in $\mode^i$.}
\label{tab:priority_example1}
\end{table}

As introduced earlier, the critical rem-job set for this mode transition is given by $\wcremjobs{i} = \left\{J_1, J_2, J_3\right\}$ with processing time $C^i_1, C^i_2$ and $C^i_3$, respectively. This will be formally proved in Corollary~\ref{cor:Multimode:worst_case_rem_jobs_set} (on page~\pageref{cor:Multimode:worst_case_rem_jobs_set}), assuming any FJP scheduler and any uniform platform. Actually, this critical rem-job set specifies \emph{only} the processing time of the jobs, \emph{not} the release time, neither the absolute deadline. Consequently, different job priority assignments can be derived from $\wcremjobs{i}$. We depict two of them in Figures~\ref{fig:Multimode:multiple_JPA_1} and~\ref{fig:Multimode:multiple_JPA_2}. In both figures the time is relative to the instant $t_{\MCR(j)}$ (i.e., $t_{\MCR(j)} = 0$). The release time and the absolute deadline of each job $J_k$ are denoted by $a_k$ and $d_k$, respectively. These two job priority assignments are obtained as follows.

\noindent \textbf{Job priority assignment 1.} If we assume that the three jobs are released \emph{exactly} at the $\MCR$ invoking time $t_{\MCR(j)}$, i.e., $a_1 = a_2 = a_3 = t_{\MCR(j)}$, then the absolute deadline of each job $J_k$ is given by $d_k \equals t_{\MCR(j)} + D_k^i$. In Figure~\ref{fig:Multimode:multiple_JPA_1}, the deadline of each job is thus: $d_1 = 15$, $d_2 = 16$ and $d_3 = 18$ and according to $\edf$, this leads to the job priority assignment $J_1 >_{\edfmin} J_2 >_{\edfmin} J_3$ (and to a makespan of $12$).  

\begin{figure}[h]
\begin{center}
\includegraphics*[width=0.5\linewidth, viewport=0 0 720 220]{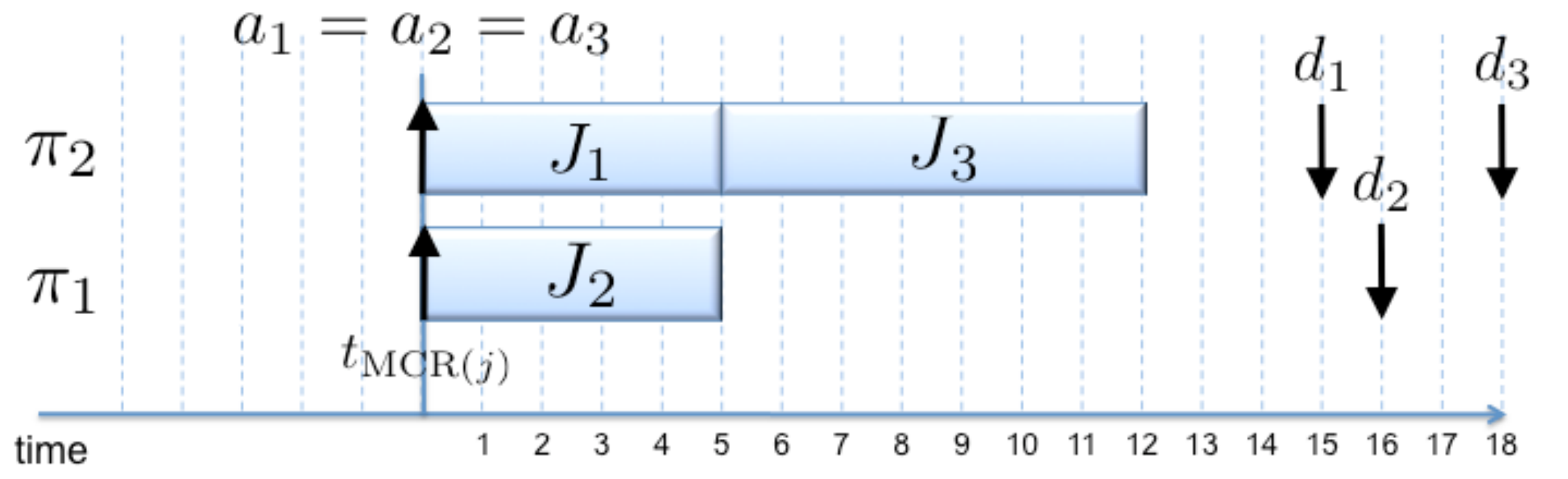}
\caption{Assuming that the three jobs are released simultaneously upon the $\MCR(j)$ allows to derive a first job priority assignment.}
\label{fig:Multimode:multiple_JPA_1}
\end{center}
\end{figure}

\noindent \textbf{Job priority assignment 2.} Starting from the previous release pattern in which all the jobs are released simultaneously at time $t_{\MCR(j)}$, one can slightly move backward the release time of job $J_3$ (for instance) in such a manner that $J_3$ is released at time $t_{\MCR(j)} - 5$ (see Figure~\ref{fig:Multimode:multiple_JPA_2}). Its absolute deadline $d_3$ is thus shifted to time $t_{\MCR(j)} + 13$ and since no assumption is made about the schedule before time $t_{\MCR(j)}$, we can suppose that $J_3$ did not execute before $t_{\MCR(j)}$. Therefore, the processing time of $J_3$ at time $t_{\MCR(j)}$ is $C^i_3 = 5$ and the job priority assignment resulting from this new release pattern is $J_3 >_{\edfmin} J_2 >_{\edfmin} J_1$ (leading to a makespan of $10$). 

\begin{figure}[h]
\begin{center}
\includegraphics*[width=0.5\linewidth, viewport=0 0 720 220]{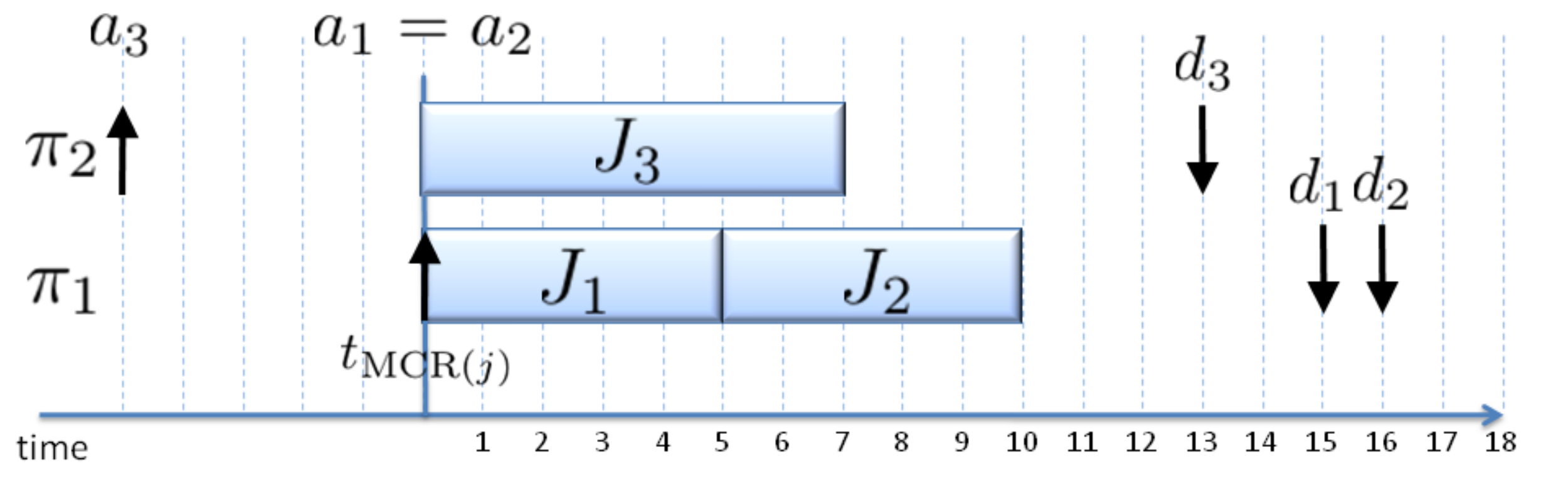}
\caption{Another job priority assignment can be derived by slightly modifying the release pattern of the jobs. Note that this modification leads to another makespan.}
\label{fig:Multimode:multiple_JPA_2}
\end{center}
\end{figure}

\noindent In the particular case of $\edf$, shifting the absolute deadline of these three jobs by distinct amplitudes can modify their relative priorities and a possibly large number of job priority assignments can be derived from the same critical rem-job set $\wcremjobs{i}$.
\end{Example}

Because the prior knowledge of the critical rem-job set does not allow determining a unique job priority assignment, FJP schedulers require to consider every possible job priority assignment in order to determine an upper-bound on the makespan. Hence, we refine the notation of $\maxmakespan(J, \pi, {\cal P}^i)$ as follows: the upper-bound on the makespan is denoted by $\maxmakespan(J, \pi, {\cal P})$ when ${\cal P}$ is explicitly specified (in the context of FTP scheduler) and by $\maxmakespan(J, \pi)$ otherwise (in the context of FJP scheduler), with the interpretation that for every job priority assignment ${\cal X}$: 
\label{null:Multimode:maxidle_instant_no_JPA}
\[ \maxmakespan(J, \pi) \geq \makespan(J, \pi, {\cal X}) \]

It goes without saying that the prior knowledge of the jobs priority assignment allows for establishing tighter upper-bounds on the makespan, i.e., the upper-bound $\maxmakespan(J, \pi, {\cal P})$ is tighter than $\maxmakespan(J, \pi)$. From these refined notations, Expression~\ref{equ:basic_SMMSO_validity_test} of Validity Test~\ref{test:Multimode:SMMSO_first} can be rewritten as
\[ \maxmakespan(\wcremjobs{i}, \pi) \leq \min_{j \neq i} \left\{ \min_{1 \le k \le n_j} \left\{ {\cal D}_k^j(M^i) \right\} \right\} \] 
for FJP schedulers, and as 
\[ \maxmakespan(\wcremjobs{i}, \pi, {\cal P}^i) \leq \min_{j \neq i} \left\{ \min_{1 \le k \le n_j} \left\{ {\cal D}_k^j(M^i) \right\}\right\} \] 
for FTP schedulers, where ${\cal P}^i$ is the job priority assignment derived from the old-mode FTP scheduler ${\cal S}^i$.

%%%%%%%%%%%%%%%%%%%%%%%%%%%%%%%%%%%%%%%%%%%%%%%%%%%%%%%%%%%%%%%%%%%%%%%%%%%%%%%%%%%%%%%%%%%%%%%%%%%%%%%%%%%%%%%%%%%%%%%
%%%%%%%%%%%%%%%%%%%%%%%%%%%%%%%%%%%%%%%%%%%%%%%%%%%%%%%%%%%%%%%%%%%%%%%%%%%%%%%%%%%%%%%%%%%%%%%%%%%%%%%%%%%%%%%%%%%%%%%
%%%%%%%%%%%%%%%%%%%%%%%%%%%%%%%%%%%%%%%%%%%%%%%%%%%%%%%%%%%%%%%%%%%%%%%%%%%%%%%%%%%%%%%%%%%%%%%%%%%%%%%%%%%%%%%%%%%%%%%
%%%%%%%%%%%%%%%%%%%%%%%%%%%%%%%%%%%%%%%%%%%%%%%%%%%%%%%%%%%%%%%%%%%%%%%%%%%%%%%%%%%%%%%%%%%%%%%%%%%%%%%%%%%%%%%%%%%%%%%
%%%%%%%%%%%%%%%%%%%%%%%%%%%%%%%%%%%%%%%%%%%%%%%%%%%%%%%%%%%%%%%%%%%%%%%%%%%%%%%%%%%%%%%%%%%%%%%%%%%%%%%%%%%%%%%%%%%%%%%

\section{The asynchronous protocol AM-MSO}
\label{sec:Multimode:AMMSO}

\subsection{Description of the protocol}
\label{sec:Multimode:AMMSO_description}

The protocol $\AMMSO$ (which stands for ``Asynchronous Multiprocessor Minimum Single Offset'' protocol) is an asynchronous version of the protocol $\SMMSO$. This protocol supports both uniform and identical platforms. The main idea of this second protocol is to reduce the delay applied to the enablement of the new-mode tasks, by enabling them as soon as possible. In contrast to SM-MSO, rem-jobs and new-mode tasks can be scheduled \emph{simultaneously} during the transition phases according to the scheduler ${\cal S}^{\operatorname{trans}}$ defined as follows: (i) the priorities of the rem-jobs are assigned according to the old-mode scheduler; (ii) the priorities of the new-mode jobs are assigned according to the new-mode scheduler, and (iii) the priority of each rem-job is higher than the priority of every new-mode job. 

Formally, suppose that the system is transitioning from mode $\mode^{\old}$ to mode $\mode^{\new}$ and let $J_i$ and $J_j$ be two active jobs during this transition. According to these notations we have $J_j >_{{\cal S}^{\trans}} J_i$ if and only if one of the following conditions is satisfied:
\begin{eqnarray}
& & (J_j \in \mode^{\old} \:\: \mbox{and} \:\: J_i \in \mode^{\new}) \nonumber \\
& \mbox{or} & (J_j \in \mode^{\old} \:\: \mbox{and} \:\: J_i \in \mode^{\old} \:\: \mbox{and} \:\: J_j >_{{\cal S}^{\old}} J_i) \nonumber \\
& \mbox{or} & (J_j \in \mode^{\new} \:\: \mbox{and} \:\: J_i \in \mode^{\new} \:\: \mbox{and} \:\: J_j >_{{\cal S}^{\new}} J_i) \nonumber
\end{eqnarray}

$\AMMSO$ proceeds as follows: upon a $\MCR(j)$, $\forall j$, all the old-mode tasks are disabled and the rem-jobs continue to be scheduled by ${\cal S}^{i}$ (assuming that $\mode^{i}$ is the old-mode). Whenever any rem-job completes (say at time $t$), if there is no more waiting rem-jobs $\AMMSO$ immediately enables some new-mode tasks, in contrast to $\SMMSO$ which waits for the completion of \emph{all} the rem-jobs. In order to select the new-mode tasks to enable at time $t$, $\AMMSO$ uses the following \emph{heuristic}: it considers every disabled new-mode task by non-decreasing order of transition deadline and enables those which can be scheduled by ${\cal S}^{j}$ upon the current available CPUs, i.e., the CPUs that are not running a rem-job and are therefore available for executing some new-mode tasks. The following example illustrates how $\AMMSO$ manages mode transitions.

\begin{Example}
\label{example:AMMSO}
Let us consider the same task sets as in Example~\ref{example_protocol_smmso}. Figure~\ref{fig:Multimode:AMMSO_example} illustrates the $\AMMSO$ transition protocol on a 2-processors platform. 
\begin{figure}[h!]
\begin{center}
\includegraphics*[width=0.5\linewidth, viewport=0 0 750 430]{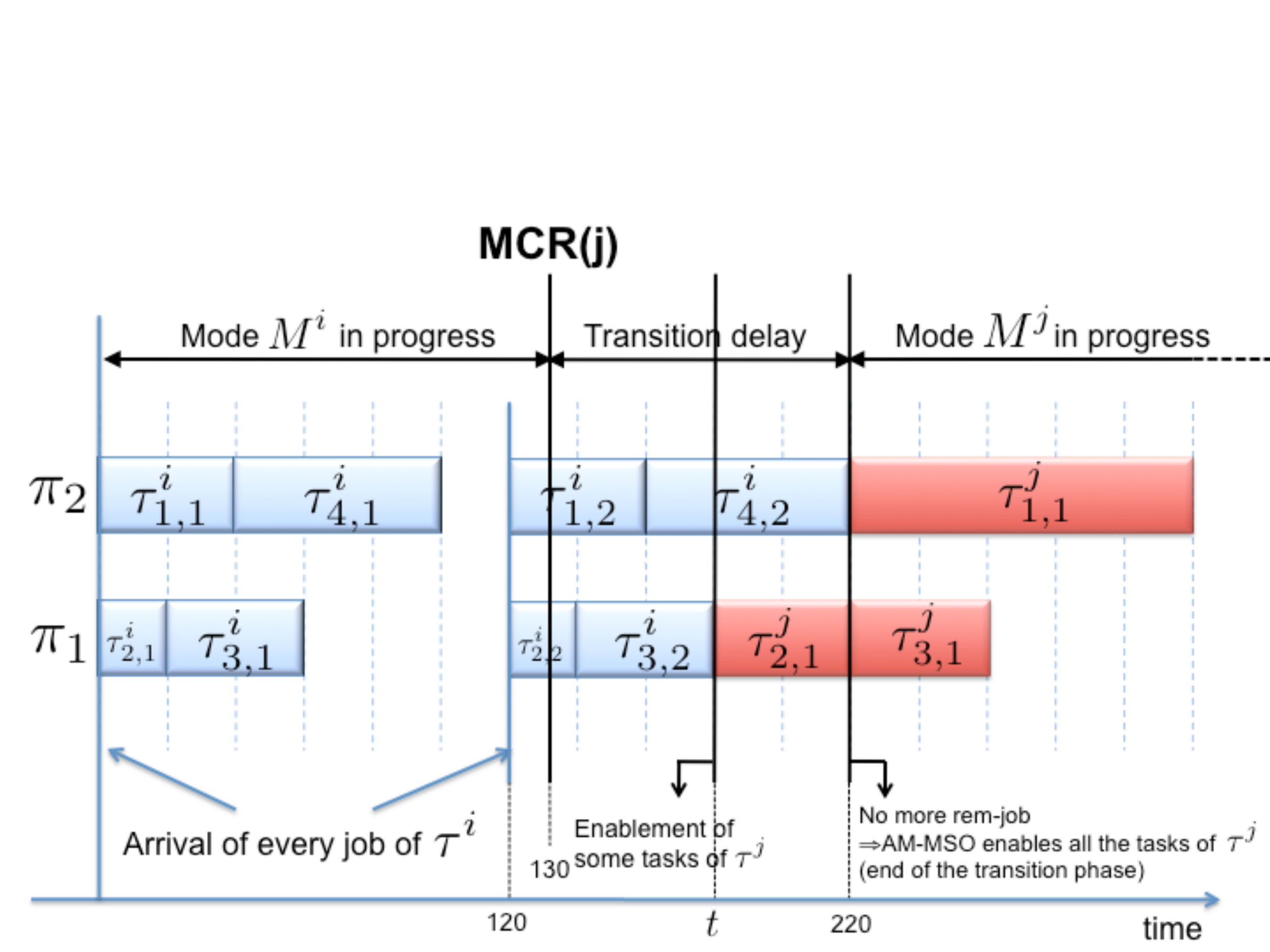}
\caption{Illustration of a mode transition handled by $\AMMSO$.}
\label{fig:Multimode:AMMSO_example}
\end{center}
\end{figure}

Similarly to protocol $\SMMSO$, $\AMMSO$ schedules the rem-jobs according to the old-mode scheduler from time $t_{\MCR(j)} = 130$ to time $t$. Then at time $t$, the rem-job $\tau_{3,2}^i$ completes on $\cpu$ $\pi_1$ and there is no more waiting rem-jobs. Here $\AMMSO$ reacts differently from $\SMMSO$: it scans every disabled task of $\tau^j$ (in non-decreasing order of transition deadline) and enables some of them in such a manner that the resulting set of enabled new-mode tasks can be scheduled by ${\cal S}^j$ upon 1 $\cpu$ (since at this time $t$, only the $\cpu$ $\pi_1$ is available for executing some new-mode tasks). We actually have no guarantee that scanning all the disabled tasks in non-decreasing order of transition deadline is optimal, but this \emph{heuristic} appears as the most intuitive choice. At time 220, $\AMMSO$ performs the same treatment as at time $t$. But since we assumed that every task set $\tau^k$, $\forall k$, is schedulable by ${\cal S}^k$ on $\pi$, we know that \emph{all} the remaining disabled new-mode tasks can be enabled at this time 220. 
\end{Example}

Notice that, in contrast to $\SMMSO$, the protocol $\AMMSO$ allows mode changes to be requested during the mode transitions \emph{only} until some new-mode tasks have been enabled (the instant $t$ in Figure~\ref{fig:Multimode:AMMSO_example}). Indeed, if the system is transitioning from any mode $\mode^i$ to any other mode $\mode^j$ and a mode change is requested to any mode $\mode^z$ before time $t$, then $\AMMSO$ can consider that the system is transitioning from mode $\mode^i$ to mode $\mode^z$ and the new-mode therefore becomes the mode $\mode^z$. However after time $t$, some tasks of mode $\mode^j$ have already been enabled and $\AMMSO$ does not allow the system to request any other mode change until the end of the transition phase from $\mode^i$ to $\mode^j$, i.e., until all the tasks of mode $\mode^j$ are enabled. 

In order to determine whether a task can be safely enabled, protocol $\AMMSO$ uses a binary function $\schedulable(\pi, {\cal S}, \tau^\ell)$ that returns $\operatorname{True}$ if and only if the task set $\tau^\ell$ is schedulable by ${\cal S}$ upon $\pi$. This function is essential as we must always guarantee that all the deadlines are met for \emph{all} the jobs in the system, including the deadlines of \emph{all} the new-mode jobs. Considering a specific scheduler ${\cal S}$, such a function can be derived from schedulability tests proposed for ${\cal S}$ in the literature\footnote{To the best of our knowledge, there is no efficient \emph{necessary and sufficient} schedulability test for any multiprocessor scheduler that complies with the requirements specified in Section~\ref{sec:Multimode:scheduler_specifications}. Theodore Baker has proposed in~\cite{Baker:07} a necessary and sufficient schedulability test for arbitrary-deadline sporadic tasks scheduled by Global-$\edf$ but its time-complexity is very high so only small applications can be tested. Fortunately, many sufficient schedulability tests have been proposed for scheduler such as Global-$\edf$ (see for instance~\cite{BakerBaruah:09,Baruah:10,BaruahGoossens:08,BaruahBaker:08,BertCiriLipari:05}) and Global-$\dm$ (see for instance~\cite{Baker:03,BaruahGoossens:08:2,BaruahFisher:07}).}. Algorithm~\ref{algo:AMMSO} provides a pseudo-code for protocol $\AMMSO$.

\begin{figure}
\begin{center}
\begin{minipage}{15cm}
\begin{algorithmic}[1]
\REQUIRE $M^i$: the old mode
\REQUIRE $M^j$: the new-mode
\REQUIRE the rem-jobs
\REQUIRE $t$: the current time during the transition
\REQUIRE $\pi$: the platform (uniform or identical)

\IF{($t$ is the $\MCR$ invoking time)}
 \STATE Disable all the tasks of $\tau^i$
 \STATE Sort the task set ``$\disabled(\tau^j, t)$'' by non-decreasing order of transition deadlines
 \STATE $\pi^{\avl} \leftarrow \emptyset$ 
\ENDIF

\STATE Schedule the rem-jobs according to ${\cal S}^{\operatorname{trans}}$
\IF{(any rem-job $J_{k}$ completes at $t$ on any $\cpu$ $\pi_{\ell}$)}
	\STATE $r \leftarrow$ number of active rem-jobs at time $t$	
	\IF{($r < m$)}		
		\STATE \textit{/* Due to the completion of $J_k$, one $\cpu$ $\notin \pi^{\avl}$ becomes available. */}
		\IF{($\pi$ is identical)}
			\STATE \textit{/* The scheduler is weakly work-conserving. Thus, the $\cpu$ that becomes available is $\pi_{\ell}$ */}
			\STATE $\pi^{\avl} \leftarrow \pi^{\avl} \cup \left\{ \pi_{\ell} \right\}$
		\ELSE
			\STATE \textit{/* The scheduler is strongly work-conserving. Thus, the $\cpu$ that becomes available is the $(m-r)^{\operatorname{th}}$ slowest $\cpu$. */}
			\STATE $\pi^{\avl} \leftarrow \pi^{\avl} \cup \left\{ \pi_{m-r} \right\}$
		\ENDIF
	\ENDIF
   	\FOR{each $\tau^j_r \in \disabled(\tau^j, t)$}
		\STATE $\tau^{\operatorname{temp}} \leftarrow \enabled(\tau^j, t) \cup \left\{ \tau^j_r \right\}$
		\IF{($\schedulable(\pi^{\avl}, {\cal S}^{j}, \tau^{\operatorname{temp}})$)}
    			\STATE enable $\tau^j_r$
		\ENDIF
	\ENDFOR
	\IF{($r = 0$)}
		\STATE enter the new-mode $M^j$
	\ELSE
		\STATE Schedule all the rem-jobs and new-mode jobs according to ${\cal S}^{\operatorname{trans}}$
  	\ENDIF
\ENDIF
\end{algorithmic}
\end{minipage}
\end{center}
\caption{$\AMMSO$ protocol}
\label{algo:AMMSO}
\end{figure}

\begin{Observation}
\label{obs:compressed_algorithm}
The whole ``if--else--endif'' block within lines 11--17 could be replaced with $\pi^{\avl} \leftarrow \pi^{\avl} \cup \left\{ \pi_{m-r} \right\}$ as adding $\pi_{m-r}$ (instead of $\pi_\ell$) to $\pi^{\avl}$ does not make any difference if $\pi$ is identical. However, we preferred to provide the reader with this longer version of the algorithm for sake of pedagogy. The shorter version explained here will be used in the Validity Algorithm~\ref{algo:AMMSO_test} presented on page~\pageref{algo:AMMSO_test}.
\end{Observation}

\subsection{Design of a validity test}
\label{sec:Multimode:AMMSO_validity_test}

For a given application $\tau$ and platform $\pi$, the main idea to determine whether $\AMMSO$ allows to meet all the transition deadlines is to run Algorithm~\ref{algo:AMMSO} for every possible mode transition, while considering the worst-case scenario for each one---the scenario in which the new-mode tasks are enabled as late as possible. From our definition of protocol $\AMMSO$, we know that every instant at which some new-mode tasks are enabled corresponds to an instant at which at least one $\cpu$ has no more rem-job to execute, i.e., an ``idle-instant'' defined as follows.

\begin{Definition}[Idle-instant $\idle{k}(J, \pi, {\cal P})$]
\label{def:Multimode:idle_instants}
Let $J = \left\{ J_1, J_2, \ldots, J_n \right\}$ be any finite set of $n$ synchronous jobs. Let $\pi$ be a uniform multiprocessor platform and let ${\cal P}$ be the job priority assignment used during the schedule of $J$ upon $\pi$. If $S$ denotes that schedule then the idle-instant $\idle{k}(J,\pi, {\cal P})$ (with $k=1, \ldots, m$) is the earliest instant in $S$ such that at least $k$ $\cpu$s idle. 
\end{Definition}

By definition of the protocol $\AMMSO$, and in particular from the definition of ${\cal S}^{\trans}$, a new-mode job never preempts a rem-job during the transition phases. Thereby, during every transition phase, new-mode tasks are enabled at each idle-instant $\idle{k}(J,\pi,{\cal P})$ ($\forall k=1, \ldots, m$) where $J$ is the set of rem-jobs at the $\MCR$ invoking time and ${\cal P}$ is the job priority assignment derived from the old-mode scheduler when the mode change is requested. For obvious reasons, the \emph{exact} values of these idle-instants depend on both the number of jobs in $J$ and their \emph{actual} execution times. Therefore, these exact value cannot be determined at system design-time and the main idea of our validity test is the following. 

{\bf First}, for every mode $\mode^i$ we determine the set $J$ of rem-jobs that leads to the largest idle-instants $\idle{k}(J, \pi, {\cal P})$ ($\forall k \in \left[ 1, m \right]$). From this point forward, we thus refine the definition of the \emph{critical rem-job set} as follows. 

\begin{Definition}[Critical rem-job set $\wcremjobs{i}$]
\label{def:Multimode:worst_case_configuration_2}
Assuming any transition from a specific mode $\mode^i$ to any other mode $\mode^j$, the critical rem-job set $\wcremjobs{i}$ is the set of jobs issued from the tasks of $\tau^i$ that leads to the largest idle-instants. 
\end{Definition}

As it will be shown in Corollary~\ref{cor:Multimode:worst_case_rem_jobs_set} (page~\pageref{cor:Multimode:worst_case_rem_jobs_set}), the critical rem-job set $\wcremjobs{i}$ of every mode $\mode^i$ is the one that contains one job $J_{\ell}$ for each task $\tau^i_{\ell}$ and such that every job $J_{\ell} \in \wcremjobs{i}$ has a processing time equals to $C^i_{\ell}$, i.e., the WCET of $\tau^i_{\ell}$. Informally speaking, the worst-case scenario during any mode transition is the one in which (i) every old-mode task releases a job \emph{exactly} when the mode change is requested and (ii) every released job executes for its WCET. 

{\bf Second}, we determine (for any given set $J$ of jobs) an upper-bound on each idle-instant $\idle{k}(J, \pi, {\cal P})$ (for $k=1, 2, \ldots, m$). As in the previous section (and for the same reason), we distinguish between FTP and FJP schedulers. That is, for FTP schedulers we focus on determining an upper-bound $\maxidle{k}(J, \pi, {\cal P})$ on each  idle-instant $\idle{k}(J, \pi, {\cal P})$ (for $k=1, 2, \ldots, m$) assuming that the job priority assignment ${\cal P}$ is known beforehand, whereas for FJP schedulers, we determine an upper-bound $\maxidle{k}(J, \pi)$ on each  idle-instant $\idle{k}(J, \pi, {\cal P})$, with the interpretation that for every job priority assignment ${\cal X}$: 
\[ \maxidle{k}(J, \pi) \geq \idle{k}(J, \pi, {\cal X}) \]

{\bf Finally}, we simulate Algorithm~\ref{algo:AMMSO} at each of these upper-bounds. That is, we verify whether all the transition deadlines are met while enabling the new-mode tasks at each instant $\maxidle{k}(\wcremjobs{i}, \pi, {\cal P})$  (or $\maxidle{k}(\wcremjobs{i}, \pi)$ depending on the family of the old-mode scheduler). Obviously, if every transition deadline is met during this simulation then it will be met during the \emph{actual} execution of the application. 

It goes without saying that the prior knowledge of the jobs priority assignment allows for establishing tighter upper-bounds on the idle-instants, i.e., the upper-bounds $\maxidle{k}(J, \pi, {\cal P})$ are tighter than $\maxidle{k}(J, \pi)$. Notice that it results from these notations that $\maxidle{\textbf{m}}(J, \pi)$ and $\maxidle{\textbf{m}}(J, \pi, {\cal P})$ correspond to the upper-bounds $\maxmakespan(\wcremjobs{i}, \pi)$ and $\maxmakespan(\wcremjobs{i}, \pi, {\cal P}^i)$ introduced in Validity Test~\ref{test:Multimode:SMMSO_first}, respectively. 

Mathematical expressions of these upper-bounds $\maxidle{k}(\wcremjobs{i}, \pi)$ and $\maxidle{k}(\wcremjobs{i}, \pi, {\cal P}^i)$ on the $k^{\operatorname{th}}$ idle-instants are defined for both identical and uniform platforms in Sections~\ref{sec:Multimode:ident_FJP}--\ref{sec:Multimode:unif_FTP}. Algorithm~\ref{algo:AMMSO_test} provides details on the validity test for $\AMMSO$, where the upper-bounds $\maxidle{k}(\wcremjobs{i}, \pi)$ must be replaced with $\maxidle{k}(\wcremjobs{i}, \pi, {\cal P}^i)$ at line~9 if the old-mode scheduler is FTP. 

\begin{figure}[h!]
\begin{center}
\begin{minipage}{10cm}
\begin{algorithmic}[1]
\REQUIRE $\tau = \left\{ \tau^1, \tau^2, \ldots, \tau^x \right\}$
\FOR{(all $i, j \in [1, x]$ such that $i \neq j$)}
	\STATE $\tau^{\disabled} \leftarrow \tau^j$
	\STATE $\tau^{\enabled} \leftarrow \emptyset$
	\STATE $\pi^{\avl} \leftarrow \emptyset$ 
	\STATE Sort $\tau^{\disabled}$ by non-decreasing order of transition deadlines
	\FOR{($k=1; k \le m; k\mbox{++}$)}
		\STATE $\pi^{\avl} \leftarrow \pi^{\avl} \cup \pi_{k}$
		\FOR{(all $\tau^j_r \in \tau^{\disabled}$)}
			\IF{(${\cal D}_r^j(\mode^i) < \maxidle{k}(\wcremjobs{i}, \pi)$)}
				\RETURN \FALSE
			\ENDIF
			\IF{($\schedulable\left(\pi^{\avl}, {\cal S}^{j}, \tau^{\enabled} \cup \left\{ \tau^j_r \right\} \right)$)}
				\STATE $\tau^{\enabled} \leftarrow \tau^{\enabled} \cup \left\{\tau^j_r\right\}$
				\STATE $\tau^{\disabled} \leftarrow \tau^{\disabled} \setminus \left\{\tau^j_r\right\}$
			\ENDIF
		\ENDFOR
	\ENDFOR
\ENDFOR
\RETURN \TRUE
\end{algorithmic}
\end{minipage}
\end{center}
\caption{Validity Test for $\AMMSO$}
\label{algo:AMMSO_test}
\end{figure}

Notice that Algorithm~\ref{algo:AMMSO_test} enables new-mode tasks only at the instants $\maxidle{k}(\wcremjobs{i}, \pi)$ (with $k = 1, 2, \ldots, m$). That is, it implicitly considers that every instant at which $\cpu$s become available to the new-mode tasks are as late as possible. As a consequence, if all the transition deadlines are met while running Algorithm~\ref{algo:AMMSO_test} then all these deadlines will be met during every transition phase at run-time\footnote{Because Algorithm~\ref{algo:AMMSO_test} considers every transition between every pair of modes of the application.}. Nevertheless, the fact that Algorithm~\ref{algo:AMMSO_test} simulates every idle-instant of every mode transition by its corresponding upper-bound $\maxidle{k}(\wcremjobs{i}, \pi)$ brings about the following situation: during the actual execution of the application, there could be some intervals of time (during any mode transition) during which the set of currently enabled new-mode tasks benefits from more (and \emph{faster}) $\cpu$s than during the execution of Algorithm~\ref{algo:AMMSO_test}. This kind of situation can occur upon identical and uniform platforms and for both FJP and FTP schedulers as shown in the following example. 

\begin{Example}
Let us consider a $5$-processors uniform platform~$\pi$ and a system which is transitioning from mode $\mode^i$ to mode $\mode^j$. Other details such as the $\cpu$ speeds, the characteristics of the jobs and the job priority assignment are not relevant in the scope of this example. Figures~\ref{fig:Multimode:early_activation_1} and~\ref{fig:Multimode:early_activation_2} illustrate a situation where during some intervals of time the set of currently enabled new-mode tasks benefits from more (and \emph{faster}) $\cpu$s than during the execution of Algorithm~\ref{algo:AMMSO_test}. 
\begin{figure}[h!]
\begin{center}
\includegraphics*[width=0.5\linewidth, viewport=0 0 650 320]{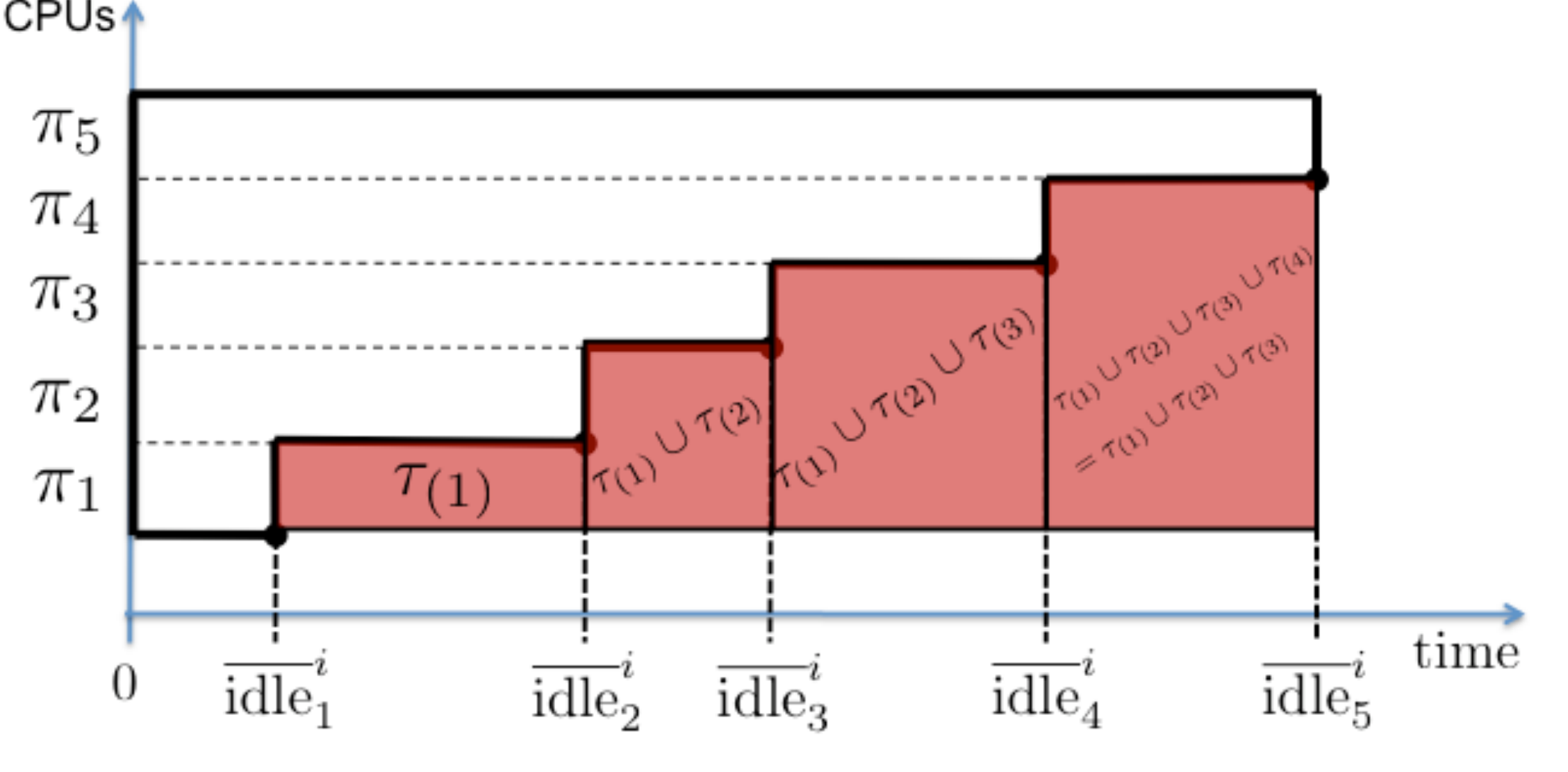}
\caption{Illustration of the schedule assumed by the execution of Algorithm~\ref{algo:AMMSO_test}. In this schedule, new-mode tasks are enabled at each instant $\maxidle{k}$, $1 \leq k \leq m$.}
\label{fig:Multimode:early_activation_1}
\end{center}
\end{figure}

\begin{figure}[h!]
\begin{center}
\includegraphics*[width=0.5\linewidth, viewport=0 0 650 380]{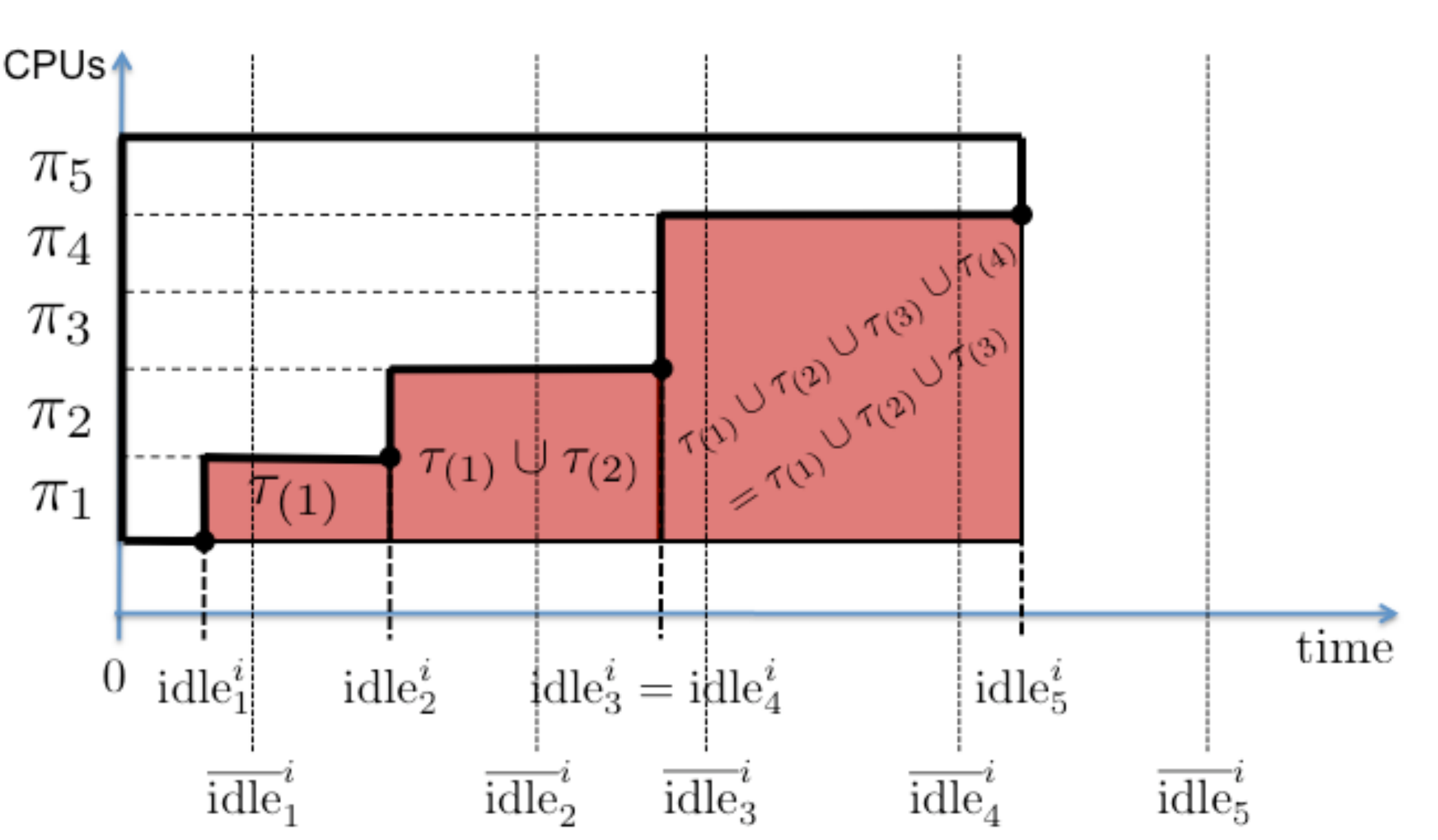}
\caption{Illustration of a possible schedule during a transition from mode $\mode^i$ to mode $\mode^j$ in the \emph{actual} execution of the application. Here, new-mode tasks are enabled at each instant $\idle{k}$, $1 \leq k \leq m$, where $\idle{k} \leq \maxidle{k}$.}
\label{fig:Multimode:early_activation_2}
\end{center}
\end{figure}

For sake of clarity, Figure~\ref{fig:Multimode:early_activation_2} uses the notations $\idle{k}^{i}$ and $\maxidle{k}^{i}$ instead of $\idle{k}(\wcremjobs{i}, \pi)$ and $\maxidle{k}(\wcremjobs{i}, \pi)$, respectively. In this latter schedule, there can be less rem-jobs and/or rem-jobs with lower processing times than in the schedule of Figure~\ref{fig:Multimode:early_activation_1} since the schedule of Figure~\ref{fig:Multimode:early_activation_1} is drawn while assuming the critical rem-job set of mode $\mode^i$. This is the reason why the schedule of Figure~\ref{fig:Multimode:early_activation_2} seems less ``loaded'' than the one of Figure~\ref{fig:Multimode:early_activation_1}. Due to the fact that (i) the validity test provided by Algorithm~\ref{algo:AMMSO_test} uses the same function $\schedulable(\pi, {\cal S}, \tau)$ as protocol $\AMMSO$ at run-time and (ii) this function $\schedulable(\pi, {\cal S}, \tau)$ is independent of the current time, we know that the set of tasks enabled at each instant $\idle{k}^{i}$ ($k=1, 2, \ldots, m$) in Figure~\ref{fig:Multimode:early_activation_2} \emph{is the same as} the set of tasks enabled at each instant $\maxidle{k}^{i}$ in Figure~\ref{fig:Multimode:early_activation_1}. Let us temporarily name this property the ``equivalence property''. Let $\tau_{(k)}$ temporarily denote the set of tasks enabled at time $\maxidle{k}^{i}$, $\forall k \in \left[ 1, m \right]$ and suppose that at time $\maxidle{3}^{i}$ in Figure~\ref{fig:Multimode:early_activation_1} some tasks are enabled (i.e., $\tau_{(3)} \neq \phi$) and at time $\maxidle{4}^{i}$ no task is enabled, i.e., $\tau_{(4)} = \phi$. Thanks to the equivalence property, we know that the tasks enabled at time $\idle{3}^{i}$ in Figure~\ref{fig:Multimode:early_activation_2} are the tasks of $\tau_{(3)}$ and those enabled at time $\idle{4}^{i}$ are the tasks of $\tau_{(4)}$. Since we assumed in Figure~\ref{fig:Multimode:early_activation_2} that $\idle{3}^{i} = \idle{4}^{i}$, it holds that the tasks enabled at time $\idle{3}^{i}$ are the tasks of $\tau_{(3)} \cup \tau_{(4)} = \tau_{(3)}$ (since $\tau_{(4)} = \phi$). It follows that in the time interval $\left[ \maxidle{3}^{i}, \maxidle{4}^{i} \right]$, only $3$ $\cpu$s are available to the task set $\tau_{(1)} \cup \tau_{(2)} \cup \tau_{(3)}$ in Figure~\ref{fig:Multimode:early_activation_1} while $4$ $\cpu$s are available to this task set in Figure~\ref{fig:Multimode:early_activation_2}. Moreover, during this time interval, the additional $\cpu$ $\pi_4$ in Figure~\ref{fig:Multimode:early_activation_2} is faster (or of equal speed) than every $\cpu$ in the subset of $\cpu$s $\{ \pi_1, \pi_2, \pi_3 \}$ available to $\tau_{(1)} \cup \tau_{(2)} \cup \tau_{(3)}$ in Figure~\ref{fig:Multimode:early_activation_1}. 
\end{Example}

Lemma~\ref{lem:Multimode:no_anomalies} proves that this kind of situation does not jeopardize the schedulability of the application during its execution. 

\begin{Lemma}[See~\cite{MeumeuNelisGoossens:10}]
\label{lem:Multimode:no_anomalies}
Any strongly work-conserving scheduler that is able to schedule a task set $\tau$ upon a uniform platform $\pi = [s_1, \ldots, s_m]$ is also able to schedule $\tau$ upon any uniform platform $\pi^{*}$ such that (i) $\pi^{*} \supseteq \pi$ and (ii) $\forall \pi_k \in \pi^{*}$ and $\pi_k \not\in \pi$ we have $s_k \geq s_m$.
\end{Lemma} 
\begin{proof}
To obtain the proof, it is sufficient to show the lemma for $\pi^{*} = [s_1, \ldots, s_m, s_{m+1}]$ where $s_{m+1} \geq s_m$. The proof is made by contradiction. Suppose there exists a task set $\tau$ that is schedulable by a strongly work-conserving scheduler ${\cal S}$ upon $\pi$, but not upon $\pi^{*} \supseteq \pi$. Consider the schedule upon $\pi^{*}$ of a particular set $J$ of jobs issued from $\tau$ that leads to a deadline miss, and let $J^{*}$ be another set of jobs derived from $J$ by reducing the processing time of each job $J_i$ by the amount of time $J_i$ executes upon the sub-platform $\pi^{*} \backslash \pi$, i.e., upon $\pi_{m+1}$. Since the scheduler is strongly work-conserving, the schedule of $J$ by ${\cal S}$ upon the $\cpu$s in common with $\pi$ is the same as the one that would be produced by ${\cal S}$ for $J^{*}$ upon platform $\pi$. Since a deadline is missed in the schedule of $J$ upon $\pi^{*}$, then a deadline is missed also in the schedule of $J^{*}$ upon $\pi$. But since the scheduler is predictable from Lemma~\ref{lem:Multimode:strongly_schedulers_predictable}, a deadline would be missed on $\pi$ even (a fortiori) with the more demanding jobs set $J$, leading to a contradiction. The lemma follows. 
\end{proof}

Lemma~\ref{lem:Multimode:no_anomalies} is proved while considering uniform platforms and strongly work-conserving schedulers but one can easily show that it also holds for identical platforms and weakly work-conserving schedulers. 

%%%%%%%%%%%%%%%%%%%%%%%%%%%%%%%%%%%%%%%%%%%%%%%%%%%%%%%%%%%%%%%%%%%%%%%%%%%%%%%%%%%%%%%%%%%%%%%%%%%%%%%%%%%%%%%%%%%%%%%
%%%%%%%%%%%%%%%%%%%%%%%%%%%%%%%%%%%%%%%%%%%%%%%%%%%%%%%%%%%%%%%%%%%%%%%%%%%%%%%%%%%%%%%%%%%%%%%%%%%%%%%%%%%%%%%%%%%%%%%
%%%%%%%%%%%%%%%%%%%%%%%%%%%%%%%%%%%%%%%%%%%%%%%%%%%%%%%%%%%%%%%%%%%%%%%%%%%%%%%%%%%%%%%%%%%%%%%%%%%%%%%%%%%%%%%%%%%%%%%
%%%%%%%%%%%%%%%%%%%%%%%%%%%%%%%%%%%%%%%%%%%%%%%%%%%%%%%%%%%%%%%%%%%%%%%%%%%%%%%%%%%%%%%%%%%%%%%%%%%%%%%%%%%%%%%%%%%%%%%
%%%%%%%%%%%%%%%%%%%%%%%%%%%%%%%%%%%%%%%%%%%%%%%%%%%%%%%%%%%%%%%%%%%%%%%%%%%%%%%%%%%%%%%%%%%%%%%%%%%%%%%%%%%%%%%%%%%%%%%

\section{Some basic results for determining validity tests}
\label{sec:Multimode:prelim_validity_tests}

\subsection{Introduction to the three required key results}
\label{sec:Multimode:three_key_results}

\emph{Three} key results are required to establish a validity test for $\SMMSO$ and $\AMMSO$. 

\begin{KeyResult}
It must be proved that disabling the old-mode tasks upon any MCR does not jeopardize the schedulability of the rem-jobs when they continue to be scheduled by the old-mode scheduler. That is, it must be guaranteed that the absolute deadline $d^i_{a,b}$ of every rem-job $\tau^i_{a,b}$ is met during any mode transition from every mode $\mode^i$.
\end{KeyResult}

\begin{KeyResult}
The critical rem-job set $\wcremjobs{i}$ for every mode $\mode^i$ must be determined. Indeed, for every mode transition from mode $\mode^i$ to any other mode $\mode^j$, our validity test (see Algorithm~\ref{algo:AMMSO_test}) determines the upper-bounds on the idle-instants by basing the computations on the corresponding critical rem-job sets $\wcremjobs{i}$ (at line~10). In all cases (i.e., identical or uniform platforms and FJP or FTP schedulers), we will provide a proof that the critical rem-job set $\wcremjobs{i}$ of every mode $\mode^i$ is the one that contains one job $J_{\ell}$ for each task $\tau^i_{\ell}$ and such that every job $J_{\ell} \in \wcremjobs{i}$ has a processing time equals to $C^i_{\ell}$, i.e., the WCET of the task $\tau^i_{\ell}$.
\end{KeyResult}

\begin{KeyResult}
A mathematical expression must be established that provides, for any given set $J$ of jobs and platform $\pi$:
\begin{enumerate}
\item an upper-bound $\maxidle{k}(J, \pi)$ ($1 \leq k \leq m$) on each idle-instant $\idle{k}(J, \pi, {\cal X})$, for every job priority assignment ${\cal X}$. This concerns FJP schedulers.
\item an upper-bound $\maxidle{k}(J, \pi, {\cal P})$ ($1 \leq k \leq m$) on each idle-instant $\idle{k}(J, \pi, {\cal P})$, for a specific job priority assignment ${\cal P}$. This concerns FTP schedulers.
\end{enumerate}
\end{KeyResult}

Note that the protocol $\SMMSO$ requires only an upper-bound on the makespan, i.e., on the $m^{\operatorname{th}}$ idle-instant $\maxidle{m}(J, \pi)$ and $\maxidle{m}(J, \pi, {\cal P})$.

\subsection{Proof of the first key result}
\label{sec:Multimode:prelim_validity_tests_first_result}

Lemma \ref{lem:Multimode:remjobs_meet_deadline} proves the first key result introduced above for any uniform platform and strongly work-conserving scheduler, as well as any identical platform and weakly work-conserving scheduler. This result, which is essential to the validity tests of both protocols $\SMMSO$ and $\AMMSO$, is based on the notion of \emph{predictability} introduced on page \pageref{def:Multimode:predictability}. It has been drawn from~\cite{NelisGoossensAndersson:09} and extended to uniform platforms.

\begin{Lemma}
\label{lem:Multimode:remjobs_meet_deadline}
Let $\mode^i$ and $\mode^j$ denote two distinct modes of the application. If the application is running in mode $\mode^i$ and a $\MCR(j)$ occurs at time $t_{\MCR(j)}$ then every rem-job meets its deadline during the transition phase while being scheduled by the old-mode scheduler ${\cal S}^i$.
\end{Lemma} 
\begin{proof}
From our first assumption on page~\pageref{null:assumption1}, the set of tasks $\tau^i$ of the mode $\mode^i$ is schedulable by ${\cal S}^i$ upon $\pi$. When the $\MCR(j)$ is invoked at time $t_{\MCR(j)}$, the transition protocol disables every old-mode task, which is equivalent to set the processing time of all their future jobs to zero. Since ${\cal S}^i$ is predictable (from Lemma~\ref{lem:Multimode:weakly_schedulers_predictable} or~\ref{lem:Multimode:strongly_schedulers_predictable} depending on the scheduler family), the deadline of every rem-job is still met in the produced schedule. The lemma follows.
\end{proof}

\subsection{Proof of the second key result}
\label{sec:Multimode:prelim_validity_tests_second_result}

Corollary~\ref{cor:Multimode:worst_case_rem_jobs_set} proves the second key result introduced above for any uniform platform and strongly work-conserving FTP (or FJP) scheduler, as well as any identical platform and weakly work-conserving FTP (or FJP) scheduler. It has been drawn from the following Lemma~\ref{lem:Multimode:worst_case_rem_jobs_set_base}.

\begin{Lemma}
\label{lem:Multimode:worst_case_rem_jobs_set_base}
Let $\pi$ be any uniform multiprocessor platforms (including identical platforms) and let $J$ and $J'$ be any fixed set of $n$ synchronous jobs such that $J = \{J_1, J_2, \ldots, J_n\}$ of processing times $c_1, c_2, \ldots, c_n$ and $J' = \{J'_1, J'_2, \ldots, J'_n\}$ of processing times $c'_1, c'_2, \ldots, c'_n$. For any job priority assignment ${\cal P}$, if there exists a bijective function between $J$ and $J'$ such that every job $J'_r \in J'$ is mapped to exactly one job $J_r \in J$ and such that $c'_r \leq c_r$, then the $k^{\operatorname{th}}$ idle-instant $\idle{k}(J, \pi, {\cal P})$ ($\forall k \in \left[ 1, m \right]$) in the schedule of $J$ upon $\pi$ is not lower than the $k^{\operatorname{th}}$ idle-instant $\idle{k}(J',\pi, {\cal P})$ in the schedule of $J'$, i.e., it holds $\forall k \in \left[1, m \right]$ that
\[ \idle{k}(J', \pi, {\cal P}) \leq \idle{k}(J, \pi, {\cal P}) \]
\end{Lemma} 
\begin{proof}
The proof is a \emph{consequence} of the predictability of work-conserving schedulers (including both weakly and strongly work-conserving schedulers). Let $S$ and $S'$ denote the schedule of $J$ and $J'$ upon $\pi$ with ${\cal P}$, respectively. We denote by $\comp{r}$ and $\comp{r}'$ the completion time of any job $J_r$ in $S$ and $J'_r$ in $S'$, respectively. It follows from the fact that $c'_r \leq c_r$ ($\forall r \in \left[ 1, n \right]$) and from the predictability of the considered schedulers (see Lemma~\ref{lem:Multimode:weakly_schedulers_predictable} or~\ref{lem:Multimode:strongly_schedulers_predictable} depending on the scheduler family) that $\forall r \in \left[ 1, n \right]$:
\begin{equation}
\label{equ:Multimode:ident_FTP_worst_case_configuration_temp1}
\comp{r}' \leq \comp{r}
\end{equation}
The proof is made by contradiction. Suppose that there exists $\ell \in \left[ 1, m \right]$ such that 
\[ \idle{\ell}(J, \pi, {\cal P}) < \idle{\ell}(J',\pi, {\cal P}) \]
Figures~\ref{fig:Multimode:worst_case_rem_job_set1_1} and~\ref{fig:Multimode:worst_case_rem_job_set1_2} illustrate an example of schedules $S$ and $S'$ on a $5$-processors \emph{uniform} platform, respectively, where $\idle{3}(J, \pi, {\cal P}) < \idle{3}(J', \pi, {\cal P})$. 
\begin{figure}[t!]
\begin{center}
\includegraphics*[width=0.5\linewidth, viewport=0 0 700 270]{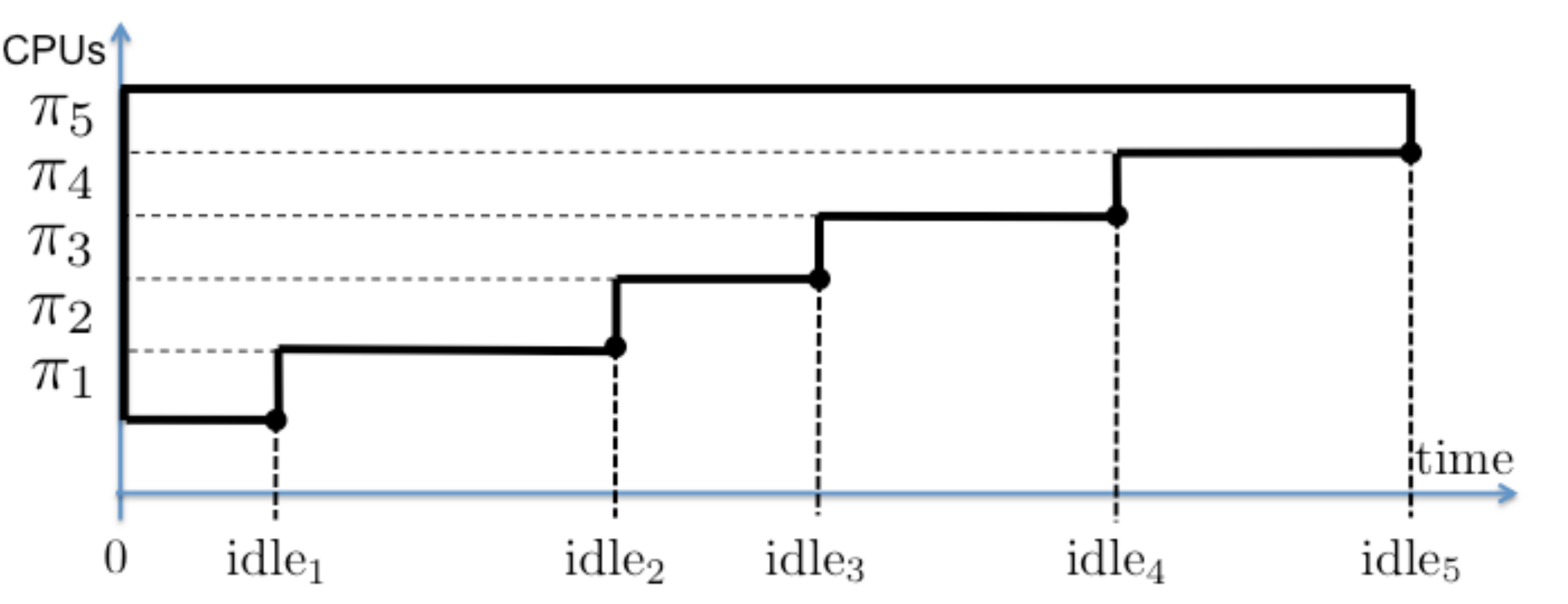}
\caption{An example of schedule $S$ upon a $5$-processors uniform platform. The idle-instants $\idle{k}(J, \pi, {\cal P})$ are denoted by $\idle{k}$ for sake of clarity.}
\label{fig:Multimode:worst_case_rem_job_set1_1}
\end{center}
\end{figure}

\begin{figure}[t!]
\begin{center}
\includegraphics*[width=0.5\linewidth, viewport=0 0 700 330]{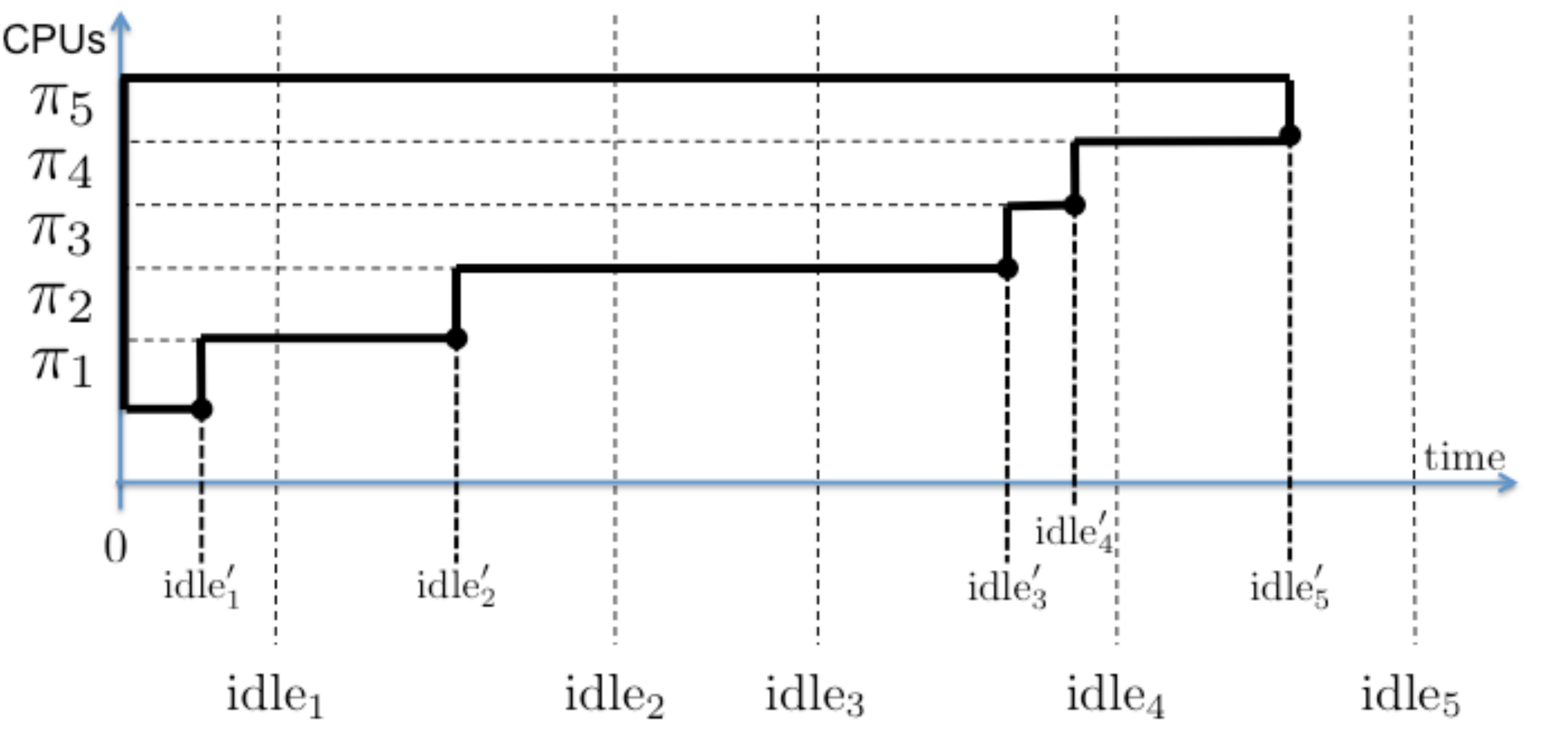}
\caption{An example of schedule $S'$ upon the same $5$-processors uniform platform. Also for sake of clarity, the idle-instants $\idle{k}(J, \pi, {\cal P})$ and $\idle{k}(J', \pi, {\cal P})$ are denoted by $\idle{k}$ and $\idle{k}'$, respectively. In this figure, we have by contradiction $\idle{3} < \idle{3}'$.}
\label{fig:Multimode:worst_case_rem_job_set1_2}
\end{center}
\end{figure}

\noindent Since the platform is uniform in these examples, the scheduler is strongly work-conserving and both schedules $S$ and $S'$ form a staircase. In both Figures~\ref{fig:Multimode:worst_case_rem_job_set1_1} and~\ref{fig:Multimode:worst_case_rem_job_set1_2}, we voluntarily omit the details about the $\cpu$ speeds, the jobs characteristics, etc. since they are useless in the scope of these examples.

Similarly, Figures~\ref{fig:Multimode:worst_case_rem_job_set2_1} and~\ref{fig:Multimode:worst_case_rem_job_set2_2} illustrate an example of schedules $S$ and $S'$ on a $5$-processors \emph{identical} platform, respectively, where $\idle{3}(J, \pi, {\cal P}) < \idle{3}(J', \pi, {\cal P})$. Since the platform is identical in these examples, the scheduler is assumed to be weakly work-conserving.  Furthermore, note that in both examples no job is released after time $0$. 

\begin{figure}[t!]
\begin{center}
\includegraphics*[width=0.5\linewidth, viewport=0 0 700 270]{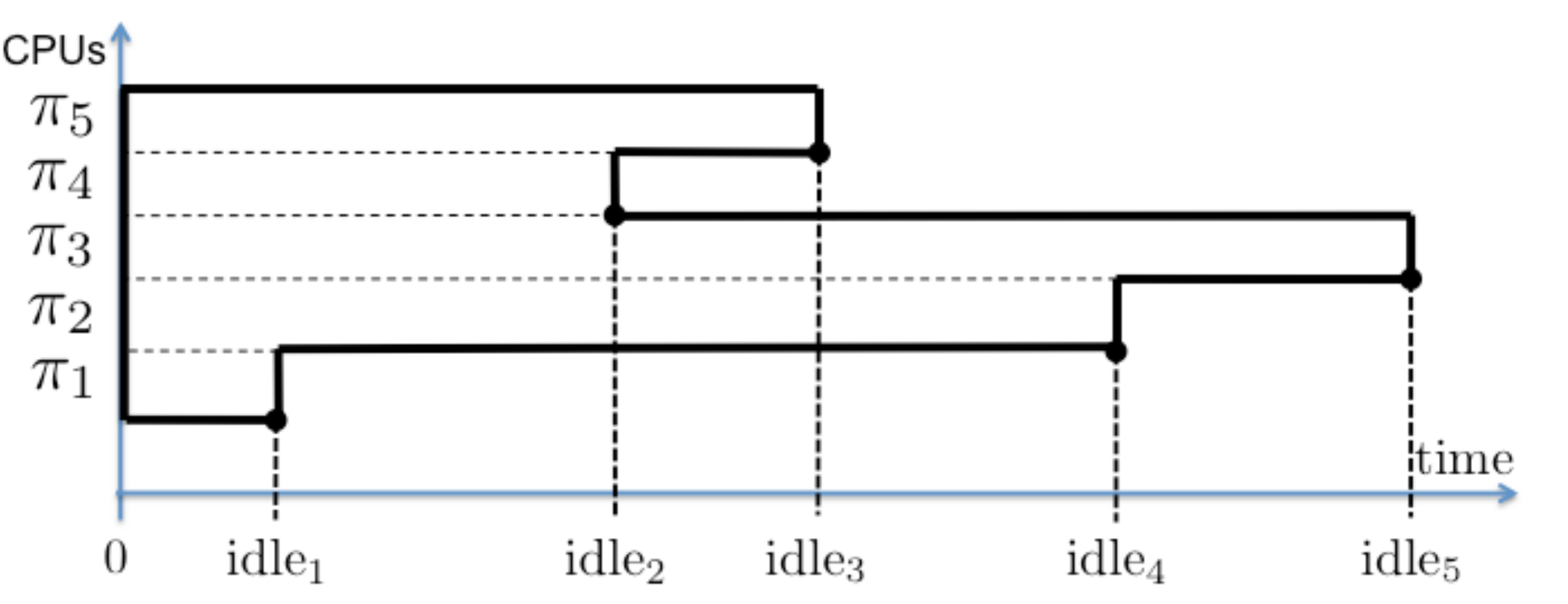}
\caption{An example of schedule $S$ upon a $5$-processors identical platform. The idle-instants $\idle{k}(J, \pi, {\cal P})$ are denoted by $\idle{k}$ for sake of clarity.}
\label{fig:Multimode:worst_case_rem_job_set2_1}
\end{center}
\end{figure}

\begin{figure}[t!]
\begin{center}
\includegraphics*[width=0.5\linewidth, viewport=0 0 700 330]{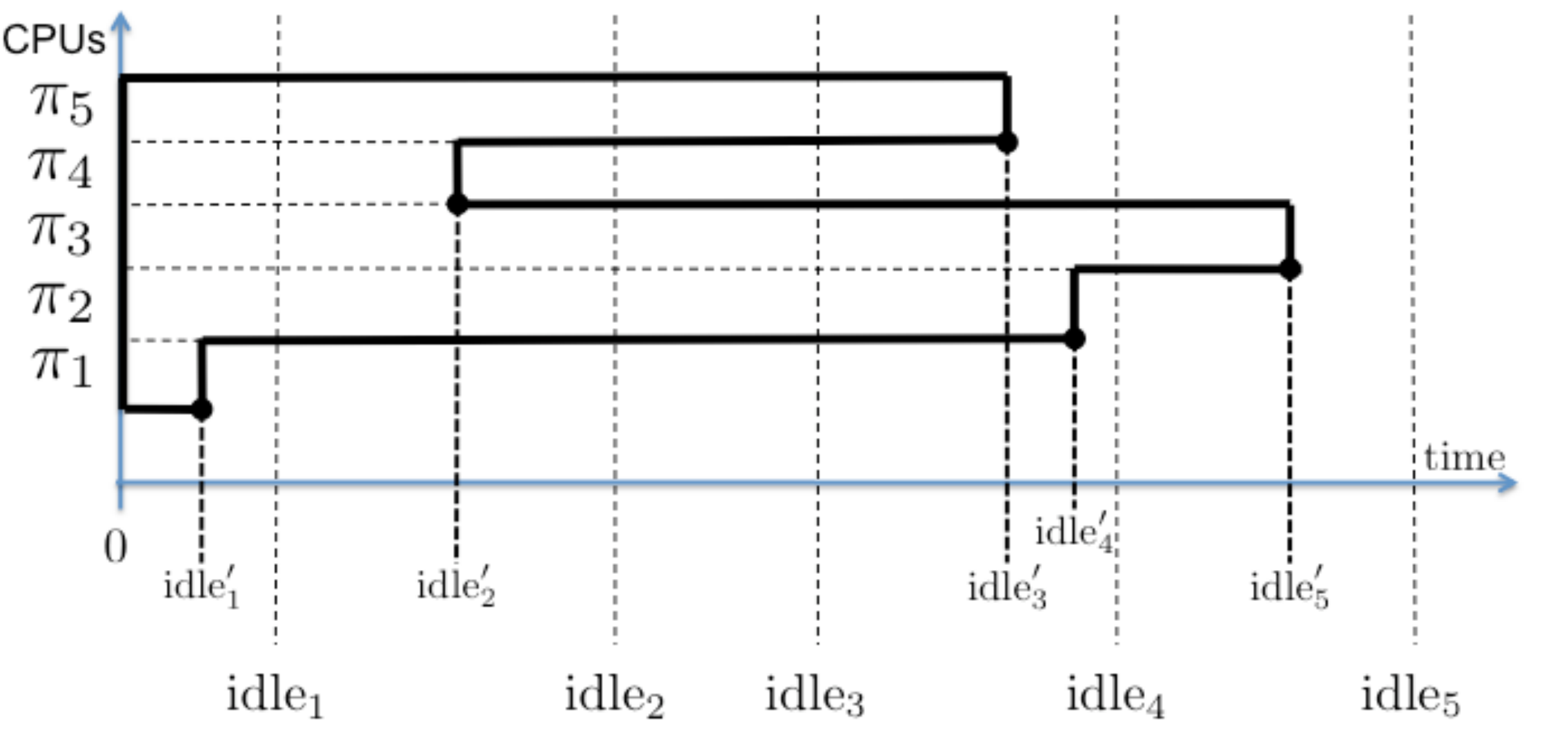}
\caption{An example of schedule $S'$ upon the same $5$-processors identical platform. Also for sake of clarity, the idle-instants $\idle{k}(J, \pi, {\cal P})$ and $\idle{k}(J', \pi, {\cal P})$ are denoted by $\idle{k}$ and $\idle{k}'$, respectively. In this figure, we have by contradiction $\idle{3} < \idle{3}'$.}
\label{fig:Multimode:worst_case_rem_job_set2_2}
\end{center}
\end{figure}

By definition of the idle-instants, the schedule of \emph{any} set ${\cal J}$ of jobs upon \emph{any} uniform or identical multiprocessor platform is such that $\forall k \in \left[ 1, m \right]$:
\begin{itemize}
\item the idle-instant $\idle{k}({\cal J}, \pi, {\cal P})$ corresponds to the completion time of a job,
\item there is no waiting job at time $\idle{k}({\cal J}, \pi, {\cal P})$ and,
\item there are \emph{at most} ($m-k$) running jobs at time $\idle{k}({\cal J}, \pi, {\cal P})$. ``At most'' since there can exist some $r > k$ such that $\idle{r}({\cal J}, \pi, {\cal P}) = \idle{k}({\cal J}, \pi, {\cal P})$.
\end{itemize}

Since every idle-instant corresponds to the completion of a job, this implies that within the time interval $\left[ \idle{\ell}(J, \pi, {\cal P}), \idle{\ell}(J', \pi, {\cal P}) \right]$ there are \emph{at most} $(m-\ell)$ running jobs in $S$ while there are \emph{at least} $(m-\ell+1)$ running jobs in $S'$. Therefore, within $\left[ \idle{\ell}(J, \pi, {\cal P}), \idle{\ell}(J', \pi, {\cal P}) \right]$, at least one job (say $J_r$) is already completed in $S$ while $J'_r$ is still running in $S'$. The fact that $J'_r$ completes later in $S'$ than $J_r$ in $S$ leads to a direct contradiction of Inequality~\ref{equ:Multimode:ident_FTP_worst_case_configuration_temp1}. As we can see in Figures~\ref{fig:Multimode:worst_case_rem_job_set1_2} and~\ref{fig:Multimode:worst_case_rem_job_set2_2}, three jobs are running in $S'$ during the time interval $\left[ \idle{3}(J, \pi, {\cal P}), \idle{3}(J', \pi, {\cal P}) \right]$ while only two jobs are running in $S$, meaning that there is one job which is completed in $S$ and still running in $S'$. The lemma follows.
\end{proof}

\begin{Corollary}
\label{cor:Multimode:worst_case_rem_jobs_set}
For any uniform multiprocessor platforms $\pi$ and for any transition of the system from mode $\mode^i$ to mode $\mode^j$, let $J^{\any}$ denote any set of rem-jobs issued from the old-mode tasks and let $\wcremjobs{i}$ be the set of rem-jobs that contains one job $J_{\ell}$ for each task $\tau^i_{\ell}$ and such that every job $J_{\ell} \in \wcremjobs{i}$ has a processing time equals to $C^i_{\ell}$. The $k^{\operatorname{th}}$ idle-instants $\idle{k}(\wcremjobs{i}, \pi)$ ($\forall k \in \left[ 1, m \right]$) in the schedule of $\wcremjobs{i}$ is never lower than the $k^{\operatorname{th}}$ idle-instant $\idle{k}(J^{\any},\pi)$ in the schedule of $J^{\any}$, i.e., it holds $\forall k \in \left[1, m \right]$ that
\[ \idle{k}(J^{\any}, \pi) \leq \idle{k}(\wcremjobs{i}, \pi) \]
\end{Corollary} 
\begin{proof}
The proof is a \emph{consequence} of Lemma~\ref{lem:Multimode:worst_case_rem_jobs_set_base}. Let $c_r^{\operatorname{wc}}$ and $c_r^{\any}$ denote the processing time of job $J_r$ in $\wcremjobs{i}$ and $J^{\any}$, respectively. By definition, $\wcremjobs{i}$ contains one job $J_r$ of processing time $C^i_r$ for each task $\tau^i_r \in \tau^i$, i.e., it holds $\forall \tau^i_r \in \tau^i$ that
\[ c_r^{\operatorname{wc}} = C^i_r \]  
and thus we know by definition of $J^{\any}$ that $\forall J_r \in J^{\any}$, 
\[ c_r^{\any} \leq c_r^{\operatorname{wc}} \]
In addition, we know that there could be some jobs $J_{\ell} \in \wcremjobs{i}$ such that $J_{\ell} \not\in J^{\any}$ (since $J^{\any}$ does not necessarily contain one job for each old-mode task). For each such job $J_{\ell}$ we add a fake job $J'_{\ell}$ in $J^{\any}$ with $c_{\ell}^{\any} = 0$. It results from this operation that the number of jobs in both $\wcremjobs{i}$ and $J^{\any}$ are the same (we denote this number by $n$) and there is a bijective function between $\wcremjobs{i}$ and $J^{\any}$ such that every job $J_r \in \wcremjobs{i}$ is mapped to by exactly one job $J_r \in J^{\any}$ and such that $c_r^{\any} \leq c_r^{\operatorname{wc}}$. Thanks to this bijection, we know from Lemma~\ref{lem:Multimode:worst_case_rem_jobs_set_base} that $\forall r \in \left[ 1, m \right]$ we have
\[ \idle{r}(J^{\any},\pi) \leq \idle{r}(\wcremjobs{i}, \pi) \]
and the corollary follows.
\end{proof}

By definition, for every mode transition from any mode $\mode^i$ upon $\pi$, each $\maxidle{k}(\wcremjobs{i}, \pi)$ is an upper-bound on the $k^{\operatorname{th}}$ idle-instant in the schedule of $\wcremjobs{i}$ (this also holds for each upper-bound $\maxidle{k}(\wcremjobs{i}, \pi, {\cal P})$ if the job priority assignment ${\cal P}$ is known beforehand). Thanks to Corollary~\ref{cor:Multimode:worst_case_rem_jobs_set}, we are now aware that each upper-bound $\maxidle{k}(\wcremjobs{i}, \pi)$ (and $\maxidle{k}(\wcremjobs{i}, \pi, {\cal P})$) is also an upper-bound on the $k^{\operatorname{th}}$ idle-instant in the schedule of \emph{any} other set of rem-jobs issued from the old-mode tasks (i.e., the tasks of $\tau^i$). That is, for every mode transition from any mode $\mode^i$ we have $\forall k \in \left[ 1, m\right]$:
$\maxidle{k}(\wcremjobs{i}, \pi) \geq \idle{k}(\wcremjobs{i}, \pi) \geq \idle{k}(J^{\any}, \pi)$ %\nonumber \\
\mbox{and} $\maxidle{k}(\wcremjobs{i}, \pi, {\cal P}) \geq \idle{k}(\wcremjobs{i}, \pi, {\cal P}) \geq \idle{k}(J^{\any}, \pi, {\cal P})$, where $J^{\any}$ denotes any set of rem-jobs issued from the tasks of $\tau^i$. As a result, the instants $\maxidle{k}(\wcremjobs{i}, \pi)$ (and $\maxidle{k}(\wcremjobs{i}, \pi, {\cal P})$), with $k=1, 2, \ldots, m$, can be considered as the \emph{largest} instants at which new-mode tasks are enabled during every transition from mode $\mode^i$ and thus, these instants can be used in our validity test given by Algorithm~\ref{algo:AMMSO_test}. 

\subsection{Organization for the third key result}
The third key result consists in determining a mathematical expression for each upper-bound $\maxidle{k}(J, \pi)$ (or $\maxidle{k}(J, \pi, {\cal P}$) depending on the scheduler family, i.e., FJP or FTP), for all $1 \leq k \leq m$. Depending on the type of the platform (uniform or identical) and on the scheduler family (FJP or FTP), we distinguish between four different cases that are studied in turn in the following four sections. More precisely:

\begin{itemize}
\renewcommand{\labelitemi}{$\triangleright$}
\item Section~\ref{sec:Multimode:ident_FJP} addresses the \emph{identical} and \emph{FJP} case.
\item Section~\ref{sec:Multimode:ident_FTP} addresses the \emph{identical} and \emph{FTP} case. 
\item Section~\ref{sec:Multimode:unif_FJP} addresses the \emph{uniform} and \emph{FJP} case.
\item Section~\ref{sec:Multimode:unif_FTP} addresses the \emph{uniform} and \emph{FTP} case. 
\end{itemize}

Recall that the protocol $\SMMSO$ requires only an upper-bound on the makespan, i.e., on the $m^{\operatorname{th}}$ idle-instant. The organization for the third key result is as follows.

%%%%%%%%%%%%%%%%%%%%%%%%%%%%%%%%%%%%%%%%%%%%%%%%%%%%%%%%%%%%%%%%%%%%%%%%%%%%%%%%%%%%%%%%%%%%%%%%%%%%%%%%%%%%%%%%%%%%%%%
%%%%%%%%%%%%%%%%%%%%%%%%%%%%%%%%%%%%%%%%%%%%%%%%%%%%%%%%%%%%%%%%%%%%%%%%%%%%%%%%%%%%%%%%%%%%%%%%%%%%%%%%%%%%%%%%%%%%%%%
%%%%%%%%%%%%%%%%%%%%%%%%%%%%%%%%%%%%%%%%%%%%%%%%%%%%%%%%%%%%%%%%%%%%%%%%%%%%%%%%%%%%%%%%%%%%%%%%%%%%%%%%%%%%%%%%%%%%%%%
%%%%%%%%%%%%%%%%%%%%%%%%%%%%%%%%%%%%%%%%%%%%%%%%%%%%%%%%%%%%%%%%%%%%%%%%%%%%%%%%%%%%%%%%%%%%%%%%%%%%%%%%%%%%%%%%%%%%%%%
%%%%%%%%%%%%%%%%%%%%%%%%%%%%%%%%%%%%%%%%%%%%%%%%%%%%%%%%%%%%%%%%%%%%%%%%%%%%%%%%%%%%%%%%%%%%%%%%%%%%%%%%%%%%%%%%%%%%%%%

\section{Identical platforms and FJP schedulers}
\label{sec:Multimode:ident_FJP}

This section is organized as follows. First, Section~\ref{sec:Multimode:ident_FJP_upper_bounds} determines an \emph{upper-bound} $\maxidle{k}(J, \pi)$ on the earliest time-instant where at least $k$ $\cpu$s are idle and derives an upper-bound $\maxmakespan(J, \pi)$ on the maximum makespan. Then, Section \ref{sec:FJP_Accuracy} shows that this upper-bound $\maxmakespan(J, \pi)$ is 2-competitive, with the interpretation that $\maxmakespan(J, \pi)$ is at most twice the exact value of the maximum makespan. Finally, Section \ref{sec:Multimode:ident_FJP_validity_test} establishes a \emph{sufficient} validity test for protocols $\SMMSO$ and $\AMMSO$. 

\subsection{Upper-bounds $\maxidle{k}(J, \pi)$ on the idle-instants}
\label{sec:Multimode:ident_FJP_upper_bounds}

Throughout this section, $J$ refers to any set of $n$ jobs. For sake of clarity, we will use the notation $\maxidle{k}$ instead of $\maxidle{k}(J, \pi)$ and similarly, we will use the notation $\idle{k}$ to denote the \emph{exact} value of the $k^{\operatorname{th}}$ idle-instant. Before introducing the computation of these upper-bounds $\maxidle{k}$, $1 \leq k \leq m$, let us introduce the following result taken from~\cite{NelisGoossensAndersson:09}. 

\begin{Lemma}[See~\cite{NelisGoossensAndersson:09}]
\label{lem:Multimode:ident_FJP_lem1}
Suppose that $J$ is sorted by non-decreasing job processing times, i.e., $c_1 \leq c_2 \leq \cdots \leq c_n$. Then, whatever the job priority assignment we have $\forall j,k \in \left[ 1,m \right]$ such that $j < k$:
\[ \idle{j} \geq \idle{k} - c_{n-m+k} \]
\end{Lemma} 

Based on this Lemma~\ref{lem:Multimode:ident_FJP_lem1}, the following result was proved in our previous work~\cite{NelisGoossensAndersson:09}. 

\begin{Lemma}[See~\cite{NelisGoossensAndersson:09}]
\label{lem:Multimode:ident_FJP_maxidle_old}
Suppose that $J$ is sorted by non-decreasing job processing times, i.e., $c_1 \leq c_2 \leq \cdots \leq c_n$. Then, whatever the job priority assignment, an upper-bound $\maxidle{k}$ on the idle-instant $\idle{k}$, $1 \leq k \leq m$, is given by $c_k$ if $n = m$ or by

\begin{small}
\begin{equation}
\label{equ:Multimode:ident_FJP_maxidle_old}
\max_{i=0}^{n-m+k-1} \left\{ \frac{\sum_{j=1}^{n} c_j - \sum_{j=i+1}^{i+m-k+1} c_j}{m} + \frac{\sum_{j=i+1}^{i+m-k+1} c_j}{m-k+1}\right\}
\end{equation}
\end{small}
otherwise ($n > m$).
\end{Lemma} 

Holding this result, we improve here this previous analysis by (i) successfully establishing \emph{another} upper-bound $\maxidle{k}(J, \pi)$ on each idle-instant $\idle{k}(J, \pi)$ and (ii) proving that these alternative upper-bounds are always tighter than those proposed in Lemma~\ref{lem:Multimode:ident_FJP_maxidle_old}. In short, we complete our previous work~\cite{NelisGoossensAndersson:09} as follows.

\begin{itemize}
\renewcommand{\labelitemi}{$\triangleright$}
\item Lemma~\ref{lem:Multimode:ident_FJP_maxidle_old_max} shows that Expression~\ref{equ:Multimode:ident_FJP_maxidle_old} of $\maxidle{k}$ is always maximal for $i = n-m+k-1$.
\item Lemma~\ref{lem:Multimode:ident_FJP_maxidle} proposes another upper-bound $\maxidle{k}(J, \pi)$ on each idle-instant $\idle{k}$.
\item Lemma~\ref{lem:Multimode:ident_FJP_maxidle_better} shows that these alternative upper-bounds $\maxidle{k}(J, \pi)$, $\forall k \in \left[ 1, m \right]$, are never larger than those provided by Expression~\ref{equ:Multimode:ident_FJP_maxidle_old}. 
\item Finally, based on these alternative upper-bounds, Corollary~\ref{cor:Multimode:ident_FJP_makespan} derives an upper-bound on the makespan. 
\end{itemize}

\begin{Lemma}
\label{lem:Multimode:ident_FJP_maxidle_old_max}
If $n > m$, Expression~\ref{equ:Multimode:ident_FJP_maxidle_old} is maximal for $i = n-m+k-1$.
\end{Lemma} 
\begin{proof}
This result is presented in Lemma~2.10 in~\cite{Nelis:10}. Due to the space limitation and because the proof is simply based on algebra, we do not repeat it here.
\end{proof}

Thanks to Lemma~\ref{lem:Multimode:ident_FJP_maxidle_old_max}, Expression~\ref{equ:Multimode:ident_FJP_maxidle_old} can be rewritten as follows: $\maxidle{k} \equals c_k$ if $n = m$ or 
\begin{equation}
\label{equ:Multimode:ident_FJP_maxidle_old_2}
\maxidle{k} \equals \displaystyle\frac{\sum_{j=1}^{n} c_j - \sum_{j=n-m+k}^{n} c_j}{m} + \displaystyle\frac{\sum_{j=n-m+k}^{n} c_j}{m-k+1}
\end{equation}
otherwise ($n > m$). 

\begin{Lemma}
\label{lem:Multimode:ident_FJP_maxidle}
Suppose that $J$ is sorted by non-decreasing job processing times, i.e., $c_1 \leq c_2 \leq \cdots \leq c_n$. Then, whatever the job priority assignment, an upper-bound $\maxidle{k}$ on the idle-instant $\idle{k}$, $1 \leq k \leq m$, is given by $\maxidle{k} \equals c_k$ if $n = m$ or by
\begin{equation}
\label{equ:Multimode:ident_FJP_maxidle}
\maxidle{k} \equals \displaystyle\frac{\sum_{i=1}^{n} c_i + (k-1) \cdot c_{n-m+k}}{m}
\end{equation}
otherwise ($n > m$).
\end{Lemma} 
\begin{proof}
The case where $n = m$ is obvious. Otherwise, the proof is made by contradiction. Suppose that there exists $k \in \left[ 1, m \right]$ such that $\idle{k} > \maxidle{k}$. The following properties hold:
\begin{itemize}
\item \textbf{Prop. (a):} $\forall j > k$: $\idle{j} \geq \idle{k}$ (by definition of the idle-instants).
\item \textbf{Prop. (b):} $\forall j < k$: $\idle{j} \geq \idle{k} - c_{n-m+k}$ (from Lemma~\ref{lem:Multimode:ident_FJP_lem1}).
\end{itemize}
\noindent The proof starts with this obvious equality:
\[ \sum_{j=1}^{m} \idle{j} = \sum_{j=1}^{k-1} \idle{j} + \idle{k} + \sum_{j=k+1}^{m} \idle{j} \]
\noindent Then, applying properties (a) and (b) to the right-hand side yields
\begin{eqnarray}
\sum_{j=1}^{m} \idle{j} & \geq & \sum_{j=1}^{k-1} (\idle{k} - c_{n-m+k}) + \idle{k} + \sum_{j=k+1}^{m} \idle{k} \nonumber \\
& \geq & (k-1) (\idle{k} - c_{n-m+k}) + \idle{k} \nonumber \\
& & + (m-k) \cdot \idle{k} \nonumber \\
& \geq &  m \cdot \idle{k} - (k-1) \cdot c_{n-m+k} \nonumber
\end{eqnarray}
Since by hypothesis $\idle{k} > \maxidle{k}$, replacing $\idle{k}$ with $\maxidle{k}$ in the above inequality leads to
\begin{eqnarray}
\sum_{j=1}^{m} \idle{j} & > & m \cdot \maxidle{k} - (k-1) \cdot c_{n-m+k} \nonumber \\
& > & m \left(\frac{\sum_{i=1}^{n} c_i + (k-1) \cdot c_{n-m+k}}{m} \right)  \nonumber \\
& & - (k-1) \cdot c_{n-m+k} \nonumber \\
& > & \sum_{i=1}^{n} c_i \nonumber
\end{eqnarray}
\noindent This leads to a contradiction since it obviously holds by definition of the idle-instants that $\sum_{j=1}^m \idle{j} = \sum_{i=1}^n c_i$. The lemma follows.
\end{proof}

\begin{Lemma}
\label{lem:Multimode:ident_FJP_maxidle_better}
The upper-bounds $\maxidle{k}$ (with $k=1, 2, \ldots, m$) provided by Expression~\ref{equ:Multimode:ident_FJP_maxidle} are never larger than those provided by Expression~\ref{equ:Multimode:ident_FJP_maxidle_old_2}.
\end{Lemma} 
\begin{proof}
The proof is made by contradiction. Let $k$ be any integer in $\left[ 1, m \right]$. Let $\maxidle{k}^{\operatorname{old}}$ and $\maxidle{k}^{\operatorname{new}}$ denote the upper-bound provided by Expressions~\ref{equ:Multimode:ident_FJP_maxidle_old_2} and~\ref{equ:Multimode:ident_FJP_maxidle}, respectively, and suppose that $\maxidle{k}^{\operatorname{new}} > \maxidle{k}^{\operatorname{old}}$. From Expressions~\ref{equ:Multimode:ident_FJP_maxidle_old_2} and~\ref{equ:Multimode:ident_FJP_maxidle} we get
\[ \frac{\sum_{j=1}^{n} c_j + (k-1) \cdot c_{n-m+k}}{m} \]
\[ > \frac{\sum_{j=1}^{n} c_j - \sum_{j=n-m+k}^{n} c_j}{m} + \frac{\sum_{j=n-m+k}^{n} c_j}{m-k+1} \]
\noindent By multiplying both sides by $m \cdot (m-k+1)$ we get
\[ (m-k+1) \cdot \left( \sum_{j=1}^{n} c_j + (k-1) \cdot c_{n-m+k} \right) \]
\[ > (m-k+1) \cdot \left( \sum_{j=1}^{n} c_j - \sum_{j=n-m+k}^{n} c_j \right) + m \sum_{j=n-m+k}^{n} c_j \]
\noindent Thus, 
\[ (m-k+1) \cdot (k-1) \cdot c_{n-m+k} > (k-1) \cdot \sum_{j=n-m+k}^{n} c_j \]
\noindent If $k=1$ then we obviously get $0 > 0$ and the lemma follows. Otherwise, if $k > 1$ then dividing both sides by $(k-1)$ yields
\[ (m-k+1) \cdot c_{n-m+k} >  \sum_{j=n-m+k}^{n} c_j \]
\noindent In this case, in the right-hand side of the above inequality, there are $m-k+1$ terms that are not lower than $c_{n-m+k}$ each. This therefore leads to a contradiction since $c_1 \leq c_2 \leq \cdots \leq c_n$. The lemma follows.
\end{proof}

The following corollary derives an upper-bound on the makespan from $\maxidle{m}$ provided by Expression~\ref{equ:Multimode:ident_FJP_maxidle}.

\begin{Corollary}
\label{cor:Multimode:ident_FJP_makespan}
Suppose that $J$ is sorted by non-decreasing job processing times, i.e., $c_1 \leq c_2 \leq \cdots \leq c_n$. Then, whatever the job priority assignment, an upper-bound $\maxmakespanIdent(J, \pi)$ on the makespan is given by $\maxmakespanIdent(J, \pi) \equals c_n$ if $n = m$, or by
\begin{equation}
\label{equ:Multimode:ident_FJP_makespan}
\maxmakespanIdent(J, \pi) \equals \displaystyle\frac{\sum_{i=1}^{n-1} c_i}{m} + c_n 
\end{equation}
otherwise.
\end{Corollary} 
\begin{proof}
Since the makespan corresponds to the $m^{\operatorname{th}}$ idle-instant, an upper-bound on the makespan is given by $\maxidle{m}$. Therefore, the proof is obtained by simply replacing $k$ with $m$ in Expression~\ref{equ:Multimode:ident_FJP_maxidle}.
\end{proof}

\subsection{Accuracy of the upper-bound $\maxmakespanIdent(J, \pi)$}
\label{sec:FJP_Accuracy}

In this section, Lemma~\ref{lem:Multimode:ident_accuracy_alpha} proves that the upper-bound $\maxmakespanIdent(J, \pi)$ is $2$-competitive, according to the following definition. 
\begin{Definition}[$\alpha$-competitive]
Any upper-bound is said to be $\alpha$-competitive if it provides \emph{at most} $\alpha$ times the \emph{exact} value of the approximated parameter. 
\end{Definition}

This is achieved under the assumption that during any mode transition all the rem-jobs execute for their WCET. Without this assumption, the minimum makespan that could be produced is always $0$ since it can always be the case that no old-mode task has an active job when the mode change is requested. For instance in Figure~\ref{fig:Multimode:SMMSO_example}, the makespan would be zero if the $\MCR(j)$ was released at time $110$. However, in order to guarantee that our approach always provides an \emph{upper-bound} on the makespan we have to consider the worst-case scenario in which every old-mode task releases a job \emph{exactly} upon the mode change request and all these jobs executes for their WCET during the transition. 

\begin{Lemma}
\label{lem:Multimode:ident_accuracy_alpha}
For any set $J$ of jobs sorted by non-decreasing job processing time and for any identical multiprocessor platform $\pi$ composed of $m$ $\cpu$s, the upper-bound $\maxmakespanIdent(J, \pi)$ is $2$-competitive.
\end{Lemma} 
\begin{proof}
Recall from Expression~\ref{equ:Multimode:ident_FJP_makespan} that, 
\[
\maxmakespanIdent(J, \pi) \equals 
\begin{cases}
	\displaystyle c_{n} & \text{if } (n \leq m) \\
	\displaystyle\frac{\sum_{i=1}^{n-1} c_i}{m} + c_n & \text{otherwise}
\end{cases}
\]
Let $\operatorname{ms}(J,m)$ denote the \emph{exact} makespan for the set $J$ of jobs and the $m$ identical $\cpu$s. Since we do not have any mathematical expression for determining this exact makespan $\operatorname{ms}(J,m)$, our analysis is performed while considering a lower-bound $\minmakespanIdent(J, m)$ on the makespan rather than its exact value, i.e., $\alpha$ is determined in such a manner that 
\[ \frac{\maxmakespanIdent(J, \pi)}{\minmakespanIdent(J, m)} \leq \alpha \] 
where
\[ \minmakespanIdent(J, m) \equals
	\begin{cases}
		c_n & \mbox{if} \:\: n \leq m \\
		\max\left\{ c_n, \frac{\sum_{i=1}^{n} c_i}{m} \right\} & \mbox{if} \:\: n > m 
	\end{cases}
\]
The case where $n \leq m$ obviously leads to $\alpha = 1$ since both $\maxmakespanIdent(J, \pi)$ and $\minmakespanIdent(J, \pi)$ return a makespan of $c_n$. Otherwise (if $n > m$) the ``max'' operator in the definition of $\minmakespanIdent(J, m)$ leads to two different cases.

\noindent\emph{Case 1:} If $c_n \geq \frac{\sum_{i=1}^{n} c_i}{m}$ then we get
\begin{eqnarray}
\frac{\maxmakespanIdent(J, \pi)}{\minmakespanIdent(J, m)} & \leq & \frac{\frac{\sum_{i=1}^{n-1} c_i}{m} + c_n}{c_n} \nonumber \\
& \leq & \frac{\frac{\sum_{i=1}^{n} c_i}{m} + c_n}{c_n} \nonumber
\end{eqnarray}
and since in this case we have $c_n \geq \frac{\sum_{i=1}^{n} c_i}{m}$, it holds that
\begin{eqnarray}
\frac{\maxmakespanIdent(J, \pi)}{\operatorname{ms}(J,m)} & \leq & \frac{c_n + c_n}{c_n} \nonumber \\
& \leq & 2 \nonumber
\end{eqnarray}

\noindent\emph{Case 2:} If $c_n < \frac{\sum_{i=1}^{n} c_i}{m}$ then
\[ \frac{\maxmakespanIdent(J, \pi)}{\minmakespanIdent(J, m)} \leq \frac{\frac{\sum_{i=1}^{n-1} c_i}{m} + c_n}{\frac{\sum_{i=1}^{n} c_i}{m}} \]
and since in this case we have $c_n < \frac{\sum_{i=1}^{n} c_i}{m}$, it holds that
\begin{eqnarray}
\frac{\maxmakespanIdent(J, \pi)}{\operatorname{ms}(J,m)} & \leq & \frac{\frac{\sum_{i=1}^{n} c_i}{m} + \frac{\sum_{i=1}^{n} c_i}{m}}{\frac{\sum_{i=1}^{n} c_i}{m}} \nonumber \\
& \leq & 2 \nonumber 
\end{eqnarray}
\noindent The lemma follows.
\end{proof}
     
It holds from Lemma~\ref{lem:Multimode:ident_accuracy_alpha} that, for any set $J$ of jobs and any identical platform composed of $m$ $\cpu$s, the upper-bound on the maximum makespan provided by $\maxmakespanIdent(J, \pi)$ is at most twice the \emph{exact} value of the maximum makespan. Additionally we can show that \emph{in some particular cases} as the one provided in the following example, the upper-bounds $\maxidle{k}(J,\pi)$ ($\forall k \in \left[ 1, m \right]$) defined on page~\pageref{lem:Multimode:ident_FJP_maxidle} are \emph{exact}. 

\begin{Example}
Let us consider the set of $12$ jobs with characteristics given in Table~\ref{tab:Multimode:tight_upper_bound_example} to be scheduled on a $3$-processors identical platform. 
\begin{table}[h!]
\centering
\begin{tabular}{| c | c | c | c | c | c |}
\hline
$c_1$ & $c_2$ & $c_3$ & $c_4$ & $c_5$ & $c_6$\\
\hline
1 & 1 & 1 & 1 & 1 & 1 \\
\hline
$c_7$ & $c_8$ & $c_9$ & $c_{10}$ & $c_{11}$ & $c_{12}$\\
\hline
3 & 3 & 6 & 6 & 9 & 12 \\
\hline
\end{tabular}
\caption{Processing times of the 12 jobs in $J$.}
\label{tab:Multimode:tight_upper_bound_example}
\end{table}

\noindent For this set of jobs,
\begin{itemize}
\item the upper-bound $\maxidle{1} = 15$ is reached with the job priority assignment $J_7 > J_9 > J_{10} > J_{12} > J_{11} > J_8 > J_1 > J_2 > J_3 > J_4 > J_5 > J_6$.
\item the upper-bound $\maxidle{2} = 18$ is reached with the job priority assignment $J_{10} > J_{9} > J_1 > J_2 > J_3 > J_4 > J_5 > J_6 > J_{12} > J_{7} > J_{8} > J_{11}$.
\item the upper-bound $\maxidle{3} = 23$ is reached with the job priority assignment $J_{7} > J_{11} > J_{10} > J_1 > J_2 > J_9 > J_8 > J_3 > J_{5} > J_{4} > J_{6} > J_{12}$.
\end{itemize}

\noindent Due to the space limitation, we did not drew the schedules corresponding to these priority assignments.
\end{Example}

\subsection{Validity tests for $\SMMSO$ and $\AMMSO$}
\label{sec:Multimode:ident_FJP_validity_test}

From Corollaries~\ref{cor:Multimode:worst_case_rem_jobs_set} and~\ref{cor:Multimode:ident_FJP_makespan}, the \emph{sufficient} validity test given by Test \ref{test:Multimode:SMMSO_first} on page \pageref{test:Multimode:SMMSO_first} can be rewritten as follows. 

\begin{validity test}[$\SMMSO$, Identical and FJP]
\label{validitytest:Multimode:ident_FJP_SMMSO}
For any multi-mode real-time application $\tau$ and any identical platform $\pi$ composed of $m$ $\cpu$s, the protocol $\SMMSO$ is valid provided that, for every mode $\mode^i$,
\[ \maxmakespanIdent(\wcremjobs{i}, \pi) \leq \min_{j \neq i} \left\{ \min_{k=1}^{n_j} \left\{ {\cal D}_k^j(\mode^i) \right\} \right\} \]
where $\maxmakespanIdent(\wcremjobs{i}, \pi)$ is defined as in Expression~\ref{equ:Multimode:ident_FJP_makespan} and $\wcremjobs{i}$ is defined as follows: 
\begin{itemize}
\renewcommand{\labelitemi}{$\triangleright$}
\item $\wcremjobs{i} \equals \left\{J_1, J_2, \ldots J_{n_i}\right\}$
\item each job $J_k \in \wcremjobs{i}$ has a processing time equal to the WCET $C_k^i$ of task $\tau_k^i$
\item $\wcremjobs{i}$ is sorted by non-decreasing processing time.
\end{itemize}
 \end{validity test}

Concerning the protocol $\AMMSO$, the upper-bounds $\maxidle{k}(\wcremjobs{i}, \pi)$ (for all $1 \leq k \leq m$) defined as in Lemma~\ref{lem:Multimode:ident_FJP_maxidle} can be used at line 10 of the validity algorithm given by Algorithm~\ref{algo:AMMSO_test} (on page~\pageref{algo:AMMSO_test}).

%%%%%%%%%%%%%%%%%%%%%%%%%%%%%%%%%%%%%%%%%%%%%%%%%%%%%%%%%%%%%%%%%%%%%%%%%%%%%%%%%%%%%%%%%%%%%%%%%%%%%%%%%%%%%%%%%%%%%%%
%%%%%%%%%%%%%%%%%%%%%%%%%%%%%%%%%%%%%%%%%%%%%%%%%%%%%%%%%%%%%%%%%%%%%%%%%%%%%%%%%%%%%%%%%%%%%%%%%%%%%%%%%%%%%%%%%%%%%%%
%%%%%%%%%%%%%%%%%%%%%%%%%%%%%%%%%%%%%%%%%%%%%%%%%%%%%%%%%%%%%%%%%%%%%%%%%%%%%%%%%%%%%%%%%%%%%%%%%%%%%%%%%%%%%%%%%%%%%%%
%%%%%%%%%%%%%%%%%%%%%%%%%%%%%%%%%%%%%%%%%%%%%%%%%%%%%%%%%%%%%%%%%%%%%%%%%%%%%%%%%%%%%%%%%%%%%%%%%%%%%%%%%%%%%%%%%%%%%%%
%%%%%%%%%%%%%%%%%%%%%%%%%%%%%%%%%%%%%%%%%%%%%%%%%%%%%%%%%%%%%%%%%%%%%%%%%%%%%%%%%%%%%%%%%%%%%%%%%%%%%%%%%%%%%%%%%%%%%%%

\section{Identical platforms and FTP schedulers}
\label{sec:Multimode:ident_FTP}

This section is organized as follows. First, Section~\ref{sec:Multimode:ident_FTP_upper_bounds} determines an \emph{upper-bound} $\maxidle{k}(J, \pi, {\cal P})$ on each idle-instant $\idle{k}(J, \pi, {\cal P})$ for any \emph{given} job priority assignment ${\cal P}$ and derives an upper-bound $\maxmakespanIdent(J, \pi, {\cal P})$ on the maximum makespan. Then, Section \ref{sec:FTP_Accuracy} shows that this upper-bound $\maxmakespanIdent(J, \pi)$ is 1-competitive, with the interpretation that $\maxmakespanIdent(J, \pi)$ corresponds to the exact value of the maximum makespan. Finally, Section \ref{sec:Multimode:ident_FTP_validity_test} establishes a \emph{sufficient} validity test for the protocols $\SMMSO$ and $\AMMSO$. 

\subsection{Upper-bounds $\maxidle{k}(J, \pi, {\cal P})$ on the idle-instants}
\label{sec:Multimode:ident_FTP_upper_bounds}

As introduced earlier, this section focuses on determining a mathematical expression for the upper-bounds $\maxidle{k}(J, \pi, {\cal P})$ where $J$ refers to any set of $n$ jobs, $\pi$ denotes any identical multiprocessor platform composed of $m$ $\cpu$s and ${\cal P}$ is a specific given job priority assignment. Indeed, for a given FTP scheduler the priority of every task (and thus of every job) is know beforehand. This prior knowledge allows us to determine tighter upper-bounds than those proposed in the previous section. Once again, for sake of clarity, we will use the notations $\idle{k}$ and $\maxidle{k}$ instead of $\idle{k}(J, \pi, {\cal P})$ and $\maxidle{k}(J, \pi, {\cal P})$, respectively. 

For any transition from a given mode $\mode^i$ to any other mode $\mode^j$, the knowledge of the critical rem-job set $\wcremjobs{i}$ and the fact that the job priority assignment is known beforehand allow us to compute the \emph{exact} maximum idle-instants $\maxidle{k}$---exact in the sense that they are actually reached if every job executes for its WCET---simply by drawing the schedule of $\wcremjobs{i}$ and by measuring the idle-instants $\idle{k}$ in that schedule. Indeed, from Corollary~\ref{cor:Multimode:worst_case_rem_jobs_set} (on page~\pageref{cor:Multimode:worst_case_rem_jobs_set}), each idle-instant $\idle{k}(\wcremjobs{i}, \pi, {\cal P})$ is an upper-bound on the idle-instant $\idle{k}(J, \pi, {\cal P})$ derived from the schedule of any other set $J$ of rem-jobs. Before expressing these exact maximum idle-instants, let us introduce the following definition. 

\begin{Definition}[Processed work $\procwork{k}{i}$]
\label{def:Multimode:processed_work}
Let $\pi$ denote any identical multiprocessor platform and let ${\cal S}$ be any global, weakly work-conserving and FTP scheduler. Let $J = \left\{ J_1, J_2, \ldots, J_n \right\}$ denote any set of $n$ jobs sorted by decreasing ${\cal S}$-priority, i.e., $J_1 >_{\cal S} J_2 >_{\cal S} \cdots >_{\cal S} J_n$ and let $S^i$ denote the schedule by ${\cal S}$ of the $i$ highest priority jobs of $J$ upon $\pi$. The processed work $\procwork{k}{i}$ ($1 \leq k \leq m$ and $0 \leq i \leq n$) denotes the amount of processing time executed on $\cpu$ $\pi_k$ in $S^{i}$.
\end{Definition}

In order to familiarize the reader with this notation $\procwork{k}{i}$, we provide the following example.

\begin{Example}
Let us consider the set $J$ of $7$ jobs with characteristics given in Table~\ref{tab:Multimode:tight_upper_bound_example2} to be scheduled on a $4$-processors identical platform, following the priority assignment: $J_1 > J_2 > \cdots > J_7$. 
\begin{table}[h!]
\begin{center}
\begin{tabular}{| c | c | c | c | c | c | c |}
\hline
$c_1$ & $c_2$ & $c_3$ & $c_4$ & $c_5$ & $c_6$ & $c_7$  \\
\hline
7 & 2 & 5 & 16 & 6 & 5 & 5 \\
\hline
\end{tabular}
\end{center}
\caption{Processing times of the 7 jobs in $J$.}
\label{tab:Multimode:tight_upper_bound_example2}
\end{table}

\begin{figure}[!h]
\begin{center}
\includegraphics*[width=0.5\linewidth, viewport=0 0 700 500]{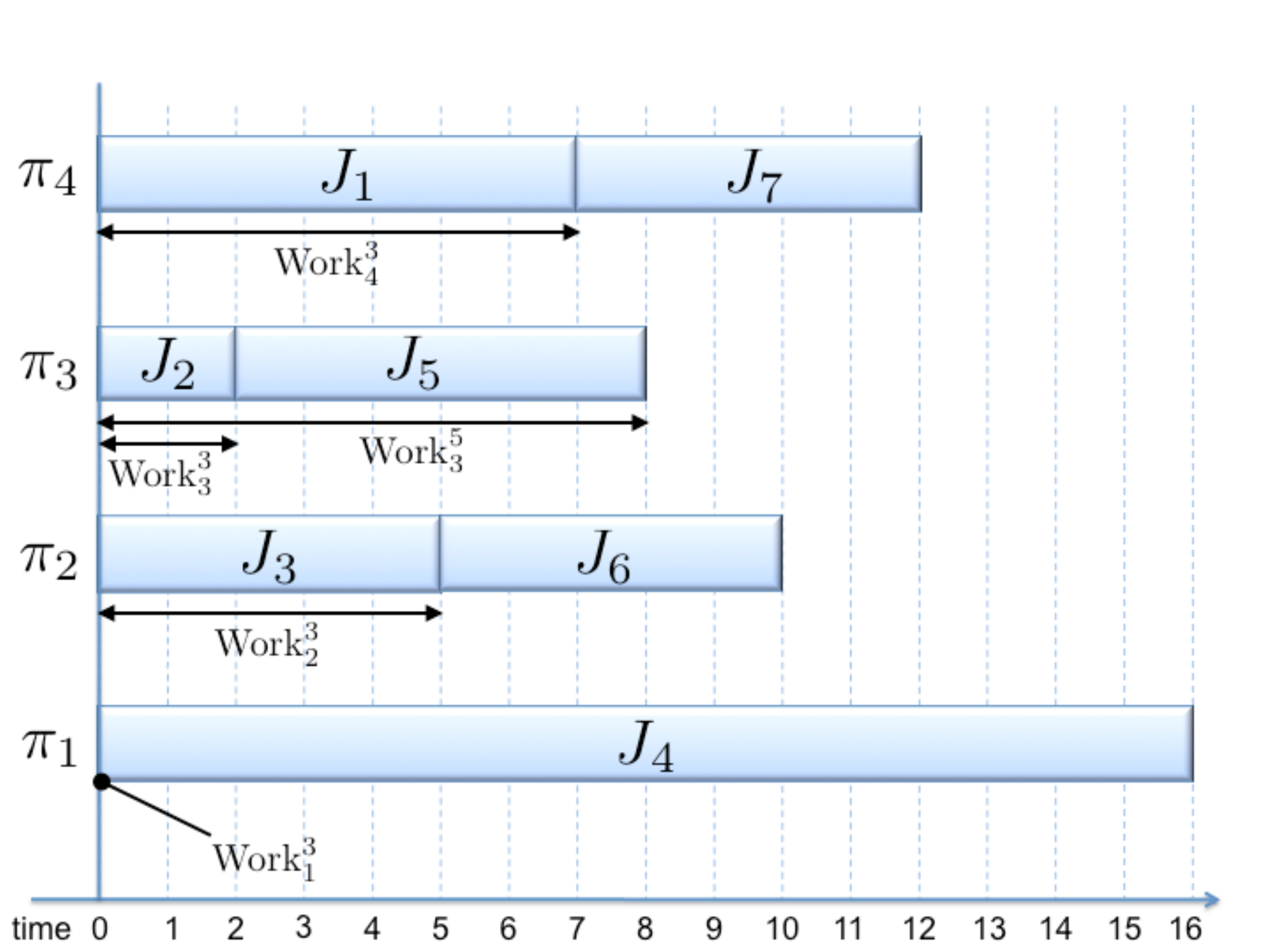}
\caption{Illustration of the notion of processed work $\procwork{k}{i}$.}
\label{fig:MM_proofs_ident_notations}
\end{center}
\end{figure}

Figure~\ref{fig:MM_proofs_ident_notations} illustrates the schedule of $J$ upon the $4$ $\cpu$s. In this schedule, we have $\procwork{3}{5} = 8$ because, in the schedule $S^{5}$ of the $5$ highest priority jobs $J_1, J_2, J_3, J_4, J_5$, the amount of processing time units executed on $\pi_3$ is $c_2 + c_5 = 8$. Similarly, $\procwork{4}{3} = 7, \procwork{3}{3} = 2, \procwork{2}{3} = 5$ and $\procwork{1}{3}~=~0$ because, in the schedule $S^3$ of jobs $J_1, J_2, J_3$, we can see that $7$ processing time units are executed on $\pi_4$ (i.e., job $J_1$), $2$ processing time units are executed on $\pi_3$ (i.e., job $J_2$), $5$ processing time units are executed on $\pi_2$ (i.e., job $J_3$) and no processing time unit is executed on $\pi_1$. Notice that $\procwork{k}{0}~=~0$ $\forall k=1, 2, \ldots, m$. 
\end{Example}

Lemma~\ref{lem:Multimode:ident_FTP_processed_work} provides the exact values of $\procwork{k}{i}$ ($\forall i\in\left[1, n\right]$ and $\forall k \in \left[1, m\right]$) when each job executes for its WCET. Then, Corollary~\ref{cor:Multimode:ident_FTP_maxidle} derives the exact maximum idle-instants $\idle{k}$ $1 \leq k \leq m$, for the scheduling of any set $J$ of $n$ jobs upon any $m$-processors identical platform. 

\begin{Lemma}
\label{lem:Multimode:ident_FTP_processed_work}
Let $\pi$ denote any identical multiprocessors platform composed of $m$ $\cpu$s. Let ${\cal S}$ be any global, weakly work-conserving and FTP scheduler and let $J$ be any set of $n$ jobs sorted by decreasing ${\cal S}$-priority, i.e., $J_1 >_{\cal S} J_2 >_{\cal S} \cdots >_{\cal S} J_n$. It holds $\forall k \in \left[1, m \right]$ and $\forall i \in \left[ 1, n \right]$ that

\begin{small}
\begin{equation}
\label{equ:Multimode:ident_FTP_processed_work}
\procwork{k}{i} = 
\begin{cases}
\procwork{k}{i-1} + c_i & \mbox{if} \:\: k = \max\left\{ \underset{\ell \: \in \: \left[1,m\right]}{\operatorname{argmin}} \left\{ \procwork{\ell}{i-1} \right\} \right\} \\
\procwork{k}{i-1} & \mbox{otherwise} \\
\end{cases}
\end{equation}
\end{small}
where $\procwork{k}{0} = 0$ $\forall k$ by definition of the processed work.
\end{Lemma} 
\begin{proof}
The proof directly follows from the definition of $\procwork{i}{k}$ $\forall i,k$ and from the second condition of our definition of a weakly work-conserving scheduler (see Definition~\ref{def:Multimode:weakly_work_conserving}, page~\pageref{def:Multimode:weakly_work_conserving}). Indeed, whenever a subset $P$ of several $\cpu$s idle (or complete a job) at the same time, ${\cal S}$ dispatches the waiting job (if any) with the highest priority to the $\cpu$ of $P$ with the highest index (this is the reason for the condition ``if $k$ is the highest value of $\ell$ that minimizes $\procwork{\ell}{i-1}$''). 
\end{proof}

\begin{Corollary}
\label{cor:Multimode:ident_FTP_maxidle}
An upper-bound $\maxidle{k}$, $1 \leq k \leq m$, is given by the $k^{\operatorname{th}}$ element of the vector $\left\{ \procwork{1}{n}, \procwork{2}{n}, \ldots, \procwork{m}{n} \right\}$ sorted by non-decreasing order.
\end{Corollary} 
\begin{proof}
The proof directly follows from the definition of the processed work $\procwork{k}{n}$, $\forall k \in \left[1, m \right]$. 
\end{proof}

\begin{Corollary}
\label{cor:Multimode:ident_FTP_makespan}
The maximum makespan $\maxmakespanIdent(J, \pi, {\cal P})$ is given by $\maxidle{m}$, where $\maxidle{m}$ is determined as in Corollary~\ref{cor:Multimode:ident_FTP_maxidle}. 
\end{Corollary}

\subsection{Accuracy of the upper-bound $\maxmakespanIdent(J, \pi, {\cal P})$}
\label{sec:FTP_Accuracy}

In this section we prove that the upper-bound $\maxmakespanIdent(J, \pi, {\cal P})$ is $1$-competitive, i.e., {\em exact}---exact in the sense that it can actually be reached if every job executes for its WCET. Again, this is achieved under the assumption that during any mode transition all the rem-jobs execute for their WCET as we have to consider the worst-case scenario in which every old-mode task releases a job \emph{exactly} upon the mode change request and all these jobs executes for their WCET during the transition. 

For any transition from a given mode $\mode^i$ to any other mode $\mode^j$, the knowledge of the critical rem-job set $\wcremjobs{i}$ and the fact that we proceed by simulation allow us to compute the \emph{exact} maximum idle-instants simply by drawing the schedule of $\wcremjobs{i}$ following ${\cal P}$ and by measuring the idle-instants in this schedule. Using this approach, the measured upper-bound $\maxmakespanIdent(J, \pi, {\cal P})$ is nothing else but $1$-competitive.

\subsection{Validity tests for $\SMMSO$ and $\AMMSO$}
\label{sec:Multimode:ident_FTP_validity_test}

From Corollary~\ref{cor:Multimode:ident_FTP_makespan}, the \emph{sufficient} validity test given by Test~\ref{test:Multimode:SMMSO_first} (on page \pageref{test:Multimode:SMMSO_first}) can be rewritten as follows. 

\begin{validity test}[$\SMMSO$, identical and FTP]
\label{validitytest:Multimode:ident_FTP_SMMSO}
For any multi-mode real-time application $\tau$ and any identical platform $\pi$ composed of $m$ $\cpu$s, the protocol $\SMMSO$ is valid provided that, for every mode $\mode^i$,
\[ \maxmakespanIdent(\wcremjobs{i}, \pi, {\cal P}^i) \leq \min_{j \neq i} \left\{ \min_{k=1}^{n_j} \left\{ {\cal D}_k^j(\mode^i) \right\} \right\} \]
where $\maxmakespanIdent(\wcremjobs{i}, \pi)$ is defined as in Corollary~\ref{cor:Multimode:ident_FTP_makespan}, ${\cal P}^i$ is obtained from the old-mode scheduler ${\cal S}^i$ and $\wcremjobs{i}$ is defined as follows: 
\begin{itemize}
\renewcommand{\labelitemi}{$\triangleright$}
\item $\wcremjobs{i} \equals \left\{J_1, J_2, \ldots J_{n_i}\right\}$
\item each job $J_k \in \wcremjobs{i}$ has a processing time equal to the WCET $C_k^i$ of task $\tau_k^i$
\item $\wcremjobs{i}$ is sorted by decreasing ${\cal S}^i$-priority. 
\end{itemize}
\end{validity test}

Concerning the protocol $\AMMSO$, the upper-bounds $\maxidle{k}(\wcremjobs{i}, \pi, {\cal P}^i)$ (for all $1 \leq k \leq m$) determined in Corollary~\ref{cor:Multimode:ident_FTP_maxidle} can be used at line 10 of the validity algorithm given by Algorithm~\ref{algo:AMMSO_test} on page~\pageref{algo:AMMSO_test}).

%%%%%%%%%%%%%%%%%%%%%%%%%%%%%%%%%%%%%%%%%%%%%%%%%%%%%%%%%%%%%%%%%%%%%%%%%%%%%%%%%%%%%%%%%%%%%%%%%%%%%%%%%%%%%%%%%%%%%%%
%%%%%%%%%%%%%%%%%%%%%%%%%%%%%%%%%%%%%%%%%%%%%%%%%%%%%%%%%%%%%%%%%%%%%%%%%%%%%%%%%%%%%%%%%%%%%%%%%%%%%%%%%%%%%%%%%%%%%%%
%%%%%%%%%%%%%%%%%%%%%%%%%%%%%%%%%%%%%%%%%%%%%%%%%%%%%%%%%%%%%%%%%%%%%%%%%%%%%%%%%%%%%%%%%%%%%%%%%%%%%%%%%%%%%%%%%%%%%%%
%%%%%%%%%%%%%%%%%%%%%%%%%%%%%%%%%%%%%%%%%%%%%%%%%%%%%%%%%%%%%%%%%%%%%%%%%%%%%%%%%%%%%%%%%%%%%%%%%%%%%%%%%%%%%%%%%%%%%%%
%%%%%%%%%%%%%%%%%%%%%%%%%%%%%%%%%%%%%%%%%%%%%%%%%%%%%%%%%%%%%%%%%%%%%%%%%%%%%%%%%%%%%%%%%%%%%%%%%%%%%%%%%%%%%%%%%%%%%%%

\section{Uniform platforms and FJP schedulers}
\label{sec:Multimode:unif_FJP}

\subsection{Some useful observations}
\label{sec:Multimode:unif_observations}

In this section, we show that the maximum makespan determination problem is highly counter-intuitive upon \emph{uniform} platforms and the methods for solving this problem cannot be straightforwardly extended from those proposed for \emph{identical} multiprocessor platforms. First, recall that the schedulers are assumed to be \emph{strongly} work-conserving here since we focus on uniform platforms. 

\begin{Observation} 
For a given set of jobs, an intuitive idea for maximizing the makespan upon any $m$-processor uniform platform is to execute, at any time, the longest job upon the slowest $\cpu$, i.e., the shorter the computation requirement of a job, the higher its priority. We name this priority assignment ``Shortest Job First'' (SJF). However, we can show by using the following example that this intuitive idea is {\em erroneous}, as SJF does not lead to the maximum makespan. 
\end{Observation}

\begin{Example}
Let us consider the set $J$ of $4$ jobs $J_1, J_2, J_3, J_4$ of respective processing times 4, 4, 16 and 22, and suppose that they are scheduled on the $2$-processors uniform platform $\pi=\left[ 1, 2 \right]$. The priority assignment SJF (i.e., $J_1 > J_2 > J_3 > J_4$) provides a makespan of $17.75$ whereas the priority assignment $J_3 > J_1 > J_2 > J_4$ leads to a makespan of~$19$. Notice that the problem of determining in a polynomial time (i.e., without trying every priority assignment) a priority assignment leading to the maximum makespan remains an open question and is out of the scope of this study. 
\end{Example}

\begin{Observation}
Another intuitive idea is to naively extend to uniform platforms the result (replicated below) of Corollary~\ref{cor:Multimode:ident_FJP_makespan} on page~\pageref{cor:Multimode:ident_FJP_makespan}, i.e., for any identical platform $\pi$ composed of $m$ $\cpu$s, an upper-bound on the makespan is given by
\begin{equation}
\label{equ:Multimode:ident_FJP_makespan_recall}
\maxmakespanIdent(J, \pi) \equals 
\begin{cases}
\displaystyle c_{n} & \text{if } (n = m) \\ 
\displaystyle\frac{\sum_{i=1}^{n-1} c_i}{m} + c_n & \text{otherwise } 
\end{cases}
\end{equation}
where $c_i$ is assumed to be such that $c_i \geq c_{i-1}$ $\forall i \in \left[ 2, n \right]$. 
\end{Observation}

Upon \emph{identical} platforms there is a sense in distinguishing the case $n=m$ from the case $n > m$, because the rem-jobs never migrate between $\cpu$s during mode transitions. Therefore, in the particular case where $n=m$, the maximum makespan does \emph{not} depend on the job priority assignment and can be determined \emph{exactly} by $\maxmakespanIdent(J, \pi) = c_n$. In contrast, we can easily show that this property does not hold upon \emph{uniform} platforms. That is, the maximum makespan in the case $n=m$ is \emph{not} independent from the job priority assignment upon uniform platforms. This is shown through the following example.

\begin{Example}
Consider the uniform platform $\pi = \left[ 1, 2\right]$ and the two jobs $J_1, J_2$ of processing time $4$ and $6$, respectively. If $J_1 > J_2$ then $J_1$ completes on $\pi_2$ at time $2$---time during which $J_2$ executes 2 execution units on $\pi_1$---and $J_2$ completes on $\pi_2$ at time $4$, thus leading to a makespan of $4$. On the other hand, if $J_2 > J_1$ then $J_2$ completes on $\pi_2$ at time $3$---time during which $J_1$ executes $3$ execution units on $\pi_1$---and $J_1$ completes on $\pi_2$ at time $3.5$, thus leading to a makespan of $3.5$. As a result, the maximum makespan in the case $n=m$ \emph{depends} on the job priority assignment on uniform platforms and the case $n= m$ can no longer be distinguished from the case $m < n$. 
\end{Example}

From the previous example, naively extending Expression~\ref{equ:Multimode:ident_FJP_makespan_recall} to uniform platforms yields the following ``1-piece'' expression\footnote{recall that $\totalspeed \equals \sum_{i=1}^m s_i$}:
\begin{equation}
\label{equ:Multimode:unif_FJP2_naive_extend}
\maxmakespanUnifZero(J, \pi) \equals \displaystyle\frac{\sum_{i=1}^{n-1} c_i}{\totalspeed} + \frac{c_n}{s_m}
\end{equation}

Unfortunately, we show in the following example that this extension does not provide an upper-bound on the maximum makespan.

\begin{figure}[h]
\begin{center}
\includegraphics*[width=0.5\linewidth, viewport=0 0 700 400]{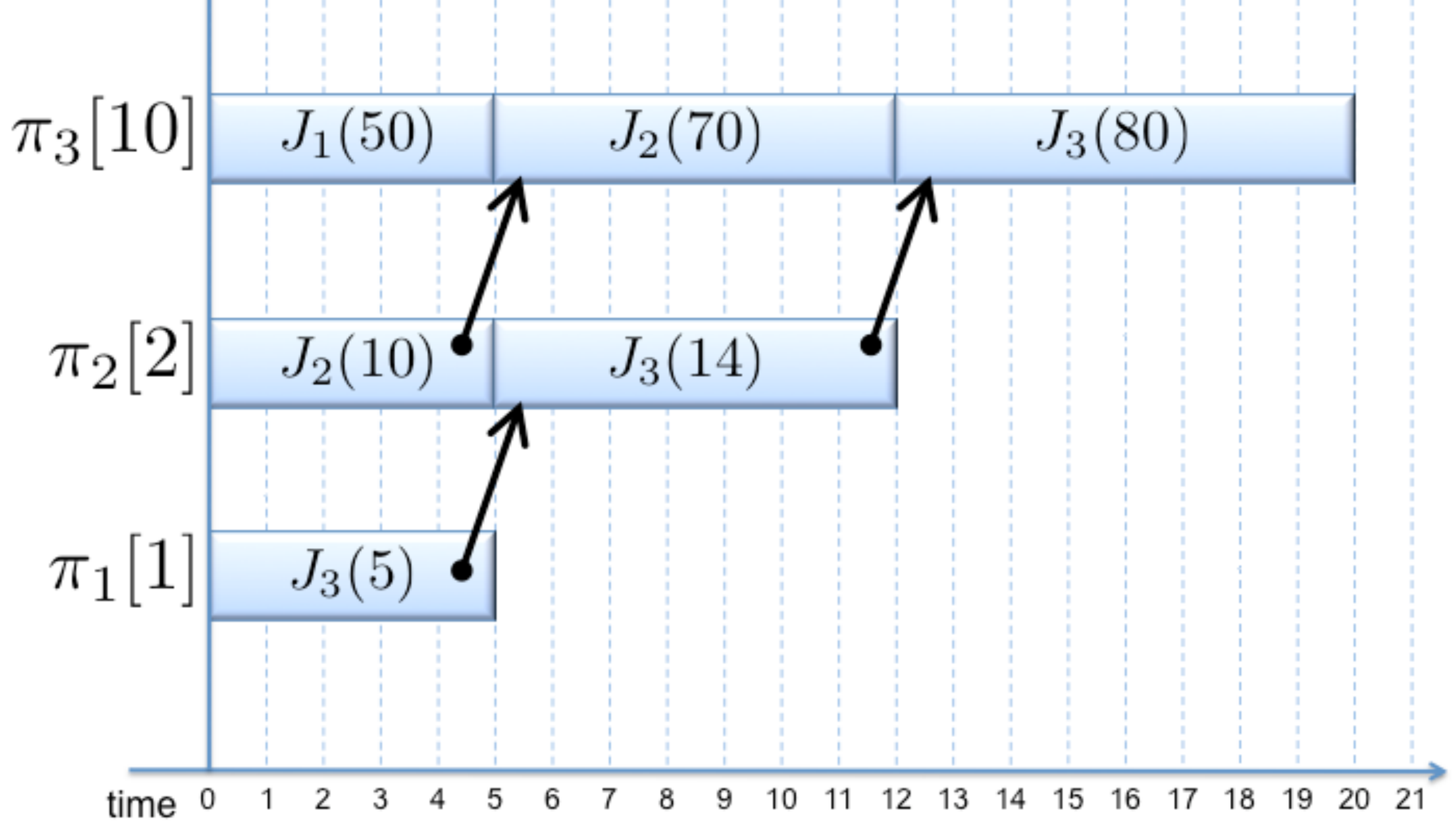}
\caption{This picture depicts a priority assignment leading to a makespan of $20$. The speed of each $\cpu$ is indicated into brackets next to its label. The numbers next to each job name $J_i$ is the amount of work processed by $J_i$ upon the allocated $\cpu$. For instance, job $J_1$ executes $50$ execution units from time 0 to 5 on $\cpu$ $\pi_3$, leading to its label $J_1(50)$.}
\label{fig:Multimode:unif_observation2_example1}
\end{center}
\end{figure}

\begin{figure}[h]
\begin{center}
\includegraphics*[width=0.5\linewidth, viewport=0 0 700 400]{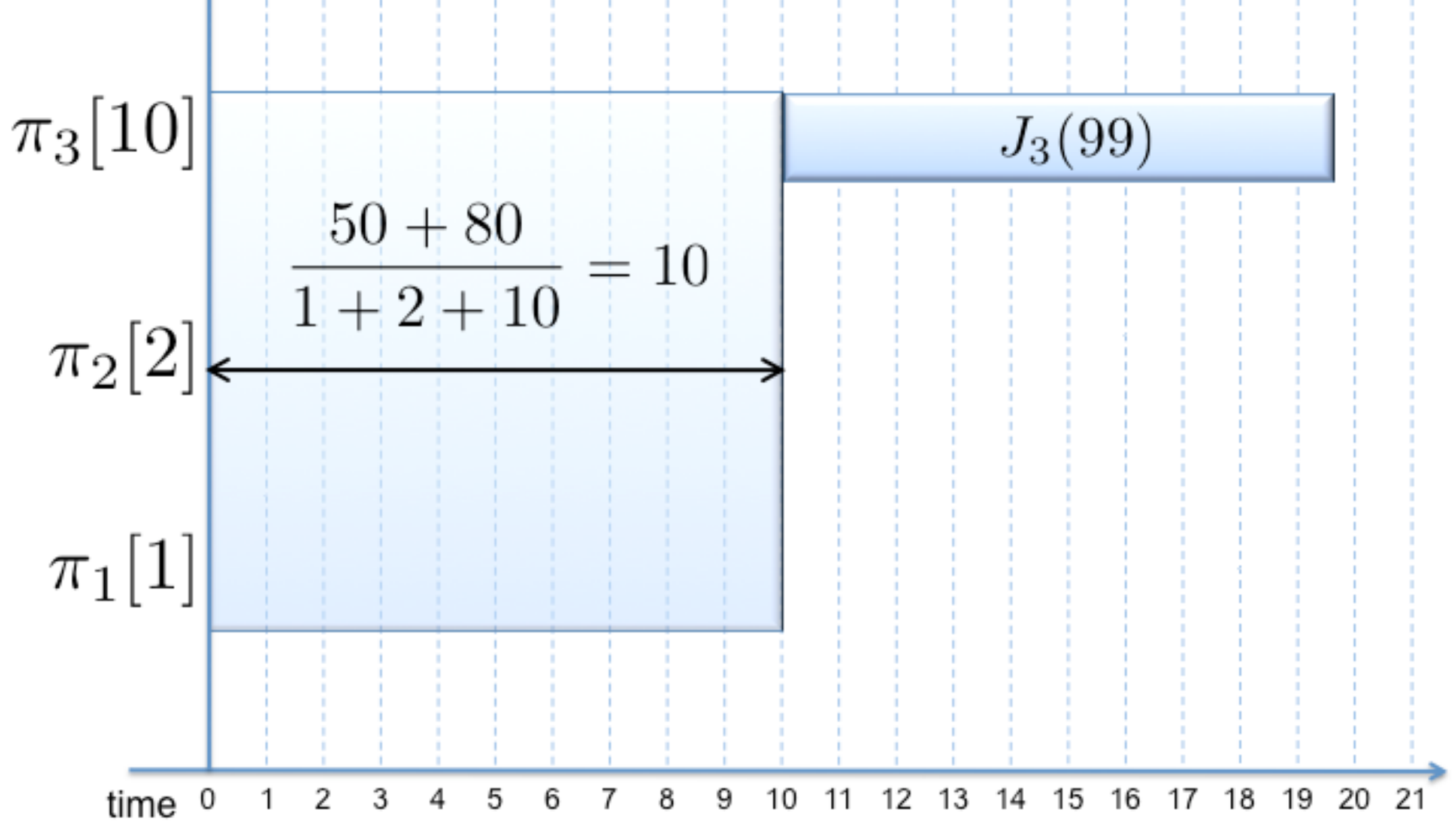}
\caption{Approximation error made by Expression~\ref{equ:Multimode:unif_FJP2_naive_extend}.}
\label{fig:Multimode:unif_observation2_example2}
\end{center}
\end{figure}

\begin{Example}
\label{ex:under_approximation}
Let us consider the set $J$ of $3$ jobs $J_1, J_2, J_3$ of respective processing times 50, 80 and 99, and suppose that they are scheduled on the $3$-processors uniform platform $\pi=\left[ 1, 2, 10 \right]$. The maximum makespan is $20$, reached using the job priority assignment $J_1 > J_2 > J_3$ (see Figure~\ref{fig:Multimode:unif_observation2_example1}). On the other hand, Expression~\ref{equ:Multimode:unif_FJP2_naive_extend} yields $\maxmakespanUnifZero(J, \pi) = \frac{50 + 80}{13} + \frac{99}{10} = 19.9$. This approximation made by Expression~\ref{equ:Multimode:unif_FJP2_naive_extend} is illustrated in Figure~\ref{fig:Multimode:unif_observation2_example2}. This simple example is \emph{much} more important than what it seems to be at first blush and we will deeply examine its impacts in Section~\ref{sec:Multimode:unif_FJP2_improvements} (page~\pageref{sec:Multimode:unif_FJP2_improvements}). 
\end{Example}

\subsection{Upper-bounds $\maxidle{k}(J, \pi)$ on the idle-instants}
\label{sec:Multimode:unif_FJP_upper_bounds}

Once more but this time for any \emph{uniform} platform $\pi$, we focus on determining a mathematical expression that provides an upper-bound $\maxidle{k}(J, \pi)$ on the $k^{\operatorname{th}}$ idle-instant, $\forall k \in \left[ 1, m \right]$. For sake of clarity, the following two lemmas use the notations $\maxidle{k}$ instead of $\maxidle{k}(J, \pi)$ and similarly, the notation $\idle{k}$ will be used to denote the \emph{exact} value of the $k^{\operatorname{th}}$ idle-instant. First, Lemma~\ref{lem:Multimode:unif_FJP_minidle} determines a \emph{lower-bound} $\minidle{k}$ on each idle-instant $\idle{k}$, $1 \leq k \leq m$. Then, Lemma~\ref{lem:Multimode:unif_FJP_maxidle} determines an \emph{upper-bound} $\maxidle{k}$ on each idle-instant $\idle{k}$. Finally, Corollary~\ref{cor:Multimode:unif_FJP_makespan} derives an upper-bound on the maximum makespan (recall that the maximum makespan is simply given by $\maxidle{m}$). 

\begin{Lemma}[See~\cite{MeumeuNelisGoossens:10}]
\label{lem:Multimode:unif_FJP_minidle}
Let $\pi = [s_1, s_2, \ldots, s_m]$ be any $m$-processors uniform platform such that $s_i \geq s_{i-1}$ $\forall i$, $2 \leq i \leq m$. Let $J = \{J_1, J_2, \ldots, J_n\}$ be any set of $n$ jobs of respective processing times $c_1, c_2, \ldots, c_n$ such that $c_1 \leq c_2 \leq \cdots \leq c_n$. Let $S$ be the schedule of $J$ upon $\pi$ following any global, strongly work-conserving and FJP scheduler. A lower bound $\minidle{k}$ on each idle-instant $\idle{k}$ ($1 \leq k \leq m$) in $S$ is given by
\begin{equation}
\label{equ:Multimode:unif_FJP_minidle}
\minidle{k} \equals \frac{\sum_{i=1}^{n-m+k} c_i}{\totalspeed}
\end{equation}
\end{Lemma} 
\begin{proof}
According to the definition of the idle-instants, at most $(m-k)$ jobs are not completed at time $\idle{k}$, meaning that \emph{at least} $(n - m + k)$ jobs are already completed. Let $J^{\any}$ be any subset of $J$ composed of $r$ jobs, where $(n - m + k) \leq r \leq n$. Obviously, a lower bound $t$ on the instant at which the $r$ jobs of $J^{\any}$ are completed is given by 
\[ t \equals \frac{\sum_{J_i \in J^{\any}} c_i}{\totalspeed} \]
and since $c_1 \leq c_2 \leq \cdots \leq c_n$, $t$ is minimal if (i) the number of jobs in $J^{\any}$ is low as possible, i.e., $r = n-m+k$, \emph{and} (ii) the processing time of each job of $J^{\any}$ is low as possible. As a result, $t$ is minimum for $J^{\any} = \{ J_1, J_2, \ldots, J_{n - m + k}\}$ and then yields a lower-bound for $\minidle{k}$. 
\end{proof}

\begin{Lemma}[See~\cite{MeumeuNelisGoossens:10}]
\label{lem:Multimode:unif_FJP_maxidle}
Using the same notations as in the previous lemma, an upper-bound $\maxidle{k}$ on each idle-instant $\idle{k}$ ($1 \leq k \leq m$) in $S$ is given by
\begin{equation}
\label{equ:Multimode:unif_FJP_maxidle}
\maxidle{k} \equals \frac{\sum_{i=1}^{n} c_i - \sum_{i=1}^{k-1} \minidle{i} \cdot s_i}{\cumulspeed{k}}
\end{equation}
where $\cumulspeed{k} \equals \sum_{i=k}^m s_i$ (as defined in Expression~\ref{equ:Multimode:cumul_speed}, page~\pageref{equ:Multimode:cumul_speed}).
\end{Lemma} 
\begin{proof}
From the ``staircase'' property derived from the definition of a strongly work-conserving scheduler on uniform platform (see page~\pageref{sec:Multimode:scheduler_specifications} for details) and from the fact that all the jobs are assumed to be synchronous at time $0$, we know that $\cpu$ $\pi_j$ becomes idle at time $\idle{j}$, $\forall j = 1, 2, \ldots, m$. Let $\work{j}$ ($1 \leq j \leq m$) denotes the amount of work executed on $\cpu$~$\pi_j$ within $[ 0, \idle{j} ]$, i.e., $\work{j} \equals \idle{j} \cdot s_j$. The proof is made by contradiction. Let $\ell$ be any integer in $\left[ 1, m \right]$ and suppose that $\idle{\ell} > \maxidle{\ell}$. By definition of $\work{j}$, we know that 
\begin{equation}\label{equ:workissumci}
\sum_{j = 1}^m \work{j} = \sum_{i = 1}^n c_i
\end{equation}
and from the definition of $\work{j}$ we know that 
\begin{eqnarray}
\sum_{j = 1}^m \work{j} & = & \sum_{j = 1}^m \idle{j} \cdot s_j \nonumber \\
& = & \sum_{j = 1}^{\ell-1} (\idle{j} \cdot s_j) + \sum_{j = \ell}^{m} (\idle{j} \cdot s_j) \nonumber
\end{eqnarray}
By definition of the idle-instants, it holds $\forall j \geq \ell$ that $\idle{j} \geq \idle{\ell}$. Therefore, replacing ``$\idle{j}$'' with ``$\idle{\ell}$'' in the second term of the right-hand side of the above equality yields
\begin{eqnarray}
\sum_{j = 1}^m \work{j} & \geq & \sum_{j = 1}^{\ell-1} (\idle{j} \cdot s_j) + \sum_{j = \ell}^{m} (\idle{\ell} \cdot s_j) \nonumber \\
& \geq & \sum_{j = 1}^{\ell-1} (\idle{j} \cdot s_j) + \idle{\ell} \cdot \sum_{j = \ell}^{m} s_j \nonumber
\end{eqnarray}
By hypothesis we have $\idle{\ell} > \maxidle{\ell}$. Therefore, replacing $\idle{\ell}$ with $\maxidle{\ell}$ in the right-hand side of the above inequality yields
\begin{small}
\begin{eqnarray}
\sum_{j = 1}^m \work{j} & > & \sum_{j = 1}^{\ell-1} (\idle{j} \cdot s_j) + \maxidle{\ell} \cdot \sum_{j = \ell}^{m} s_j \nonumber \\
& > & \sum_{j = 1}^{\ell-1} (\idle{j} \cdot s_j) + \frac{\sum_{i=1}^{n} c_i - \sum_{i=1}^{\ell-1} \minidle{i} \cdot s_i}{\sum_{i=\ell}^m s_i} \cdot \sum_{j = \ell}^{m} s_j \nonumber \\
& > & \sum_{j = 1}^{\ell-1} (\idle{j} \cdot s_j) + \sum_{i=1}^{n} c_i - \sum_{i=1}^{\ell-1} \minidle{i} \cdot s_i \nonumber \\
& > & \sum_{i=1}^{n} c_i + \sum_{j = 1}^{\ell-1} \left((\idle{j} - \minidle{j} ) \cdot s_j \right) \nonumber
\end{eqnarray}
\end{small}
\noindent Since from Lemma~\ref{lem:Multimode:unif_FJP_minidle} it holds that $\minidle{i} \leq \idle{i}$ $\forall i=1, 2, \ldots, m$, it holds that 
\[ \sum_{j = 1}^{\ell-1} \left((\idle{j} - \minidle{j} ) \cdot s_j \right) \geq 0 \]
and thus
\[ \sum_{j = 1}^m \work{j} > \sum_{i=1}^{n} c_i \] 
leading to a contradiction with Equality~\ref{equ:workissumci}. The lemma follows. 
\end{proof}

\begin{Corollary}[See~\cite{MeumeuNelisGoossens:10}]
\label{cor:Multimode:unif_FJP_makespan}
Whatever the job priority assignment, an upper-bound $\maxmakespanUnifOne(J, \pi)$ on the makespan is given by
\begin{equation}
\label{equ:Multimode:unif_FJP_makespan}
\maxmakespanUnifOne(J, \pi) \equals \frac{1}{s_m} \cdot \left( \sum_{i=1}^n c_i - \sum_{i=1}^{m-1} \minidle{i} \cdot s_i \right)
\end{equation}
\end{Corollary} 
\begin{proof}
Since the makespan corresponds to the idle-instant $\idle{m}$, an upper-bound on the makespan is given by $\maxidle{m}$. Therefore, the proof is obtained by simply replacing $k$ with $m$ in Expression~\ref{equ:Multimode:unif_FJP_maxidle}. \\
\end{proof}

\subsection{Accuracy of the upper-bound $\maxmakespanUnifOne(J, \pi)$}

In this section we prove that the upper-bound $\maxmakespanUnifOne(J, \pi)$ is $\frac{\totalspeed}{s_m}$-competitive, with the interpretation that the value returned by $\maxmakespanUnifOne(J, \pi)$ is {\em at most} $\frac{\totalspeed}{s_m}$ times the {\em exact} value of the maximum makespan for any given set $J$ of jobs and uniform platform $\pi$. Once again, this is achieved under the assumption that during any mode transition all the rem-jobs execute for their WCET as we have to consider the worst-case scenario in which every old-mode task releases a job exactly upon the mode change request and all these jobs executes for their WCET during the transition.

\begin{Lemma}
\label{lem:Multimode:unif_accuracy_alpha}
For any set $J$ of jobs sorted by non-decreasing job processing time and any uniform platform $\pi = \left[ s_1, s_2, \ldots, s_m \right]$ with $s_i \geq s_{i-1}$ $\forall i$, $\maxmakespanUnifOne(J, \pi)$ is $\alpha_1(\pi)$-competitive, where $\alpha_1(\pi) \equals \frac{\totalspeed}{s_m}$.
\end{Lemma} 
\begin{proof}
Recall from Expression~\ref{equ:Multimode:unif_FJP_makespan} that
\[ \maxmakespanUnifOne(J, \pi) \equals \frac{\sum_{i=1}^n c_i}{s_m} - \frac{\sum_{k=1}^{m-1} \left( \sum_{i=1}^{n-m+k} c_i \cdot s_k \right)}{s_m \cdot \totalspeed} \]
Let $\operatorname{ms}(J,\pi)$ denote the \emph{exact} makespan for any given set $J$ of jobs and any uniform platform $\pi$. Since we do not have any mathematical expression for determining this exact makespan $\operatorname{ms}(J,\pi)$, our analysis of $\alpha_1(\pi)$ is performed while considering a lower-bound $\widetilde{\operatorname{ms}}(J,\pi)$ on the makespan rather than its exact value, i.e., $\alpha_1(\pi)$ is determined in such a manner that 
\[ \frac{\maxmakespanUnifOne(J, \pi)}{\widetilde{\operatorname{ms}}(J,\pi)} \leq \alpha_1(\pi) \]
Obviously, we know that $\operatorname{ms}(J,\pi) \geq \frac{\sum_{i=1}^{n} c_i}{\totalspeed}$ and this implies that $\widetilde{\operatorname{ms}}(J,\pi) \equals \frac{\sum_{i=1}^{n} c_i}{\totalspeed}$ is a lower-bound on the makespan. This yields
\begin{eqnarray}
\frac{\maxmakespanUnifOne(J, \pi)}{\operatorname{ms}(J,\pi)} & \leq & \frac{\maxmakespanUnifOne(J, \pi)}{\widetilde{\operatorname{ms}}(J, \pi)} \nonumber 
\end{eqnarray}
and thus,
\[ \frac{\maxmakespanUnifOne(J, \pi)}{\operatorname{ms}(J,\pi)} \leq \frac{\frac{\sum_{i=1}^n c_i}{s_m} - \frac{\sum_{k=1}^{m-1} \left( \sum_{i=1}^{n-m+k} c_i \cdot s_k \right)}{s_m \cdot \totalspeed}}{\frac{\sum_{i=1}^{n} c_i}{\totalspeed}} \]
\begin{small}
\begin{eqnarray}
\label{lem:Multimode:unif_accuracy_alpha_exp1}  & \leq & \left( \frac{\displaystyle\sum_{i=1}^n c_i}{s_m} - \frac{\displaystyle\sum_{k=1}^{m-1} \left( \displaystyle\sum_{i=1}^{n-m+k} c_i \cdot s_k \right)}{s_m \cdot \totalspeed} \right) \cdot \frac{\totalspeed}{\sum_{i=1}^{n} c_i} \\
\label{lem:Multimode:unif_accuracy_alpha_exp2} & \leq & \left( \frac{\sum_{i=1}^n c_i}{s_m} \right) \cdot \frac{\totalspeed}{\sum_{i=1}^{n} c_i} \\
& \leq & \frac{\totalspeed}{s_m} \nonumber
\end{eqnarray}
\end{small}
Notice the important loss of accuracy that this inequality underwent when we ignored the term $\left(- \frac{\sum_{k=1}^{m-1} \left( \sum_{i=1}^{n-m+k} c_i \cdot s_k \right)}{s_m \cdot \totalspeed}\right)$ while passing from Inequality~\ref{lem:Multimode:unif_accuracy_alpha_exp1} to Inequality~\ref{lem:Multimode:unif_accuracy_alpha_exp2}. The lemma follows.
\end{proof}

\subsection{Another analysis of the maximum makespan}
\label{sec:Multimode:unif_FJP2_improvements}

In Example~\ref{ex:under_approximation} on page \pageref{ex:under_approximation}, we have showed that the naive extension of $\maxmakespanIdent(J, \pi)$ (given by $\maxmakespanUnifZero(J, \pi)$ in Expression \ref{equ:Multimode:unif_FJP2_naive_extend}, page~\pageref{equ:Multimode:unif_FJP2_naive_extend}) does not provide an upper-bound on the maximum makespan considering uniform platforms. Essentially, in addition to refute the fact that $\maxmakespanUnifZero(J, \pi)$ provides an upper-bound on the maximum makespan, this example also refutes the \emph{main concept behind the expression of $\maxmakespanIdent(J, \pi)$}. Indeed, in the expression of $\maxmakespanIdent(J, \pi)$, it can be easily shown that the term $\frac{\sum_{i=1}^{n-1} c_i}{m}$ is an upper-bound on the time at which $J_n$ starts its execution, i.e., its \emph{dispatching} time. Therefore, the whole expression can be interpreted as follows: upper-bound on the makespan = upper-bound on the dispatching time of $J_n$ + $c_n$, where $J_n$ is the (or any) job with the largest processing time. That is, this expression of $\maxmakespanIdent(J,\pi)$ is based on the intuition that the maximum makespan is reached when the longest job is dispatched as late as possible and executes for its WCET. This intuition has revealed to be true for the case of identical platforms, but not for the uniform case (as shown by Example~\ref{ex:under_approximation})\footnote{Indeed, we can also easily show that the term $\left(\sum_{i=1}^{n-1} c_i\right) / \totalspeed$ in Expression~\ref{equ:Multimode:unif_FJP2_naive_extend} is an upper-bound on the dispatching time of $J_n$ and at that time $J_n$ is dispatched to the fastest $\cpu$ $\pi_m$, leading to a WCET of $\frac{c_m}{s_m}$.}. The whole concept is not extendable to uniform platforms and in order to figure out the underlying cause, let us focus on Example~\ref{ex:under_approximation}.

\begin{figure}
\begin{center}
\includegraphics*[width=0.5\linewidth]{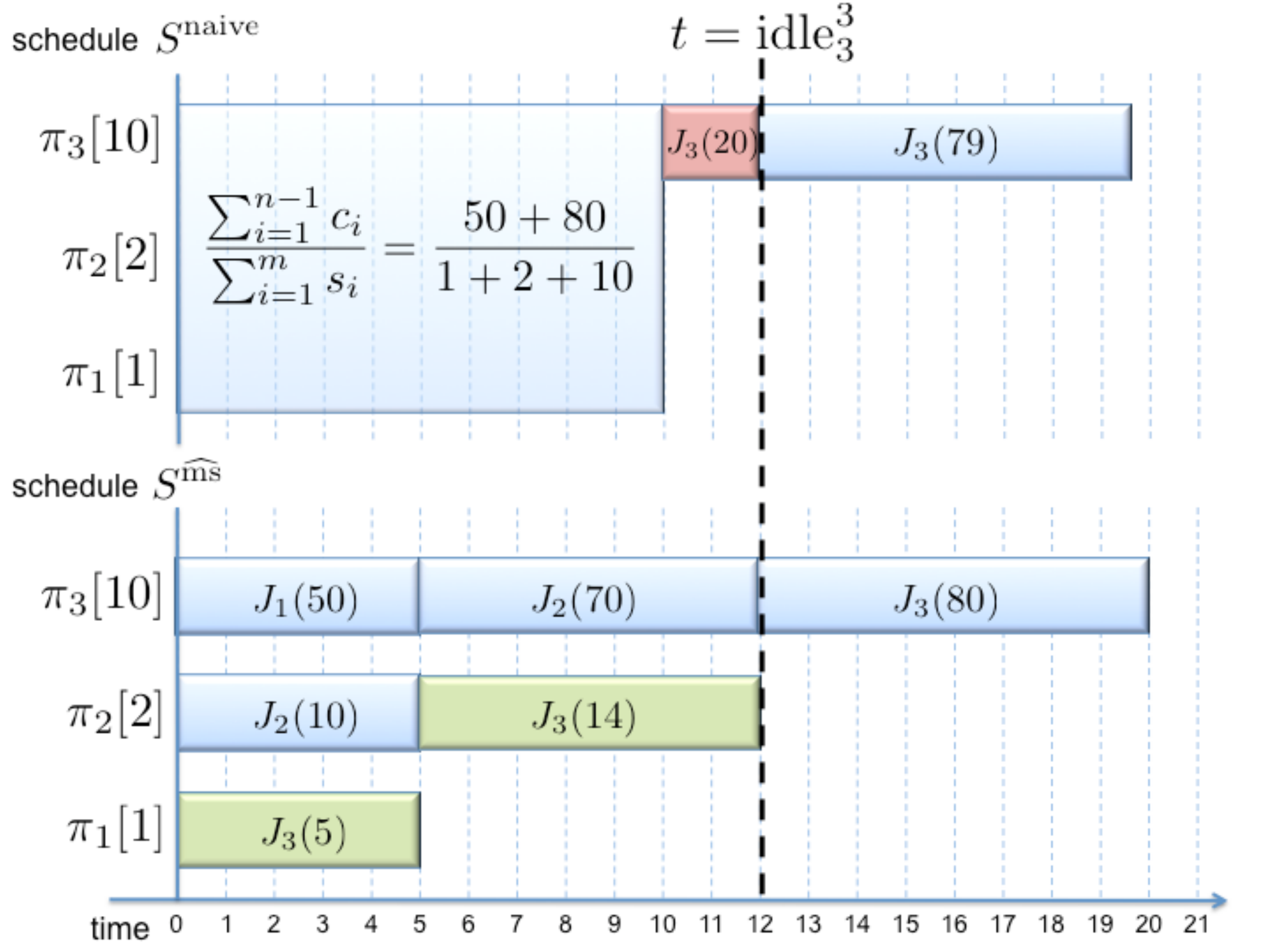}
\caption{Example of schedule in which the makespan is larger than that returned by $\maxmakespanUnifZero(J, \pi)$.}
\label{fig:Multimode:unif_FJP2_improvement_principle}
\end{center}
\end{figure}

Let $S^{\operatorname{naive}}$ and $S^{\maxmakespan}$ denote the two schedules depicted in Figure~\ref{fig:Multimode:unif_FJP2_improvement_principle}, issued from the approximation $\maxmakespanUnifZero(J, \pi)$ and from the priority assignment $J_1 > J_2 > J_3$ which leads to the maximum makespan, respectively. The reason why $\maxmakespanUnifZero(J, \pi)$ \emph{under}-approximates the maximum makespan comes from the following fact: \emph{if $t$ denotes the instant at which job $J_3$ is dispatched to $\cpu$ $\pi_3$ in $S^{\maxmakespan}$ (here, $t=12$), then during the time interval $\left[ 0, t \right]$, $J_3$ has executed a lower amount of execution units in the stairs of $S^{\maxmakespan}$ than upon $\pi_3$ in $S^{\operatorname{naive}}$}. In other words the \emph{cumulated} green areas  in Figure~\ref{fig:Multimode:unif_FJP2_improvement_principle} represent a lower amount of execution units than the red area. Indeed, $J_3$ executes $5+14=19$ execution units within $\left[ 0, t \right]$ in $S^{\maxmakespan}$ whereas it executes $20$ execution units on $\pi_3$ in $S^{\operatorname{naive}}$. As a result, the remaining processing time of $J_3$ at time $t$ is higher in $S^{\maxmakespan}$ (here, 80) than in $S^{\operatorname{naive}}$ (here, 79), implying that $J_3$ completes later in $S^{\maxmakespan}$ than in $S^{\operatorname{naive}}$. This is the reason why the expression $\maxmakespanUnifZero(J, \pi)$ does not provide the maximum makespan in the example above: on uniform platforms, \emph{the schedule in which any job $J_i$ reaches its maximum completion time is not necessarily the schedule in which $J_i$ is dispatched as late as possible}. 

Based on this fundamental observation, we propose and prove correct in \cite{Nelis:10} (pages 138--163 and 351--367) two additional upper-bounds $\maxmakespanUnifTwo(J, \pi)$ and $\maxmakespanUnifThree(J, \pi)$ on the maximum makespan, considering uniform platforms and FJP schedulers. These upper-bounds are replicated below.
\begin{equation}
\label{equ:Multimode:unif_FJP_makespan2}
\maxmakespanUnifTwo(J, \pi) \equals \frac{1}{s_m} \cdot \sum_{i=1}^{n}  \left(c_{i} + s_1 \cdot \frac{\sum_{j=1}^{i -1} c_j}{\totalspeed}\right) \cdot K_{n - i}
\end{equation}
where $K_j$ is such that $\forall j$,
\[
K_j \equals 
\begin{cases}
	1 & \mbox{if} \:\: s_1 = s_m \:\: \mbox{and} \:\: j=0\\
	\left(1 - \frac{s_1}{s_m}\right)^j & \mbox{otherwise}
\end{cases}
\]
\noindent and
\begin{equation}
\label{equ:Multimode:unif_FJP_makespan3}
\maxmakespanUnifThree(J, \pi) \equals \frac{1}{s_m} \cdot \sum_{\ell=1}^{n} \left( c_{\ell} + \frac{s_x \cdot s_m \cdot \sum_{j=1}^{\ell-1} c_j}{\totalspeed \cdot \sum_{j=1}^{x} s_j} \right) \cdot H_{n-\ell}
\end{equation}
where 
\[ x = \underset{i \: \in \: \left[1,m\right]}{\operatorname{argmin}} \left\{ \frac{s_i}{\sum_{j=1}^i s_j} \right\} \]
and $H_j$ is such that $\forall j$,
\[
H_j \equals 
\begin{cases}
	1 & \mbox{if} \:\: s_x = \sum_{i=1}^{x} s_i \:\: \mbox{and} \:\: j=0\\
	\left(1 - \frac{s_x}{\sum_{i=1}^{x} s_i}\right)^j & \mbox{otherwise}
\end{cases}
\]

Each of these two upper-bounds is based on a distinct upper-bound on the amount of execution units that can be executed in the green areas (see~Figure~\ref{fig:Multimode:unif_FJP2_improvement_principle}), and then derives an upper-bound on the completion time of every job, and finally on the makespan.

\subsection{Validity tests for $\SMMSO$ and $\AMMSO$}
\label{sec:Multimode:unif_FJP_validity_test}

\noindent From Expressions~\ref{equ:Multimode:unif_FJP_makespan}, \ref{equ:Multimode:unif_FJP_makespan2}, \ref{equ:Multimode:unif_FJP_makespan3}, and Corollary~\ref{cor:Multimode:worst_case_rem_jobs_set}, a \emph{sufficient} validity test for the protocol $\SMMSO$ can therefore be formalized as follows. 

\begin{validity test}[$\SMMSO$, uniform and FJP]
\label{validitytest:Multimode:unif_FJP_SMMSO}
For any multi-mode real-time application $\tau$ and any uniform platform $\pi = [s_1, s_2, \ldots, s_m]$ composed of $m$ $\cpu$s, the protocol $\SMMSO$ is valid provided that, for every mode $\mode^i$,
\[ \maxmakespanUnifMin(\wcremjobs{i}, \pi) \leq \min_{j \neq i} \left\{ \min_{k=1}^{n_j} \left\{ {\cal D}_k^j(\mode^i) \right\} \right\}\]
where $\maxmakespanUnifMin(\wcremjobs{i}, \pi)$ is defined as $\maxmakespanUnifMin(\wcremjobs{i}, \pi) \equals$
\begin{equation}
\label{maxmakespanUnifMin}
\footnotesize \min\left\{ \maxmakespanUnifOne(\wcremjobs{i}, \pi), \maxmakespanUnifTwo(\wcremjobs{i}, \pi), \maxmakespanUnifThree(\wcremjobs{i}, \pi) \right\} 
\end{equation}
\noindent and $\maxmakespanUnifOne(\wcremjobs{i}, \pi)$, $\maxmakespanUnifTwo(\wcremjobs{i}, \pi)$ and $\maxmakespanUnifThree(\wcremjobs{i}, \pi)$ are defined as in Expressions~\ref{equ:Multimode:unif_FJP_makespan}, \ref{equ:Multimode:unif_FJP_makespan2} and~\ref{equ:Multimode:unif_FJP_makespan3}, respectively. This is performed considering the set $\wcremjobs{i}$ composed of $n_i$ jobs of processing time $C_1^i, C_2^i, \ldots, C_{n_i}^i$ such that $C_j^i \geq C_{j-1}^i$ $\forall j=2, 3, \ldots, n_i$. 
\end{validity test}

Concerning the protocol $\AMMSO$, the upper-bounds $\maxidle{k}(\wcremjobs{i}, \pi)$ (for all $1 \leq k \leq m$) defined as in Lemma~\ref{lem:Multimode:unif_FJP_maxidle} can be used at line 10 of the validity algorithm given by Algorithm~\ref{algo:AMMSO_test} (on page~\pageref{algo:AMMSO_test}).

\subsection{Simulation results}
\label{sec:Multimode:simulations}

Because our analysis of the competitive factor did not lead to a constant $\alpha$ for the upper-bound $\maxmakespanUnifOne(J, \pi)$ (as well as for the upper-bounds $\maxmakespanUnifTwo(J, \pi)$ and $\maxmakespanUnifThree(J, \pi)$ as shown in \cite{Nelis:10}), this section reports on the results of simulations in order to quantify the precision of the three upper-bounds $\maxmakespanUnifOne(J, \pi)$, $\maxmakespanUnifTwo(J, \pi)$ and $\maxmakespanUnifThree(J, \pi)$. These simulations are performed considering a single set $J$ of jobs scheduled and multiple uniform platforms. We consider only a single set $J$ of jobs for which the \emph{exact} processing times are given in Table~\ref{tab:Multimode:simulations_jobs_processing_time}. We will explain below where these parameters are drawn from and why we consider only a single set of jobs rather than generating numerous job sets. \\ 

\begin{table}[h!]
\centering
\begin{tabular}{| c | c | c | c | c |}
\hline
$c_1$ & $c_2$ & $c_3$ & $c_4$ & $c_5$ \\
\hline
3896 & 3964 & 878 & 1378 & 2228 \\
\hline
$c_6$ & $c_7$ & $c_8$ & $c_9$ & $c_{10}$ \\
\hline
3612 & 1230 & 1232 & 1668 & 4672 \\
\hline
\end{tabular}
\caption{Processing times of the 10 jobs in $J$.}
\label{tab:Multimode:simulations_jobs_processing_time}
\end{table}

For experimental purposes, let us introduce the parameter $\lambda_\pi$ defined in \cite{FunkGoossensBaruah:01} for any $m$-processor uniform platform $\pi = [s_1, s_2, \ldots, s_m]$, 
\[ \lambda_\pi \equals \max_{j=1}^{m} \left\{\frac{\sum_{k=1}^{j-1} s_k}{s_j} \right\} \] 
Informally speaking, this parameter $\lambda_\pi$ measures the ``degree'' by which $\pi$ differs from an identical multiprocessor platform, i.e., its ``degree of heterogeneity''. For any identical platform composed of $m$ $\cpu$s, it holds that $s_1 = s_2 = \cdots = s_m$ and thus, $\lambda_\pi \equals \max_{j=1}^{m} \left\{\frac{\sum_{k=1}^{j-1} s_k}{s_j} \right\}$ is maximum for $j=m$, leading to $\lambda_\pi \equals \frac{\sum_{k=1}^{m-1} s_k}{s_m} = m-1$. The more homogeneous the platform $\pi$ is, the closer to $(m-1)$ is its corresponding $\lambda_{\pi}$. For instance, the uniform platform $\pi = \left[ 1, 500, 1000\right]$ has a corresponding $\lambda_{\pi} = \frac{501}{1000} \approx 0.5$ whereas $\lambda_{\pi} = \frac{501}{600} = 0.835$ for the uniform platform $\pi = \left[ 1, 500, 600\right]$ and $\lambda_{\pi} = \frac{1000}{600} \approx 1.67$ for the platform $\pi = \left[ 500, 500, 600\right]$. In short, $\lambda_\pi = (m-1)$ if $\pi$ is comprised of $m$ identical $\cpu$s and becomes progressively smaller as the speeds of the $\cpu$s differ from each other by greater amounts. 

The platform $\pi$ considered in our simulations is composed of $m=4$ $\cpu$s for which we make their computing speed varying within $\left[1, 101 \right]$ with an increment of $10$. More precisely, we consider all possible combinations of the $\cpu$ speeds in the range $\left[1, 101 \right]$ with an increment of $10$, i.e., the first simulation is performed considering $\pi = \left[ 1, 1, 1, 1 \right]$, the second simulation considers $\pi = \left[ 1, 1, 1, 11 \right]$, the third one considers $\pi = \left[ 1, 1, 1, 21 \right]$, and so on until reaching the speed assignment $\pi = \left[ 101, 101, 101, 101 \right]$. For every speed assignment, we determine the corresponding parameter $\lambda_{\pi}$ as well as the \emph{exact} value $\makespan(J,\pi)$ of the maximum makespan. This exact maximum makespan $\makespan(J,\pi)$ is determined by building the schedule of $J$ upon $\pi$ for \emph{every} job priority assignment and by retaining only the maximum generated makespan. This is a highly computational-intensive operation that requires the exhaustive enumeration of every possible job priority assignment. This is the reason why we consider only a single set $J$ of jobs in our simulations. Indeed, according to this approach, our simulation process considers $11$ different speeds for each $\cpu$, leading to a total of $11^m = 11^4 = 14,641$ different platforms $\pi$. For each platform $\pi$, the computation of the exact makespan requires to generate the schedules derived from every job priority assignment. Since there are $10$ jobs, the number of considered priority assignments is $10! = 3,628,800$. Multiplied by the number of platforms, this leads to $53,129,260,800$ operations. Our simulations were performed on HYDRA, the Scientific Computer Configuration at the VUB/ULB Computing Centre, where we fully distributed the computations among 15 processors AMD Opteron dual-core @ 2.8GHz. Distributing the computations allowed us to complete the simulation in about 2 hours but unfortunately, the computation time grows exponentially with the number of $\cpu$s and in a factorial manner with the number of jobs. For instance, considering $13$ jobs would result in $91,169,811,532,800$ operations, $14$ jobs to approximately $20 \cdot 10^{15}$ operations, resulting in a computation time of about $82$ years. The processing times of the jobs have been drawn from~\cite{IainBate:98} where the authors present realistic parameters that concern the avionic domain. But since the number of operations of our algorithm is strongly restricted by the number of jobs, we arbitrarily selected $10$ WCETs from these parameters. 

For each speed assignment of the platform we computed the error $\mkerrorUnifOne(J,\pi)$ corresponding to the difference (in percent) between $\maxmakespanUnifOne(J, \pi)$ and $\makespan(J,\pi)$. Formally, 
\[ \mkerrorUnifOne(J,\pi) \equals \frac{\maxmakespanUnifOne(J, \pi) - \makespan(J,\pi)}{\makespan(J,\pi)} \cdot 100 \]
and in a similar way we also computed the errors $\mkerrorUnifTwo(J,\pi)$ and $\mkerrorUnifThree(J,\pi)$.

\begin{figure}
\includegraphics[width=\linewidth]{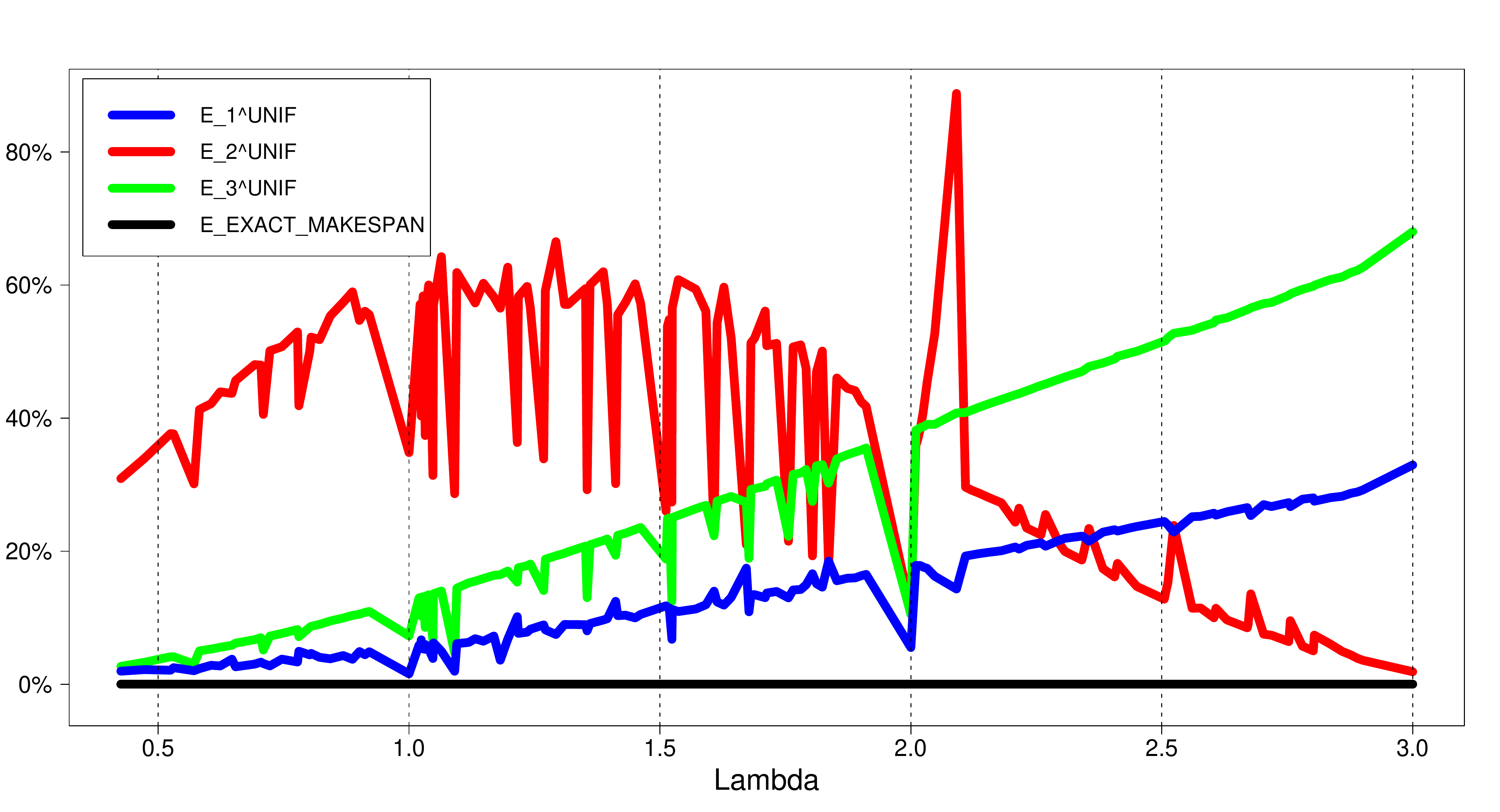}
\caption{The three estimation errors $\mkerrorUnifOne(J,\pi)$, $\mkerrorUnifTwo(J,\pi)$ and $\mkerrorUnifThree(J,\pi)$ displayed relative to the corresponding $\lambda_{\pi}$.}
\label{fig:Multimode:simulations_result_1}
\end{figure}

\begin{figure}
\includegraphics[width=\linewidth]{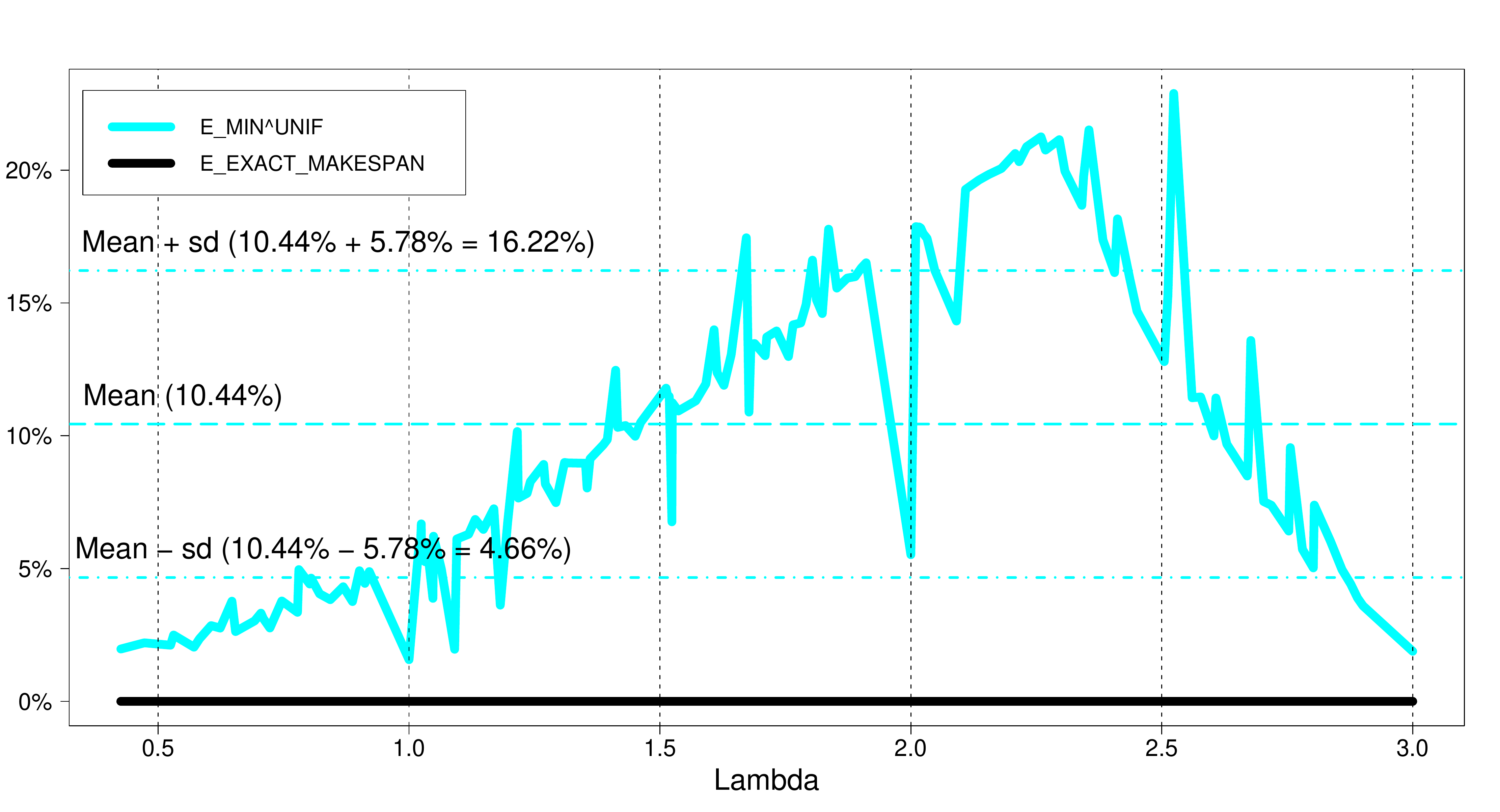}
\caption{The estimation error $\mkerrorUnifMin(J, \pi)$ displayed relative to the corresponding $\lambda_{\pi}$.}
\label{fig:Multimode:simulations_result_2}
\end{figure}

The errors $\mkerrorUnifOne(J,\pi)$, $\mkerrorUnifTwo(J,\pi)$ and $\mkerrorUnifThree(J,\pi)$ are displayed in Figure~\ref{fig:Multimode:simulations_result_1} relative to the corresponding $\lambda_{\pi}$. The horizontal black line is the error ``E\_EXACT\_MAKESPAN'' of $\makespan(J,\pi)$ over the exact value of the maximum makespan. Obviously, this error is always $0$. Also, for every speed assignment of $\pi$, we define the estimator $\maxmakespanUnifMin(J, \pi)$ as in Expression~\ref{maxmakespanUnifMin} and its associated error $\mkerrorUnifMin(J,\pi)$. This error is displayed in Figure~\ref{fig:Multimode:simulations_result_2} relative to the corresponding $\lambda_{\pi}$. Finally, Table~\ref{tab:Multimode:simulations_outline} provides the reader with some statistics issued from the simulation. 

\begin{table}[h!]
\centering
\begin{tabular}{| c | c | c | c | c |}
\hline
& $\mkerrorUnifOne(J,\pi)$ & $\mkerrorUnifTwo(J,\pi)$ & $\mkerrorUnifThree(J,\pi)$ & $\mkerrorUnifMin(J, \pi)$ \\
\hline
Min. & 1.57\% & 1.89\% & 2.7\% & 1.57\% \\
\hline
1st Qu. & 6\% & 21.74\% & 13.28\% & 5.3\% \\
\hline
Median & 12.72\% & 41.07\% & 27.11\% & 9.92\% \\
\hline
Mean & 13.68\% & 37.91\% & 29.25\% & 10.44\% \\
\hline
3rd Qu. & 20.72\% & 55.5\% & 43.99\% & 15.08\% \\
\hline
Max. & 32.96\% & 88.78\% & 68.01\% & 22.89\% \\
\hline
Variance & 69.76 & 359.37 & 320.47 & 33.36 \\
\hline
SD\footnote{Squared distance} & 8.35\% & 18.96\% & 17.9\% & 5.78\% \\
\hline
Bias & 42.35\% & 133.69\% & 94.05\% & 26.93\% \\
\hline
MSE\footnote{Mean Square Error} & 1863.25 & 18233.15 & 9165.97 & 758.43 \\
\hline
\end{tabular}
\caption{Statistics issued from the simulation}
\label{tab:Multimode:simulations_outline}
\end{table}

For obvious reason, the most accurate estimator (i.e., the most accurate upper-bound on the maximum makespan) is $\maxmakespanUnifMin(J, \pi)$. As presented in Table~\ref{tab:Multimode:simulations_outline}, the most important error that we obtained for $\maxmakespanUnifMin(J, \pi)$ is $22.89\%$ and the minimal one is $1.57\%$. The average error is $10.44\%$ with a squared distance of $5.78\%$. Hence, we believe that this is a promising path to go for more competitive bounds and for practical use. An open question remains however. For $\lambda_{\pi} \in \left[ 0, 2\right]$, we can see in Figure~\ref{fig:Multimode:simulations_result_1} that $\maxmakespanUnifOne(J, \pi)$ is clearly lower than both $\maxmakespanUnifTwo(J, \pi)$ and $\maxmakespanUnifThree(J, \pi)$, i.e., $\maxmakespanUnifMin(J, \pi) = \maxmakespanUnifOne(J, \pi)$ for $\lambda_{\pi} \in \left[ 0, 2\right]$. Within this interval $\left[ 0, 2\right]$, when the parameter $\lambda_{\pi}$ reaches an integer value (here, $1$ and $2$), something happens that considerably improves the accuracy of $\maxmakespanUnifMin(J, \pi)$. But up to now, we did not find any interpretation to that phenomenon. 

%%%%%%%%%%%%%%%%%%%%%%%%%%%%%%%%%%%%%%%%%%%%%%%%%%%%%%%%%%%%%%%%%%%%%%%%%%%%%%%%%%%%%%%%%%%%%%%%%%%%%%%%%%%%%%%%%%%%%%%
%%%%%%%%%%%%%%%%%%%%%%%%%%%%%%%%%%%%%%%%%%%%%%%%%%%%%%%%%%%%%%%%%%%%%%%%%%%%%%%%%%%%%%%%%%%%%%%%%%%%%%%%%%%%%%%%%%%%%%%
%%%%%%%%%%%%%%%%%%%%%%%%%%%%%%%%%%%%%%%%%%%%%%%%%%%%%%%%%%%%%%%%%%%%%%%%%%%%%%%%%%%%%%%%%%%%%%%%%%%%%%%%%%%%%%%%%%%%%%%
%%%%%%%%%%%%%%%%%%%%%%%%%%%%%%%%%%%%%%%%%%%%%%%%%%%%%%%%%%%%%%%%%%%%%%%%%%%%%%%%%%%%%%%%%%%%%%%%%%%%%%%%%%%%%%%%%%%%%%%
%%%%%%%%%%%%%%%%%%%%%%%%%%%%%%%%%%%%%%%%%%%%%%%%%%%%%%%%%%%%%%%%%%%%%%%%%%%%%%%%%%%%%%%%%%%%%%%%%%%%%%%%%%%%%%%%%%%%%%%

\section{Uniform platforms and FTP schedulers}
\label{sec:Multimode:unif_FTP}

This section follows the same reasoning as the one for identical platforms and FTP schedulers. For any transition from a specific mode $\mode^i$ to any other mode $\mode^j$, the knowledge of the critical rem-job set $\wcremjobs{i}$ and the fact that the priorities are known beforehand enable us to compute the \emph{exact} maximum idle-instants $\idle{k}$, $1 \leq k \leq m$, simply by simulating the scheduling of the critical rem-job set and by measuring the idle-instants $\idle{k}$, $1 \leq k \leq m$, in that schedule (from Corollary~\ref{cor:Multimode:worst_case_rem_jobs_set} presented on page~\pageref{cor:Multimode:worst_case_rem_jobs_set}). Thus, each idle-instant $\idle{k}$ measured in the schedule of the critical rem-job set is an upper-bound on the idle-instants $\idle{k}$ in the schedule derived from any other set of rem-jobs. In conclusion, FTP schedulers enable us to determine the \emph{exact}\footnote{Exact in the sense that this value is actually reached if every job executes for its WCET.} maximum idle-instants $\maxidle{k}$, $1 \leq k \leq m$, rather than over-approximating them (as done for the FJP schedulers). 

\subsection{Upper-bounds $\maxidle{k}(J, \pi, {\cal P})$ on the idle-instants}
\label{sec:Multimode:unif_FTP_upper_bounds}

Lemma~\ref{lem:Multimode:unif_FTP_idle_ki} provides the exact values of $\idle{j}(J_i, \pi, {\cal P})$ $\forall j \in \left[ 1, m \right], i \in \left[ 1, n\right]$ and $\forall {\cal P}$, assuming that every job $J_i$ executes for its WCET. However, in this particular case of FTP scheduler, we redefine the idle-instants $\idle{j}(J_i, \pi, {\cal P})$ as follows.

\begin{Definition}[Idle-instant $\idle{j}(J_i, \pi, {\cal P})$]
\label{def:Multimode:unif_FTP_idle_ij_redefined}
If $S^i$ denotes the schedule upon $\pi$ of only the jobs with a higher (or equal) priority than $J_i$ according to ${\cal P}$, then $\idle{j}(J_i, \pi, {\cal P})$ is the earliest instant in $S^i$ at which at least $j$ $\cpu$s idle. 
\end{Definition}

The only difference w.r.t. the previous one resides in the ``higher \emph{(or equal)} priority than (...)''. The reason for this redefinition is that, with the previous one, it was not possible to express the idle-instants $\idle{j}(J, \pi, {\cal P})$ (for $j=1, \ldots, m$) as in Definition~\ref{def:Multimode:idle_instants} (page~\pageref{def:Multimode:idle_instants}). Indeed, these idle-instants $\idle{j}(J, \pi, {\cal P})$ consider that every job of $J$ are scheduled while the previous definition of the idle-instants $\idle{j}(J_i, \pi, {\cal P})$ requires a job index $i$ and considers that only the jobs with a higher priority than $J_i$ are scheduled. Thereby, this previous definition always excludes the job $J_i$ in the computation of the idle-instants. Now, thanks to this new definition, the idle-instants $\idle{j}(J, \pi, {\cal P})$ (for $j=1, \ldots, m$) can be expressed by $\idle{j}(J_{\low}, \pi, {\cal P})$, where $J_{\low}$ is the lowest priority job according to ${\cal P}$. Once again, we use in Corollary~\ref{cor:Multimode:unif_FTP_maxidle} the notations $\idle{j}^i$ to refer to the idle-instants $\idle{j}(J_i, \pi, {\cal P})$ defined as in Definition~\ref{def:Multimode:unif_FTP_idle_ij_redefined}. 

\begin{figure}
\includegraphics[width=\linewidth, viewport=0 0 800 500]{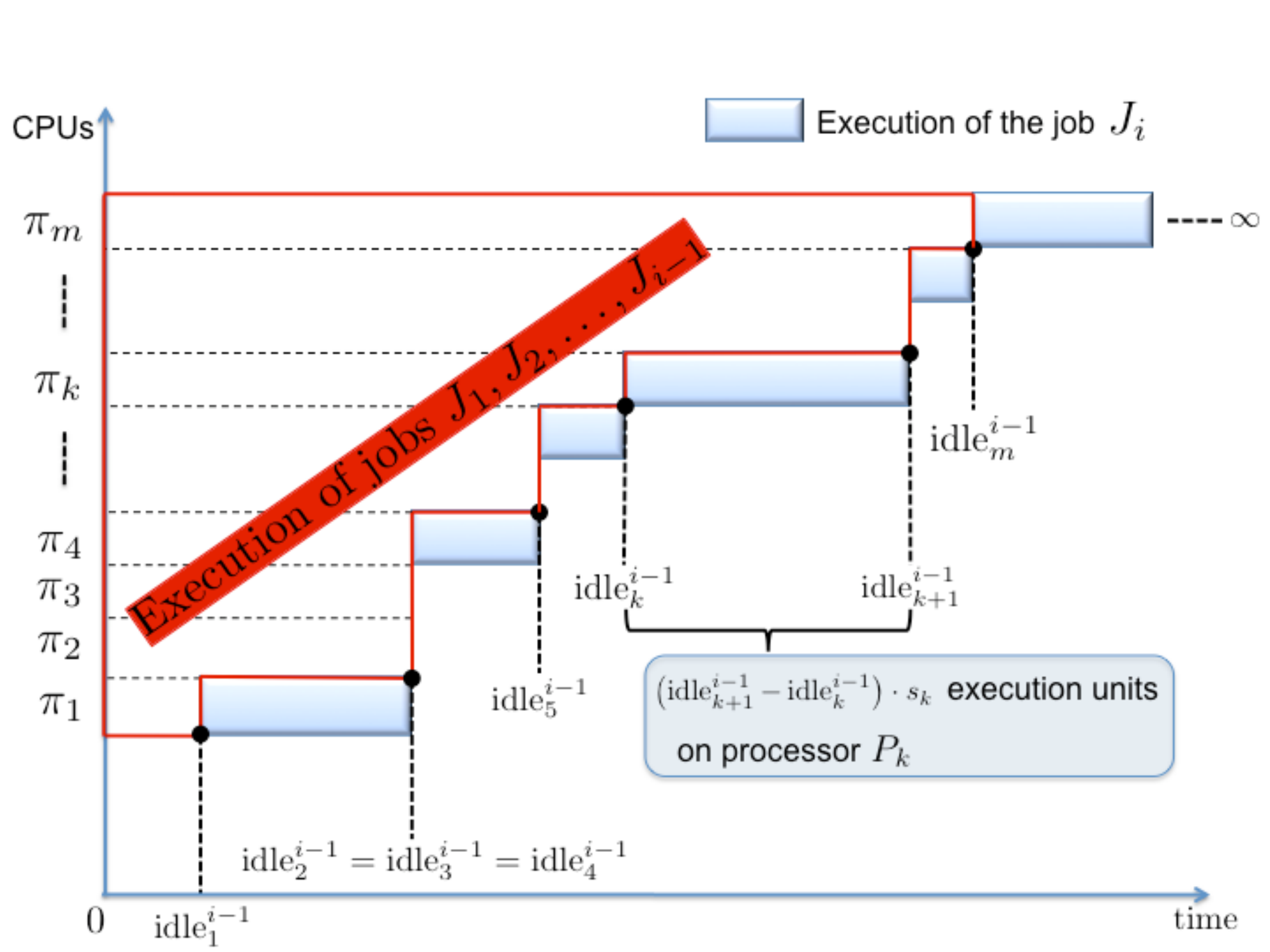}
\caption{Staircase defined by the $\idle{j}^{i-1}$}
\label{fig:Multimode:unif_FTP_idle_ki}
\end{figure}

\begin{Lemma}[See~\cite{MeumeuNelisGoossens:10}]
\label{lem:Multimode:unif_FTP_idle_ki}
Let $\pi = \left[ s_1, s_2, \ldots, s_m \right]$ denote any uniform multiprocessor platform composed of $m$ $\cpu$s and assume that $s_i \geq s_{i-1}$, $\forall i= 2, 3, \ldots, m$. Let $J = \{J_1, J_2, \ldots, J_n\}$ be any set of $n$ jobs, all released at time $t = 0$, with respective computation time $c_1, c_2, \ldots, c_n$. Let ${\cal S}$ denote any global, FTP and strongly work-conserving scheduler and suppose that $J$ is sorted by decreasing ${\cal S}$-priority, i.e., $J_{i} >_{\cal S} J_{i+1}$. If these jobs are scheduled by ${\cal S}$ upon $\pi$, then $\idle{j}^{i}$ is inductively defined as follows: \\

\noindent Initialization:
\begin{eqnarray}
\forall 1 \leq j \leq m: & \idle{j}^{0} \equals 0 \nonumber \\
\forall 1 \leq i \leq n: & \idle{m+1}^{i} \equals \infty \nonumber
\end{eqnarray}

\noindent Iteration:

for ($i = 1$ to $n$)

\hspace{0.2cm} for ($j = m$ to $1$) \\

\noindent  
\begin{equation}
\label{equ:Multimode:unif_FTP_idle_ki}
\idle{j}^{i} \equals 
\begin{cases}
\idle{j}^{i-1} \:\: \mbox{if} \:\: \idle{j}^{i-1} = \idle{j+1}^{i-1} \\
\idle{j+1}^{i-1} \:\: \mbox{else if} \:\: c_{i} \geq \sum_{k=1}^{j} (\idle{k+1}^{i-1} - \idle{k}^{i-1}) \cdot s_k \\
\idle{j}^{i-1} + \frac{ \left(c_{i} - \sum_{k=1}^{j-1} (\idle{k+1}^{i-1} - \idle{k}^{i-1})\cdot s_k \right)}{s_j} \:\: \mbox{otherwise} 
\end{cases}
\end{equation}
\end{Lemma} 
\begin{proof} 
Initially, the $m$ $\cpu$s idle and thus, $\idle{j}^{0} = 0$, $\forall j$, $1 \leq j \leq m$. We find convenient to define $\idle{m+1}^{i} \equals \infty, \forall i$, which means that we have at most $m$ $\cpu$s available. In the following, we prove the correctness of the value of $\idle{j}^{i}$ ($\forall j$, $1 \leq j \leq m$) assuming that $\idle{j}^{i-1}$ are defined ($\forall j \leq m+1$). The idle-instants $\idle{j}^{i-1}$ define a staircase as illustrated in Figure~\ref{fig:Multimode:unif_FTP_idle_ki} for the scheduling of jobs $J_{1}, \ldots, J_{i-1}$. Thus, job $J_i$ can only progress into the blue areas and two cases have to be distinguished:

\paragraph{Case 1} $\idle{j}^{i-1} = \idle{j+1}^{i-1}$, meaning that \emph{at least} one $\cpu$ faster than $\pi_j$ becomes available at time $\idle{j}^{i-1}$ (the blue area on $\cpu$ $\pi_j$ is void in that case). This situation is depicted in Figure~\ref{fig:Multimode:unif_FTP_idle_ki} where $\idle{2}^{i-1} = \idle{3}^{i-1} = \idle{4}^{i-1}$ and the blue area is void on $\cpu$s $\pi_2$ and $\pi_3$. In this kind of situation, the job $J_{i}$ is executed (if not completed) upon a faster $\cpu$ and the first instant at which at least $j$ $\cpu$s idle remains unchanged after having scheduled the job $J_i$, i.e., $\idle{j}^{i} = \idle{j}^{i-1}$.

\paragraph{Case 2} Otherwise, $J_{i}$ is dispatched to $\cpu$ $\pi_j$ at instant $\idle{j}^{i-1}$ and keeps executing on $\pi_j$ as long as (i) no faster $\cpu$s become idle or (ii) $J_{i}$ completes. In the first case, $J_i$ executes on $\pi_j$ until the next idle-instant $\idle{j+1}^{i-1}$, leading to the first sub-case $\idle{j}^{i} = \idle{j+1}^{i-1}$. In the second case, $J_i$ executes on $\cpu$ $\pi_j$ but completes before time $\idle{j+1}^{i-1}$. Thus, the idle-instant $\idle{j}^{i}$ is the instant $\idle{j}^{i-1}$ at which $J_i$ was dispatched to $\pi_j$ \emph{plus} its remaining processing time on $\cpu$ $\pi_j$ at time $\idle{j}^{i-1}$. Since $\sum_{k=1}^{j} (\idle{k}^{i-1} - \idle{k-1}^{i-1})\cdot s_{k}$ corresponds to the amount of work that $J_i$ has executed in the interval of time $\left[ 0, \idle{j}^{i-1} \right]$, its remaining processing time on $\cpu$ $\pi_j$ at time $\idle{j}^{i-1}$ is given by $\frac{c_i - \sum_{k=1}^{j} (\idle{k}^{i-1} - \idle{k-1}^{i-1})\cdot s_{k}}{s_j}$, leading to the second sub-case. 
\end{proof}

\begin{Corollary}[See~\cite{MeumeuNelisGoossens:10}]
\label{cor:Multimode:unif_FTP_maxidle}
The maximum idle-instant $\maxidle{k}(J, \pi, {\cal P})$ ($\forall k \in \left[ 1, m \right]$) is given by $\idle{k}^n$ computed as in Lemma~\ref{lem:Multimode:unif_FTP_idle_ki}. 
\end{Corollary}

\begin{Corollary}
\label{cor:Multimode:unif_FTP_makespan}
The maximum makespan $\maxmakespanUnifOne(J, \pi)$ is given by $\idle{m}^n$ computed as in Lemma~\ref{lem:Multimode:unif_FTP_idle_ki}.
\end{Corollary}

\subsection{Validity tests for $\SMMSO$ and $\AMMSO$}
\label{sec:Multimode:unif_FTP_validity_test}

From Corollary~\ref{cor:Multimode:unif_FTP_makespan}, a \emph{sufficient} validity test for the protocol $\SMMSO$ can therefore be formalized as follows. 

\begin{validity test}[$\SMMSO$, uniform and FTP]
\label{validitytest:Multimode:unif_FTP_SMMSO}
For any multi-mode real-time application $\tau$ and any identical platform $\pi$ composed of $m$ $\cpu$s, the protocol $\SMMSO$ is valid provided that, for every mode $\mode^i$,
\[ \idle{m}^{n_i}(\wcremjobs{i}, \pi, {\cal S}^i) \leq \min_{j \neq i} \left\{ \min_{k=1}^{n_j} \left\{ {\cal D}_k^j(\mode^i) \right\} \right\}\]
where $\idle{m}^{n_i}(\wcremjobs{i}, \pi, {\cal S}^i)$ is computed as $\idle{m}^{n_i}$ in Lemma~\ref{lem:Multimode:unif_FTP_idle_ki}, considering the critical rem-job set $\wcremjobs{i}$ composed of $n_i$ jobs $J_1, J_2, \ldots J_{n_i}$ of respective processing time $C_1^i, C_2^i, \ldots, C_{n_i}^i$ and such that $\wcremjobs{i}$ is sorted by decreasing ${\cal S}^i$-priority.
\end{validity test}

Similarly, the upper-bounds $\maxidle{k}(J, \pi, {\cal P})$ (where $1 \leq k \leq m$ and ${\cal P}$ corresponds to the job priority assignment of the old-mode scheduler ${\cal S}^i$) determined in Lemma~\ref{lem:Multimode:unif_FTP_idle_ki} can be used at line 10 of the validity algorithm of $\AMMSO$ (see Algorithm~\ref{algo:AMMSO_test} page~\pageref{algo:AMMSO_test}), as long as these upper-bounds are computed while assuming the critical rem-job set $\wcremjobs{i}$ for the transitions from every mode $\mode^i$. 
 
%%%%%%%%%%%%%%%%%%%%%%%%%%%%%%%%%%%%%%%%%%%%%%%%%%%%%%%%%%%%%%%%%%%%%%%%%%%%%%%%%%%%%%%%%%%%%%%%%%%%%%%%%%%%%%%%%%%%%%%
%%%%%%%%%%%%%%%%%%%%%%%%%%%%%%%%%%%%%%%%%%%%%%%%%%%%%%%%%%%%%%%%%%%%%%%%%%%%%%%%%%%%%%%%%%%%%%%%%%%%%%%%%%%%%%%%%%%%%%%
%%%%%%%%%%%%%%%%%%%%%%%%%%%%%%%%%%%%%%%%%%%%%%%%%%%%%%%%%%%%%%%%%%%%%%%%%%%%%%%%%%%%%%%%%%%%%%%%%%%%%%%%%%%%%%%%%%%%%%%
%%%%%%%%%%%%%%%%%%%%%%%%%%%%%%%%%%%%%%%%%%%%%%%%%%%%%%%%%%%%%%%%%%%%%%%%%%%%%%%%%%%%%%%%%%%%%%%%%%%%%%%%%%%%%%%%%%%%%%%
%%%%%%%%%%%%%%%%%%%%%%%%%%%%%%%%%%%%%%%%%%%%%%%%%%%%%%%%%%%%%%%%%%%%%%%%%%%%%%%%%%%%%%%%%%%%%%%%%%%%%%%%%%%%%%%%%%%%%%%

\section{Conclusion and open problems}
\label{sec:Multimode:Conclusion}

In this paper, we addressed the scheduling problem of multi-mode real-time applications upon identical and uniform multiprocessor platforms. We assumed that every mode of the application was scheduled by following a global and Fixed-Task-Priority or Fixed-Job-Priority scheduler. Under these assumptions, we proposed two protocols for managing every transition between every pair of modes of the system, namely $\SMMSO$ and $\AMMSO$. For both protocols, we established validity tests that allow the system designer to predict whether the given application can meet all the expected timing requirements upon the given platform. We prove the correctness of our schedulability analyses by extending the theory about the makespan determination problem. 

In our future work, we aim at taking into account mode-independent tasks, i.e., tasks whose the periodic (or sporadic) activation pattern is not affected by the mode changes. Moreover, instead of scheduling the rem-jobs by using the scheduler of the old-mode during the transitions, it could be better, in term of the enablement delays applied to the new-mode tasks, to propose a \emph{dedicated priority assignment} which meets the deadline of every rem-job, \emph{while} minimizing the makespan. To the best of our knowledge, the problem of minimizing the makespan while meeting job deadlines is not yet addressed in the literature and remains open.  Table~\ref{tab:open_problem} outlines a brief overview of all different problems, considering the task and platform model introduced in this paper. For each problem, we indicated either the reference(s) where solutions have been proposed or T.W. (\textbf{T}his \textbf{W}ork) or F.W. (\textbf{F}uture \textbf{W}ork) or O.P. (\textbf{O}pen \textbf{P}roblem).

\begin{table}[h!]
\centering
\begin{tabular}{| c | c | c | c |}
\hline
\multicolumn{4}{| c |}{Protocols without periodicity} \\
\hline
Protocol & Platform & Scheduler & Existing results \\
\hline
Synchronous & identical & FJP & \cite{NelisGoossens:08, NelisGoossensAndersson:09, Nelis:10}, T.W. \\
\hline
Synchronous & identical & FTP &  \cite{NelisGoossens:08, NelisGoossensAndersson:09, Nelis:10}, T.W. \\
\hline
Synchronous & uniform & FJP & \cite{MeumeuNelisGoossens:10, Nelis:10}, T.W. \\
\hline
Synchronous & uniform & FTP & \cite{MeumeuNelisGoossens:10, Nelis:10}, T.W. \\
\hline
Asynchronous & identical & FJP &  \cite{NelisGoossensAndersson:09, Nelis:10}, T.W.\\
\hline
Asynchronous & identical & FTP & \cite{NelisGoossensAndersson:09, Nelis:10}, T.W.\\
\hline
Asynchronous & uniform & FJP & \cite{MeumeuNelisGoossens:10, Nelis:10}, T.W. \\
\hline
Asynchronous & uniform & FTP &  \cite{MeumeuNelisGoossens:10, Nelis:10}, T.W. \\
\hline
\hline
\multicolumn{4}{| c |}{Protocols with periodicity} \\
\hline
Protocol & Platform & Scheduler & Existing results \\
\hline
Synchronous & identical & FJP &  \cite{NelisAnderssonGoossens:09}, F.W. \\
\hline
Synchronous & identical & FTP &  \cite{NelisAnderssonGoossens:09}, F.W. \\
\hline
Synchronous & uniform & FJP &  F.W. \\
\hline
Synchronous & uniform & FTP &  F.W. \\
\hline
Asynchronous & identical & FJP &  O.P. \\
\hline
Asynchronous & identical & FTP &  O.P. \\
\hline
Asynchronous & uniform & FJP & O.P. \\
\hline
Asynchronous & uniform & FTP & O.P. \\
\hline
\end{tabular}
\caption{State-of-the-art at a glance.}
\label{tab:open_problem}
\end{table}

% if have a single appendix:
%\appendix[Proof of the Zonklar Equations]
% or
%\appendix  % for no appendix heading
% do not use \section anymore after \appendix, only \section*
% is possibly needed

% use appendices with more than one appendix
% then use \section to start each appendix
% you must declare a \section before using any
% \subsection or using \label (\appendices by itself
% starts a section numbered zero.)
%

%\appendices
%\section{Proof of the First Zonklar Equation}
%Appendix one text goes here.

% you can choose not to have a title for an appendix
% if you want by leaving the argument blank
%\section{}
%Appendix two text goes here.

% use section* for acknowledgement
%\section*{Acknowledgment}

%The authors would like to thank...

% trigger a \newpage just before the given reference
% number - used to balance the columns on the last page
% adjust value as needed - may need to be readjusted if
% the document is modified later
%\IEEEtriggeratref{8}
% The "triggered" command can be changed if desired:
%\IEEEtriggercmd{\enlargethispage{-5in}}

% references section

% can use a bibliography generated by BibTeX as a .bbl file
% BibTeX documentation can be easily obtained at:
% http://www.ctan.org/tex-archive/biblio/bibtex/contrib/doc/
% The IEEEtran BibTeX style support page is at:
% http://www.michaelshell.org/tex/ieeetran/bibtex/
\bibliographystyle{acm}
% argument is your BibTeX string definitions and bibliography database(s)
\bibliography{biblio}

\begin{thebibliography}{10}

\bibitem{Andersson:03}
{\sc Andersson, B.}
\newblock {\em Static-priority scheduling on multiprocessors}.
\newblock PhD thesis, Chalmers University of Technology, 2003.

\bibitem{Andersson:08}
{\sc Andersson, B.}
\newblock Uniprocessor {EDF} scheduling with mode change.
\newblock In {\em Proceedings of the 12th International Conference on
  Principles of Distributed Systems\/} (Berlin, Heidelberg, 2008), OPODIS'08,
  Springer-Verlag, pp.~572--577.

\bibitem{Bailey:93}
{\sc Bailey, C.~M.}
\newblock Hard real-time operating system kernel. investigation of mode change.
\newblock Tech. rep., Task 14 Deliverable on {ESTSEC} Contract 9198/90/NL/SF,
  British Aerospace Systems Ltd., 1993.

\bibitem{Baker:03}
{\sc Baker, T.~P.}
\newblock Multiprocessor {EDF} and deadline monotonic schedulability analysis.
\newblock In {\em Proceedings of the 24th IEEE International Real-Time Systems
  Symposium (RTSS '03)\/} (Washington, DC, USA, 2003), IEEE Computer Society,
  pp.~120--129.

\bibitem{BakerBaruah:09}
{\sc Baker, T.~P., and Baruah, S.~K.}
\newblock An analysis of global {EDF} schedulability for arbitrary-deadline
  sporadic task systems.
\newblock {\em Real-Time Systems 43}, 1 (2009), 3--24.

\bibitem{Baker:07}
{\sc Baker, T.~P., and Cirinei, M.}
\newblock Brute-force determination of multiprocessor schedulability for sets
  of sporadic hard-deadline tasks.
\newblock In {\em Proceedings of the 11th international conference on
  Principles of distributed systems\/} (Berlin, Heidelberg, 2007), OPODIS'07,
  Springer-Verlag, pp.~62--75.

\bibitem{Baruah:10}
{\sc Baruah, S.}
\newblock An improved global {EDF} schedulability test for uniform
  multiprocessors.
\newblock In {\em Proceedings of the 16th IEEE Real-Time and Embedded
  Technology and Applications Symposium (RTAS '10)\/} (Los Alamitos, CA, USA,
  2010), IEEE Computer Society, pp.~184--192.

\bibitem{Baruah:03}
{\sc Baruah, S., and Anderson, J.}
\newblock Energy-aware implementation of hard-real-time systems upon
  multiprocessor platform.
\newblock In {\em Proceedings of the 16th International Conference on Parallel
  and Distributed Computing Systems\/} (August 2003), pp.~430--435.

\bibitem{Baruah:04}
{\sc Baruah, S., and Anderson, J.}
\newblock Energy-efficient synthesis of periodic task systems upon identical
  multiprocessor platforms.
\newblock In {\em Proceedings of the Twenty-Fourth International Conference on
  Distributed Computing Systems\/} (Tokyo, Japan, March 2004), IEEE Computer
  Society Press, pp.~428--435.

\bibitem{BaruahBaker:08}
{\sc Baruah, S., and Baker, T.}
\newblock Global {EDF} schedulability analysis of arbitrary sporadic task
  systems.
\newblock In {\em Proceedings of the 2008 Euromicro Conference on Real-Time
  Systems (ECRTS '08)\/} (Washington, DC, USA, 2008), IEEE Computer Society,
  pp.~3--12.

\bibitem{BaruahFisher:07}
{\sc Baruah, S., and Fisher, N.}
\newblock Global deadline-monotonic scheduling of arbitrary-deadline sporadic
  task systems.
\newblock In {\em Proceedings of the 11th international conference on
  Principles of distributed systems (OPODIS '07)\/} (Guadeloupe, French West
  Indies, 2007), Springer-Verlag, pp.~204--216.

\bibitem{BaruahGoossens:08:2}
{\sc Baruah, S., and Goossens, J.}
\newblock Deadline monotonic scheduling on uniform multiprocessors.
\newblock In {\em Proceedings of the 12th International Conference on
  Principles of Distributed Systems (OPODIS '08)\/} (Berlin, Heidelberg, 2008),
  Springer-Verlag, pp.~89--104.

\bibitem{BaruahGoossens:08}
{\sc Baruah, S., and Goossens, J.}
\newblock The {EDF} scheduling of sporadic task systems on uniform
  multiprocessors.
\newblock In {\em Proceedings of the 2008 Real-Time Systems Symposium (RTSS
  '08)\/} (Washington, DC, USA, 2008), IEEE Computer Society, pp.~367--374.

\bibitem{BaruahGoossens:03}
{\sc Baruah, S.~K., and Goossens, J.}
\newblock Rate-monotonic scheduling on uniform multiprocessors.
\newblock {\em IEEE Transactions on Computers 52}, 7 (2003), 966--970.

\bibitem{IainBate:98}
{\sc Bate, I.~J.}
\newblock {\em Scheduling and Timing Analysis for Safety Critical Real-Time
  Systems}.
\newblock PhD thesis, University of York, November 1998.

\bibitem{BertCiriLipari:05}
{\sc Bertogna, M., Cirinei, M., and Lipari, G.}
\newblock Improved schedulability analysis of {EDF} on multiprocessor
  platforms.
\newblock In {\em Proceedings of the 17th Euromicro Conference on Real-Time
  Systems (ECRTS '05)\/} (Washington, DC, USA, 2005), IEEE Computer Society,
  pp.~209--218.

\bibitem{CucuGoossens:10}
{\sc Cucu-Grosjean, L., and Goossens, J.}
\newblock Predictability of fixed-job priority schedulers on heterogeneous
  multiprocessor real-time systems.
\newblock {\em Information Processing Letters 110}, 10 (2010), 399--402.

\bibitem{FunkGoossensBaruah:01}
{\sc Funk, S., Goossens, J., and Baruah, S.}
\newblock On-line scheduling on uniform multiprocessors.
\newblock In {\em Proceedings of the 22nd IEEE Real-Time Systems Symposium
  (RTSS '01)\/} (Washington, DC, USA, 2001), IEEE Computer Society,
  pp.~183--192.

\bibitem{Garey:90}
{\sc Garey, M.~R., and Johnson, D.~S.}
\newblock {\em Computers and Intractability; A Guide to the Theory of
  NP-Completeness}.
\newblock W. H. Freeman \& Co., New York, NY, USA, 1990.

\bibitem{GoossensFunkBaruah:02}
{\sc Goossens, J., Funk, S., and Baruah, S.}
\newblock Real-time scheduling on uniform multiprocessors.
\newblock In {\em Proceedings of the 10th international conference on real-time
  systems\/} (Paris France, March 2002), In Teknea, editors, pp.~189--204.

\bibitem{Goyal:05}
{\sc Goyal, S.~V.}
\newblock 15-854: {Approximations Algorithms, Lecturer: R. Ravi, Topic: Greedy
  Algorithms: Minimizing Makespan, Multiway Cut}, September 2005.

\bibitem{Ha:95}
{\sc Ha, R.}
\newblock {\em Validating Timing Constraints in Multiprocessor and Distributed
  Systems}.
\newblock PhD thesis, Department of Computer Science, University of Illinois at
  Urbana-Champaign, 1995.

\bibitem{HaLiu:93}
{\sc Ha, R., and Liu, J.~W.}
\newblock Validating timing constraints in multiprocessor and distributed
  real-time systems.
\newblock Tech. rep., Department of Computer Science, University of Illinois at
  Urbana-Champaign, Champaign, IL, USA, 1993.

\bibitem{HaLiu:94}
{\sc Ha, R., and Liu, J. W.~S.}
\newblock Validating timing constraints in multiprocessor and distributed
  real-time systems.
\newblock In {\em Proceedings of the 14th IEEE International Conference on
  Distributed Computing Systems\/} (Los Alamitos, CA, USA, 1994), IEEE Computer
  Society Press, pp.~162--171.

\bibitem{Henia:07}
{\sc Henia, R., and Ernst, R.}
\newblock Scenario aware analysis for complex event models and distributed
  systems.
\newblock In {\em Proceedings of the 28th IEEE International Real-Time Systems
  Symposium (RTSS '07)\/} (Washington, DC, USA, 2007), IEEE Computer Society,
  pp.~171--180.

\bibitem{Liu:73}
{\sc Liu, C.~L., and Layland, J.~W.}
\newblock Scheduling algorithms for multiprogramming in a hard-real-time
  environment.
\newblock {\em Journal of ACM 20}, 1 (1973), 46---61.

\bibitem{MeumeuNelisGoossens:10}
{\sc {Meumeu~Yomsi}, P., Nelis, V., and Goossens, J.}
\newblock Scheduling multi-mode real-time systems upon uniform multiprocessor
  platforms.
\newblock In {\em The 15th IEEE International Conference on Emerging
  Technologies and Factory Automation\/} (2010), IEEE Computer Society Press.

\bibitem{Nelis:10}
{\sc Nelis, V.}
\newblock {\em Energy-Aware Real-Time Scheduling in Multiprocessor Embedded
  Systems}.
\newblock PhD thesis, Universit\'e Libre de Bruxelles, 2010.

\bibitem{NelisAnderssonGoossens:09}
{\sc Nelis, V., Andersson, B., and Goossens, J.}
\newblock A synchronous transition protocol with periodicity for global
  scheduling of multimode real-time systems on multiprocessors.
\newblock In {\em The 30th IEEE Real-Time Systems Symposium (RTSS)
  Work-in-progress session\/} (Washington D.C. USA, December 2009), D.~Zhu,
  Ed., pp.~13--16.

\bibitem{NelisGoossens:08}
{\sc Nelis, V., and Goossens, J.}
\newblock Mode change protocol for multi-mode real-time systems upon identical
  multiprocessors.
\newblock In {\em Proceedings of the 29th IEEE Real-Time Systems Symposium
  (Work in Progress session - RTSS08-WiP)\/} (Barcelona Spain, December 2008),
  pp.~9--12.

\bibitem{NelisGoossensAndersson:09}
{\sc Nelis, V., Goossens, J., and Andersson, B.}
\newblock Two protocols for scheduling multi-mode real-time systems upon
  identical multiprocessor platforms.
\newblock In {\em 21st Euromicro Conference on Real-Time Systems (ECRTS'09)\/}
  (Dublin Ireland, July 2009), IEEE Computer Society, pp.~151--160.

\bibitem{Pedro:99}
{\sc Pedro, P.}
\newblock {\em Schedulability of mode changes in flexible real-time distributed
  systems}.
\newblock PhD thesis, University of York, Department of Computer Science, 1999.

\bibitem{Pedro:98}
{\sc Pedro, P., and Burns, A.}
\newblock Schedulability analysis for mode changes in flexible real-time
  systems.
\newblock In {\em Proceedings of the 10th Euromicro Workshop on Real-Time
  Systems\/} (Los Alamitos, CA, USA, 1998), IEEE Computer Society,
  pp.~172--179.

\bibitem{JoAlfons:04}
{\sc Real, J., and Crespo, A.}
\newblock Mode change protocols for real-time systems: A survey and a new
  proposal.
\newblock {\em Real-Time Systems 26}, 2 (2004), 161--197.

\bibitem{Sha:89}
{\sc Sha, L., Rajkumar, R., Lehoczky, J., and Ramamritham, K.}
\newblock Mode change protocols for priority-driven preemptive scheduling.
\newblock {\em Real-Time Systems 1\/} (1989), 243--264.

\bibitem{Sha:88}
{\sc Sha, L., Sha, L., Rajkumar, R., Rajkumar, R., Lehoczky, J., Lehoczky, J.,
  Ramamritham, K., and Ramamritham, K.}
\newblock Mode change protocols for priority-driven preemptive scheduling.
\newblock {\em Real-Time Systems 1\/} (1988), 243--264.

\bibitem{Stoimenov:09}
{\sc Stoimenov, N., Perathoner, S., and Thiele, L.}
\newblock Reliable mode changes in real-time systems with fixed priority or
  {EDF} scheduling.
\newblock In {\em Proceedings of the Conference on Design, Automation and Test
  in Europe (DATE '09)\/} (2009), pp.~99--104.

\bibitem{Tindell:96}
{\sc Tindell, K., and Alonso, A.}
\newblock A very simple protocol for mode changes in priority preemptive
  systems.
\newblock Tech. rep., Universidad Polit\'ecnica de Madrid, 1996.

\bibitem{Tindell:92}
{\sc Tindell, K., Burns, A., and Wellings, A.~J.}
\newblock Mode changes in priority pre-emptively scheduled systems.
\newblock In {\em Proceedings of the 13th Real-Time Systems Symposium\/}
  (Phoenix, Arizona, 1992), pp.~100--109.

\end{thebibliography}
\end{document}